\documentclass[10pt,a4paper]{amsart}

\usepackage{amsthm,amssymb,amsmath,epic,eepic,float}
\usepackage{rotating,epsfig,indentfirst,array,varioref}

\usepackage{color}

\def\michaelwrites#1{{\color{black} #1}}
\def\m{\michaelwrites}

\def\ll{\left\lgroup}
\def\rr{\right\rgroup}

\def\leq{\leqslant}
\def\geq{\geqslant}

\def\x{x^{(1)}}
\def\xx{x^{(2)}}
\def\b{b^{(1)}}
\def\bb{b^{(2)}}

\newtheorem{myLemma}{Lemma}
\newcommand{\myProof}{\noindent \emph{Proof}.\ }

\newcommand{\N}{\mathcal{N}}

\def\bref{\bf\ref}

\def\proofend{\hfill$\square$} 

\restylefloat{figure}

\newcommand{\cG}{\mathcal{G}}

\newcommand{\cK}{\mathcal{K}}

\newcommand{\cS}{\mathcal{S}}

\def\ll{ \left\lgroup}
\def\rr{\right\rgroup}

\newcommand{\1}{{\bf 1.}}
\newcommand{\2}{{\bf 2.}}

\def\det{\operatorname{det}}

\begin{document}

\title{Colour-independent partition functions in coloured vertex models}

\author[O Foda and M Wheeler]{O Foda \!$^1$ and M Wheeler \!$^2$}

\address{
\!\!\!\!\!\!\!$^1$ Dept of Mathematics and Statistics,
University of Melbourne,
Parkville, VIC 3010, Australia
\newline
$^2$ Laboratoire de Physique Th\'eorique et Hautes Energies,
CNRS UMR 7589 and Universit\'e Pierre et Marie Curie - Paris 6,
4 place Jussieu, 75252 Paris cedex 05, France
}

\email{omar.foda@unimelb.edu.au, mwheeler@lpthe.jussieu.fr}


\begin{abstract}
We study lattice configurations related to $\cS_n$, the scalar 
product of an off-shell state and an on-shell state in rational 
$A_n$ integrable vertex models, $n \in \{1, 2\}$. 
The lattice lines are colourless and oriented. The state variables 
are $n$ conserved colours that flow along the line orientations, 
but do not necessarily cover every bond in the lattice. 

Choosing boundary conditions such that the positions 
where the colours flow into the lattice are fixed, and
where they        flow out              are summed over, 
we show that the partition functions of these configurations, 
with these boundary conditions, are $n$-independent. 
Our results extend to trigonometric $A_n$ models, and to 
all $n$.

This $n$-independence explains, in vertex-model terms, results from 
recent studies of $\cS_2$ \cite{caetano.su3, wheeler.su3}. Namely, 
{\bf 1.} $\cS_2$,
which depends on two sets of Bethe roots, $\{b_1\}$ and $\{b_2\}$, 
and cannot (as far as we know) be expressed in single determinant 
form, degenerates in the limit 
$\{b_1\} \rightarrow \infty$, and/or 
$\{b_2\} \rightarrow \infty$, 
into a product of determinants,
{\bf 2.} Each of the latter determinants is an $A_1$ vertex-model 
partition function.  
\end{abstract}
\maketitle
\setcounter{section}{0}

\section{Introduction and Motivation}
\label{introduction}

The subject of this paper is integrable periodic spin chains and 
vertex models with rational and trigonometric $A_n$ $R$-matrices 
$n \in \{1, 2, \dots\}$. We focus on the rational models, 
and on $n \in \{1, 2\}$, to simplify the presentation and the 
proofs but, as we will explain, our conclusions extend without 
modification to the trigonometric models, and to all $n$. 

We wish to show that there are $A_n$-model partition functions 
that are independent of the number of colours $n$. In 
other words, if one considers a set of these $A_n$ configurations, 
regards the $n$ colours as identical, and re-evaluates the partition 
function as if these were $A_1$ configurations, the result would
be the same. 

\subsection{Two types of \michaelwrites{Bethe} states} 
There are two types of \michaelwrites{Bethe} states in the space of states of 
an integrable $A_n$ rational or trigonometric spin chain 
or vertex model \cite{korepin.book.1, korepin.book.2, baxter.book}. 
{\bf 1.} Off-shell \michaelwrites{Bethe} states 
$\{ \alpha \}$
which are characterized by rapidity variables that are free, 
and are not eigenstates of the transfer matrix.
{\bf 2.} On-shell \michaelwrites{Bethe} states 
$\{ \beta \}$
which are characterized by rapidity variables that satisfy Bethe 
equations, and are eigenstates of the transfer matrix. 

\subsection{Three types of scalar products} 
There are three types of scalar products \michaelwrites{between Bethe} states. 
{\bf 1.} The off-shell/off-shell scalar product 
${\cK}_n (\alpha_i, \alpha_j)$ $=$ 
$\langle \alpha_i | \alpha_j \rangle$.
A sum expression was obtained in \cite{korepin} for ${\cK}_1$,
and in \cite{reshetikhin} for ${\cK}_2$.
${\cK}_n$ cannot be expressed in single determinant form.
{\bf 2.} The off-shell/on-shell scalar product
${\cS}_n (\alpha_i, \beta_j)$ 
$=$
$\langle \alpha_i | \beta_j \rangle$.
${\cS}_1$ was evaluated in determinant form in \cite{slavnov}.
A second determinant expression was obtained in \cite{kostov.matsuo}, 
and a third was obtained in \cite{foda.wheeler.variations}.
{\bf 3.} The on-shell/on-shell scalar product 
${\cG}_n (\beta_i, \beta_j)$
$=$
$\langle \beta_i | 
         \beta_j \rangle$,
which vanishes for $i \neq j$, since the on-shell \michaelwrites{Bethe} states 
\michaelwrites{are orthogonal} \cite{gaudin.book, korepin.book.1},
and gives the (square of the) norm of $| \beta_i \rangle$,  
${\cG}_n (\beta_i, \beta_i)$
$=$
$\langle \beta_i |
         \beta_i \rangle$, for $i\!=\!j$. 
A determinant expression was obtained 
in \cite{gaudin, gaudin.book} for ${\cG}_1$,
in \cite{reshetikhin} for ${\cG}_2$, 
and conjectured 
in \cite{E1} for all ${\cG}_n$.

\subsection{Structure constants of operators in $A_1$ scalar 
sub-sectors of SYM$_4$}
Building on applications of integrability in supersymmetric 
Yang-Mills theories 
\footnote{
For a comprehensive overview of integrability in Yang-Mills 
theories with particular emphasis on the AdS/CFT correspondence, 
see \cite{beisert.review}. 
For a more compact review, see \cite{serban.review}.},
the off-shell/off-shell scalar product ${\cK}_1$, as well 
as the norm ${\cG}_1$, were used in \cite{E1, E2} to compute 
the tree-level structure constants 
$\mathcal{C}^{(0)}_{ijk}$ of 3-point functions of gauge-invariant 
local composite operators in the $A_1$ scalar 
sub-sectors of planar $\mathcal{N}\!=\!4$ supersymmetric Yang-Mills 
theory, SYM$_4$. In these computations, all three operators involved 
correspond to non-BPS states.
In \cite{GSV}, the semi-classical limit of $\mathcal{C}^{(0)}_{ijk}$ 
was computed in the case where two of the operators 
involved correspond to BPS states. 

In \cite{foda}, the off-shell/on-shell scalar product ${\cS}_1$, 
as well as the norm ${\cG}_1$, were used to express 
$\mathcal{C}^{(0)}_{ijk}$, with three non-BPS operators, 
in determinant form. 
The result of \cite{foda} was extended to $A_1$ sub-sectors in 
planar Yang-Mills theories with fewer supersymmetries in 
\cite{foda.wheeler.jimbo.fest}, as well as planar QCD in 
\cite{ahn.foda.nepomechie}.
In \cite{foda.wheeler.jimbo.fest}, we further showed that 
$\mathcal{C}^{(0)}_{ijk}$ is a discrete KP $\tau$-function in 
the free auxiliary rapidities that characterize an off-shell 
spin-chain state used to represent one of the operators in 
the 3-point function.

Determinant expressions are ideally suited to numerical evaluations 
as well as to computing the asymptotics of the quantities that they 
represent. 
In \cite{kostov.short.paper, kostov.long.paper}, Kostov wrote 
the determinant expression of $\mathcal{C}^{(0)}_{ijk}$, with 
three non-BPS states, in terms of free fermions, and from that 
obtained its semi-classical limit. 

In \cite{gromov.vieira.theta}, Gromov and Vieira showed that given 
$\mathcal{C}^{(0)}_{ijk}$ in spin-chain terms and in the presence 
of inhomogeneous quantum rapidities, the 1-loop (and possibly 2-loop) 
corrections are obtained by applying a certain differential operator 
in the quantum rapidities. In \cite{serban}, Serban discussed 
an extension of this statement to all loops. 
In \cite{foda.wheeler.variations}, we showed that the 1-loop-corrected 
version of $\mathcal{C}^{(0)}_{ijk}$ of \cite{gromov.vieira.theta} can 
be put in determinant form.

\subsection{Structure constants of operators in $A_2$ scalar sub-sectors 
of SYM$_4$}
Following the developments in computing $A_1$ scalar sub-sector structure 
constants outlined above, it is natural to look for analogous results in 
the $A_2$ scalar sub-sectors as the next step towards evaluating SYM$_4$ 
structure constants in all generality. However, while there is 
a sum expression for $\cK_2$ due to Reshetikhin \cite{reshetikhin}, 
there is no determinant expression for $\cS_2$. 
Moreover, recent results suggest that no such determinant form exists 
\cite{belliard.1, belliard.2}.

\subsection{Degenerate $A_2$ scalar products and colour independence}
There are determinant expressions for degenerations of $\cS_2$ 
obtained by taking some or all of the Bethe roots to infinity
\cite{caetano.su3, wheeler.su3}. 
This is surprising, not only because these expressions are products 
of determinants, but also because each of these determinants is an 
$A_1$ vertex model partition function.

In combinatorial terms, one starts with $\cS_2$ and regards that as 
a partition function of configurations in three state variables. 
Degenerating $\cS_2$ by taking one set or both sets of Bethe roots 
to infinity, one obtains what can be thought of as the product of
partition functions of configurations of two, rather than three state 
variables. If the initial state variables are 
$\{$\textit{white}, \textit{black}, \textit{blue}$\}$, 
the final state variables are 
$\{$\textit{white}, \textit{black}$\}$ and all dependence on the colour 
blue has disappeared.

\subsection{Aim of this work} 
We wish to show that the results of \cite{caetano.su3, wheeler.su3} 
follow from the fact that there are $A_2$, or more generally $A_n$ 
($n = 2, 3, 4, \dots$) vertex-model lattice configurations whose 
partition functions are $n$-independent, and from that we show that 
\1 The partition function corresponding to $\cS_2$ factorizes, and 
\2 The factors are $A_1$ expressions that can be evaluated 
in determinant form.

\subsection{Outline of contents}
In Section {\bf \ref{section.vertex.model}}, 
we recall the basics of rational and trigonometric $A_n$ vertex models. 
In {\bf \ref{section.dwpf}} and {\bf \ref{section.scalar.product}}, 
we recall the $A_1$ domain wall and scalar product configurations. 
In {\bf \ref{section.colouring.00}}, we introduce an extra colour 
in $A_1$ configurations, thereby turning them into $A_2$ configurations. 
This can be further extended to obtain $A_n$ configurations.
In 
{\bf \ref{section.colouring.01}} and 
{\bf \ref{section.colouring.02}} we study the \lq coloured\rq\ version 
of the $A_1$ domain wall and scalar product, which are $A_2$ configurations. 
In {\bf \ref{section.applications}}, we apply the results obtained 
in previous sections to $A_2$ scalar product configurations. Section 
{\bf \ref{section.comments}} contains various remarks.
An appendix contains various technical details.

\section{The $A_n$ vertex model}
\label{section.vertex.model}

\subsection{Rapidity variables, lattice lines and coloured state 
variables}

$A_n$ vertex model partition functions depend on up to $2n$ 
sets of auxiliary rapidities, and $n$ sets of quantum rapidities. 
The auxiliary rapidities may satisfy Bethe equations, so they are 
\lq Bethe roots\rq, or they are free. 
The quantum rapidities, or \lq inhomogeneities\rq, are free. 
We work in the inhomogeneous setting where the quantum rapidities 
are not necessarily equal.
We use $\{\alpha\}$ for the set of elements $\alpha_1, \alpha_2, \dots$, 
and denote its cardinality by $|\alpha|$.

All lattice lines are oriented, and the rapidities can be viewed as 
flowing along the lattice lines in the same direction as the line
orientations. The state variables of the $A_n$ model, $\iota$, take 
values in the set $\{0, 1, \dots, n\}$. We shall think of the state 
variable assignments $\iota \in \{1, 2, \dots, n\}$ as \emph{colours} 
that flow in the same directions as the rapidities along the line 
orientations, and of the assignment $\iota = 0$ as a colourless, 
or \emph{white} background. 
Among the colours $\{1,2, \dots, n\}$ we distinguish $\iota = 1$ 
as \emph{black}. With this definition, $A_1$ model configurations 
consist of only black and white. $A_{n\geq 2}$ model configurations 
are genuinely coloured. The aim of this work is to show that certain 
coloured configurations in $A_{n \geq 2}$ models are 
equivalent to black and white ones in the $A_1$ model.

For $n=1$, the auxiliary rapidities flow in horizontal lines from 
left to right and the quantum rapidities flow in vertical lines 
from bottom to top. For $n= 2, 3, \dots$, the rapidities flow in 
well-defined directions that are indicated on the diagrams.

We use $\{x^{(i)} \}$, $i \in \{1, \dots, n\}$, 
for the $i$-th set of auxiliary rapidities that are always free,  
and define $|x| = |x^{(1)}| + \cdots + |x^{(n)}|$.
We use $\{b^{(i)} \}$, $i \in \{1, \dots, n\}$, 
for the $i$-th set of auxiliary rapidities that are sometimes assumed to 
satisfy Bethe equations (depending on context), and 
define $|b|=|b^{(1)}| + \cdots + |b^{(n)}|$.
In fact the only time we require the $\{b^{(i)}\}$ to be Bethe roots is when we 
seek a determinant evaluation of a partition function that depends on them. 
The rest of the time they can be considered free.
We use $\{y\}$ for the quantum rapidities, or inhomogeneities, 
so that $|y| = L$, where $L$ is the length of the spin chain. 
Finally, we use the notation

\begin{align}
\Delta\{x\}_N = \prod_{1\leq i<j \leq N} (x_j-x_i),
\ \ 
\Delta\{-x\}_N = \prod_{1 \leq i<j \leq N} (x_i-x_j)
\end{align}
for the Vandermonde determinant in the set 
$\{x\} = \{x_1,\dots,x_N\}$. 

\subsection{$R$-matrices and vertex weights}

\begin{figure}
\begin{center}
\begin{minipage}{4.3in}
\setlength{\unitlength}{0.00040cm}

\begin{picture}(20000,8000)(-14000,-4000)
\put(-12000,-200){
$[R_{\alpha\beta}(x,y)]^{i_{\alpha} j_{\alpha}}_{i_{\beta} j_{\beta}} =$}
\path(-2000,0000)(2000,0000)
\put(-2800,250){\tiny{$i_{\alpha}$}}
\put(2200,250){\tiny{$j_{\alpha}$}}
\put(-3750,0){\scriptsize{$x$}}
\blacken\path(-1250,250)(-1250,-250)(-750,000)(-1250,250)
\path(0000,-2000)(0000,2000)
\put(0000,-2600){\tiny{$i_{\beta}$}}
\put(0000,2450){\tiny{$j_{\beta}$}}
\put(0,-3650){\scriptsize{$y$}}
\blacken\path(250,-1250)(-250,-1250)(000,-750)(250,-1250)
\end{picture}
\end{minipage}
\end{center}

\caption{
In 
$[R_{\alpha\beta}(x,y)]^{i_{\alpha} j_{\alpha}}_{i_{\beta} j_{\beta}}$, 
the $(i_{\alpha},j_{\alpha})$-th component of the $R$-matrix 
acts on $V_{\alpha}$, and 
the $(i_{\beta},j_{\beta})$-th component of the $R$-matrix 
acts on $V_{\beta}$. 
We associate 
$[R_{\alpha\beta}(x,y)]^{i_{\alpha} j_{\alpha}}_{i_{\beta} j_{\beta}}$ 
with the vertex shown.} 
\label{fig-vert}
\end{figure}
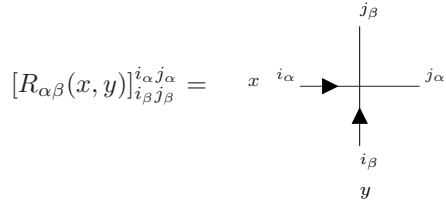

We are interested in solutions of the Yang-Baxter equation

\begin{align}
R_{\alpha\beta}(x,y) R_{\alpha\gamma}(x,z) R_{\beta\gamma}(y,z)
=
R_{\beta\gamma}(y,z) R_{\alpha\gamma}(x,z) R_{\alpha\beta}(x,y)
\label{qyb}
\end{align}
that are based on $A_n$ algebras. These are $(n+1)^2 \times (n+1)^2$ 
$R$-matrices of the form
\begin{multline}
\label{Rmat}
R_{\alpha\beta}^{(n)}(x,y)
=
a(x,y)
\sum_{0 \leq i \leq n} 
E^{(ii)}_{\alpha} E^{(ii)}_{\beta}
\\
+
\sum_{0 \leq i < j \leq n}
\ll
b_{+}(x,y)
E^{(ii)}_{\alpha} E^{(jj)}_{\beta}
+
b_{-}(x,y)
E^{(jj)}_{\alpha} E^{(ii)}_{\beta}
\rr
\\
+
\sum_{0 \leq i<j \leq n}
\ll
c_{+}(x,y)
E^{(ij)}_{\alpha} E^{(ji)}_{\beta}
+
c_{-}(x,y)
E^{(ji)}_{\alpha} E^{(ij)}_{\beta}
\rr
\end{multline}
where $E^{(ij)}_{\alpha}$ is an $(n+1) \times (n+1)$ elementary 
matrix acting on the vector space $V_{\alpha} = \mathbb{C}^{n+1}$, whose 
$(i, j)$-th entry is 1 while all remaining entries are 0. The functions 
$a,b_{\pm},c_{\pm}$ are defined to be

\begin{align}
a(x,y) = 1,
\quad\quad
b_{\pm}(x,y) = \frac{x-y}{x-y+1},
\quad\quad
c_{\pm}(x,y) = \frac{1}{x-y+1}
\label{rat-wt}
\end{align}

\noindent for rational models, and

\begin{align}
a(x,y) = 1,
\quad\quad
b_{\pm}(x,y) = e^{\mp \gamma} \frac{[x-y]}{[x-y+\gamma]},
\quad\quad
c_{\pm}(x,y) = e^{\pm (x-y)} \frac{[\gamma]}{[x-y+\gamma]}
\label{trig-wt}
\end{align}

\noindent for trigonometric models, where we use the notation 
$[x] = \sinh (x)$. 
The $R$-matrices (\ref{Rmat}) have $(n+1)(2n+1)$ non-zero entries, which 
can be identified with vertices according to the convention shown in 
Figure {\bf\ref{fig-vert}}. We call the resulting $(n+1)(2n+1)$-vertex 
model the $A_n$ vertex model.

\subsection{The $R$-matrix of the $A_1$ model}
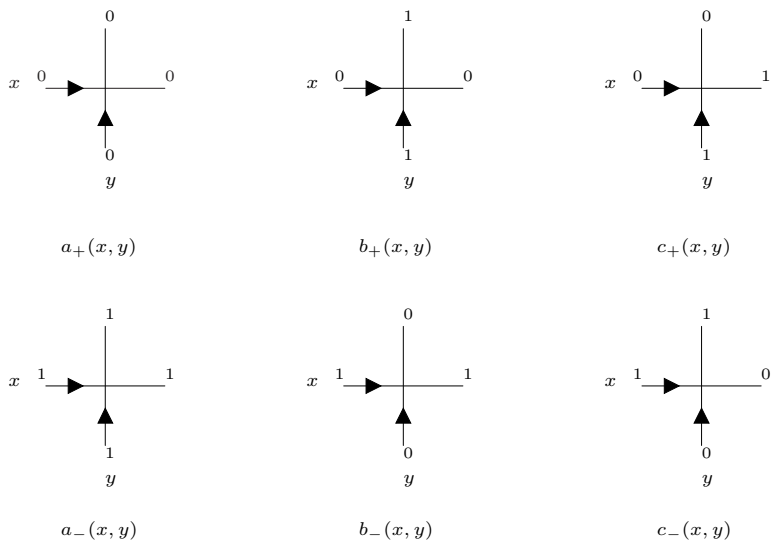
\begin{figure}
\begin{center}
\begin{minipage}{4.3in}
\setlength{\unitlength}{0.00040cm}
\begin{picture}(20000,18000)(-4500,-13000)
%
\path(-2000,2000)(2000,2000)
\put(-2300,2250){\tiny 0}
\put(2000,2250){\tiny 0}
\put(-3250,2000){\scriptsize{$x$}}
\blacken\path(-1250,2250)(-1250,1750)(-750,2000)(-1250,2250)
\path(0000,0000)(0000,4000)
\put(0000,-400){\tiny 0}
\put(0000,4250){\tiny 0}
\put(-1500,-3500){\scriptsize$a_{+}(x,y)$}
\put(0,-1250){\scriptsize{$y$}}
\blacken\path(250,750)(-250,750)(000,1250)(250,750)
%
\path(8000,2000)(12000,2000)
\put(7700,2250){\tiny 0}
\put(12000,2250){\tiny 0}
\put(6750,2000){\scriptsize{$x$}}
\blacken\path(8750,2250)(8750,1750)(9250,2000)(8750,2250)
\path(10000,0000)(10000,4000)
\put(10000,-400){\tiny 1}
\put(10000,4250){\tiny 1}
\put(8500,-3500){\scriptsize$b_{+}(x,y)$}
\put(10000,-1250){\scriptsize{$y$}}
\blacken\path(10250,750)(9750,750)(10000,1250)(10250,750)
%
\path(18000,2000)(22000,2000)
\put(17700,2250){\tiny 0}
\put(22000,2250){\tiny 1}
\put(16750,2000){\scriptsize{$x$}}
\blacken\path(18750,2250)(18750,1750)(19250,2000)(18750,2250)
\path(20000,0000)(20000,4000)
\put(20000,-400){\tiny 1}
\put(20000,4250){\tiny 0}
\put(18500,-3500){\scriptsize$c_{+}(x,y)$}
\put(20000,-1250){\scriptsize{$y$}}
\blacken\path(20250,750)(19750,750)(20000,1250)(20250,750)
%
\path(-2000,-8000)(2000,-8000)
\put(-2300,-7750){\tiny 1}
\put(2000,-7750){\tiny 1}
\put(-3250,-8000){\scriptsize{$x$}}
\blacken\path(-1250,-7750)(-1250,-8250)(-750,-8000)(-1250,-7750)
\path(0000,-10000)(0000,-6000)
\put(0000,-10400){\tiny 1}
\put(0000,-5750){\tiny 1}
\put(-1500,-13000){\scriptsize$a_{-}(x,y)$}
\put(0,-11250){\scriptsize{$y$}}
\blacken\path(250,-9250)(-250,-9250)(000,-8750)(250,-9250)
%
\path(8000,-8000)(12000,-8000)
\put(7700,-7750){\tiny 1}
\put(12000,-7750){\tiny 1}
\put(6750,-8000){\scriptsize{$x$}}
\blacken\path(8750,-7750)(8750,-8250)(9250,-8000)(8750,-7750)
\path(10000,-10000)(10000,-6000)
\put(10000,-10400){\tiny 0}
\put(10000,-5750){\tiny 0}
\put(8500,-13000){\scriptsize$b_{-}(x,y)$}
\put(10000,-11250){\scriptsize{$y$}}
\blacken\path(10250,-9250)(9750,-9250)(10000,-8750)(10250,-9250)
%
\path(18000,-8000)(22000,-8000)
\put(17700,-7750){\tiny 1}
\put(22000,-7750){\tiny 0}
\put(16750,-8000){\scriptsize{$x$}}
\blacken\path(18750,-7750)(18750,-8250)(19250,-8000)(18750,-7750)
\path(20000,-10000)(20000,-6000)
\put(20000,-10400){\tiny 0}
\put(20000,-5750){\tiny 1}
\put(18500,-13000){\scriptsize$c_{-}(x,y)$}
\put(20000,-11250){\scriptsize{$y$}}
\blacken\path(20250,-9250)(19750,-9250)(20000,-8750)(20250,-9250)
\end{picture}
\end{minipage}
\end{center}

\caption{Identifying the entries of the $R$-matrix with vertices. 
For the purpose of making each vertex weight unique, we have defined 
$a_{+}(x,y)=a_{-}(x,y) = a(x,y)$.} 
\label{fig-6v}
\end{figure}

The simplest of the models discussed above, corresponding to the case 
$n=1$, is the six-vertex model. Using the general formula (\ref{Rmat}), 
the $A_1$ $R$-matrix can be written as

\begin{align}
R_{\alpha\beta}(x,y)
=
\ll
\begin{array}{cccc}
a(x,y) & 0 & 0 & 0
\\
0 & b_{+}(x,y) & c_{+}(x,y) & 0
\\
0 & c_{-}(x,y) & b_{-}(x,y) & 0
\\ 
0 & 0 & 0 & a(x,y)
\end{array}
\rr_{\alpha\beta}
\end{align}

\noindent and we match its entries with the vertices shown in Figure 
{\bf\ref{fig-6v}}. In $A_{n \geq 2}$ models, we continue to refer to 
vertices as  $a$, $b$ or $c$ vertices when they are of the form shown 
in Figure {\bf\ref{fig-6v}} but with 0 and/or 1 replaced by more 
general colours.

\section{$A_1$ domain wall partition function}
\label{section.dwpf}

\subsection{Definition of the DWPF}
\label{ssec-dwpf}

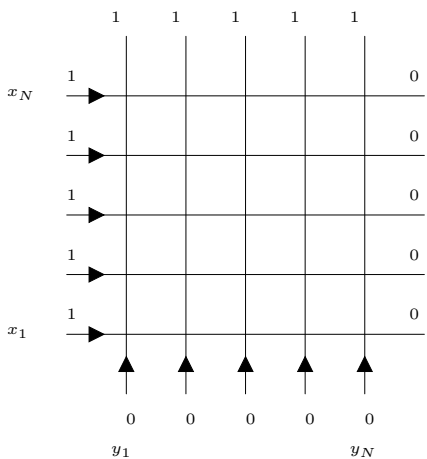
\begin{figure}

\begin{center}
\begin{minipage}{4.3in}

\setlength{\unitlength}{0.00040cm}
\begin{picture}(40000,18000)(6000,7000)


\blacken\path(14750,20250)(14750,19750)(15250,20000)(14750,20250)
\path(14000,20000)(26000,20000) \put(12000,20000){\tiny $x_N$} 
\put(14000,20500){\tiny 1}  
\put(25500,20500){\tiny 0}

\blacken\path(14750,18250)(14750,17750)(15250,18000)(14750,18250)
\path(14000,18000)(26000,18000)
\put(14000,18500){\tiny 1}  
\put(25500,18500){\tiny 0}

\blacken\path(14750,16250)(14750,15750)(15250,16000)(14750,16250)
\path(14000,16000)(26000,16000)
\put(14000,16500){\tiny 1}  
\put(25500,16500){\tiny 0}

\blacken\path(14750,14250)(14750,13750)(15250,14000)(14750,14250)
\path(14000,14000)(26000,14000)
\put(14000,14500){\tiny 1}  
\put(25500,14500){\tiny 0}

\blacken\path(14750,12250)(14750,11750)(15250,12000)(14750,12250)
\path(14000,12000)(26000,12000) \put(12000,12000){\tiny $x_{1}$}
\put(14000,12500){\tiny 1} 
\put(25500,12500){\tiny 0}


\blacken\path(15750,10750)(16250,10750)(16000,11250)(15750,10750)
\path(16000,10000)(16000,22000) \put(15500,8000){\tiny $y_1$}
\put(16000,9000){\tiny 0} \put(15500,22500){\tiny 1}

\blacken\path(17750,10750)(18250,10750)(18000,11250)(17750,10750)
\path(18000,10000)(18000,22000)
\put(18000,9000){\tiny 0} \put(17500,22500){\tiny 1}

\blacken\path(19750,10750)(20250,10750)(20000,11250)(19750,10750)
\path(20000,10000)(20000,22000)
\put(20000,9000){\tiny 0} \put(19500,22500){\tiny 1}

\blacken\path(21750,10750)(22250,10750)(22000,11250)(21750,10750)
\path(22000,10000)(22000,22000)
\put(22000,9000){\tiny 0} \put(21500,22500){\tiny 1}

\blacken\path(23750,10750)(24250,10750)(24000,11250)(23750,10750)
\path(24000,10000)(24000,22000) \put(23500,8000){\tiny $y_N$}
\put(24000,9000){\tiny 0} \put(23500,22500){\tiny 1}

\end{picture}

\end{minipage}
\end{center}

\caption{Lattice representation of $Z(\{x\}_N | \{y\}_N)$. Every 
intersection of a horizontal and vertical line is a vertex, as 
defined in Figure {\bf\ref{fig-6v}}. The colours on all external 
segments are fixed to the values shown, while all internal segments 
are summed over.}
\label{fig-dwpf}

\end{figure}

The domain wall partition function is a function in two sets of variables 
$\{x\}_N = \{x_1,\dots,x_N\}$ and $\{y\}_N = \{y_1,\dots,y_N\}$, and we 
denote it by $Z(\{x\}_N | \{y\}_N)$. We define this quantity to be the 
partition function of the lattice shown in Figure {\bf\ref{fig-dwpf}}.

\subsection{Properties of the DWPF}
\label{ssec-pf-prop}

Following \cite{korepin}, the rational DWPF satisfies a set of four 
properties which determine it uniquely. 

{\bf A.} $Z(\{x\}_N | \{y\}_N)$ is a meromorphic function of the form
\begin{align}
Z\ll\{x\}_N \Big| \{y\}_N\rr
=
\frac{P(\{x\}_N | \{y\}_N)}{\prod_{i,j=1}^{N} (x_i-y_j+1)}
\label{pf-a}
\end{align}
where $P(\{x\}_N | \{y\}_N)$ is a polynomial of degree $N-1$ in 
the variable $x_N$.

{\bf B.} $Z(\{x\}_N | \{y\}_N)$ is symmetric in the set of variables 
$\{y_1,\dots,y_N\}$.

{\bf C.} By setting $x_N = y_N$, we obtain the recursion relation
\begin{align}
Z\ll\{x\}_N \Big| \{y\}_N\rr
\Big|_{x_N = y_N}
=
Z\ll\{x\}_{N-1} \Big| \{y\}_{N-1}\rr
\label{pf-c} 
\end{align}

{\bf D.} In the case $N=1$, we have $Z(x_1|y_1) = c_{-}(x_1,y_1)$.

\medskip

\myProof We prove these properties using the lattice representation of 
the DWPF in Figure {\bf\ref{fig-dwpf}}.

{\bf A.} We assume that the factor $\frac{1}{(x_i-y_j+1)}$ is common to 
any weight at the intersection of the $i$-th horizontal and $j$-th vertical 
lines
\footnote{
By writing $a(x_i,y_j) = 
\frac{(x_i-y_j+1)}{(x_i-y_j+1)}$.}, so the denominator in (\ref{pf-a}) 
is explained. For the numerator, it is easy to see that the top line 
of the lattice in Figure {\bf \ref{fig-dwpf}} (which contributes all 
$x_N$ dependence to the DWPF) must contain exactly one $c_{-}$ vertex, 
which has degree 0 in $x_N$. This explains the fact that 
$P(\{x\}_N | \{y\}_N)$ is degree $N-1$ in $x_N$.

\begin{figure}

\begin{center}
\begin{minipage}{4.3in}

\setlength{\unitlength}{0.00032cm}
\begin{picture}(40000,20000)(14000,6500)



\blacken\path(14750,20250)(14750,19750)(15250,20000)(14750,20250)
\path(14000,20000)(26000,20000) \put(12000,20000){\tiny $x_N$} 
\put(14000,20500){\tiny 1}  
\put(25500,20500){\tiny 0}

\blacken\path(14750,18250)(14750,17750)(15250,18000)(14750,18250)
\path(14000,18000)(26000,18000)
\put(14000,18500){\tiny 1}  
\put(25500,18500){\tiny 0}

\blacken\path(14750,16250)(14750,15750)(15250,16000)(14750,16250)
\path(14000,16000)(26000,16000)
\put(14000,16500){\tiny 1}  
\put(25500,16500){\tiny 0}

\blacken\path(14750,14250)(14750,13750)(15250,14000)(14750,14250)
\path(14000,14000)(26000,14000)
\put(14000,14500){\tiny 1}  
\put(25500,14500){\tiny 0}

\blacken\path(14750,12250)(14750,11750)(15250,12000)(14750,12250)
\path(14000,12000)(26000,12000) \put(12000,12000){\tiny $x_1$}
\put(14000,12500){\tiny 1} 
\put(25500,12500){\tiny 0}


\blacken\path(15750,10750)(16250,10750)(16000,11250)(15750,10750)
\path(16000,10000)(16000,22000) \put(15500,8000){\tiny $y_1$}
\put(16000,9000){\tiny 0} \put(15500,22500){\tiny 1}

\blacken\path(17750,10750)(18250,10750)(18000,11250)(17750,10750)
\path(18000,10000)(18000,22000)
\put(18000,9000){\tiny 0} \put(17500,22500){\tiny 1}

\blacken\path(19750,10750)(20250,10750)(20000,11250)(19750,10750)
\path(20000,10000)(20000,22000) \put(19500,6000){\tiny $y_{j+1}$}
\put(20000,7000){\tiny 0} \put(19500,22500){\tiny 1}

\path(20000,10000)(22000,8000)
\path(22000,10000)(20000,8000)

\blacken\path(21750,10750)(22250,10750)(22000,11250)(21750,10750)
\path(22000,10000)(22000,22000) \put(22000,6000){\tiny $y_j$}
\put(22000,7000){\tiny 0} \put(21500,22500){\tiny 1}

\blacken\path(23750,10750)(24250,10750)(24000,11250)(23750,10750)
\path(24000,10000)(24000,22000) \put(23500,8000){\tiny $y_{N}$}
\put(24000,9000){\tiny 0} \put(23500,22500){\tiny 1}


\put(29000,16000){$=$}



\blacken\path(34750,20250)(34750,19750)(35250,20000)(34750,20250)
\path(34000,20000)(46000,20000) \put(32000,20000){\tiny $x_N$} 
\put(34000,20500){\tiny 1}  
\put(45500,20500){\tiny 0}

\blacken\path(34750,18250)(34750,17750)(35250,18000)(34750,18250)
\path(34000,18000)(46000,18000)
\put(34000,18500){\tiny 1}  
\put(45500,18500){\tiny 0}

\blacken\path(34750,16250)(34750,15750)(35250,16000)(34750,16250)
\path(34000,16000)(46000,16000)
\put(34000,16500){\tiny 1}  
\put(45500,16500){\tiny 0}

\blacken\path(34750,14250)(34750,13750)(35250,14000)(34750,14250)
\path(34000,14000)(46000,14000)
\put(34000,14500){\tiny 1}  
\put(45500,14500){\tiny 0}

\blacken\path(34750,12250)(34750,11750)(35250,12000)(34750,12250)
\path(34000,12000)(46000,12000) \put(32000,12000){\tiny $x_1$}
\put(34000,12500){\tiny 1} 
\put(45500,12500){\tiny 0}


\blacken\path(35750,10750)(36250,10750)(36000,11250)(35750,10750)
\path(36000,10000)(36000,22000) \put(35500,8000){\tiny $y_1$}
\put(36000,9000){\tiny 0} \put(35500,22500){\tiny 1}

\blacken\path(37750,10750)(38250,10750)(38000,11250)(37750,10750)
\path(38000,10000)(38000,22000)
\put(38000,9000){\tiny 0} \put(37500,22500){\tiny 1}

\blacken\path(39750,10750)(40250,10750)(40000,11250)(39750,10750)
\path(40000,10000)(40000,22000) \put(39500,8000){\tiny $y_{j+1}$}
\put(40000,9000){\tiny 0} \put(39500,24500){\tiny 1}

\path(40000,22000)(42000,24000)
\path(42000,22000)(40000,24000)

\blacken\path(41750,10750)(42250,10750)(42000,11250)(41750,10750)
\path(42000,10000)(42000,22000) \put(42000,8000){\tiny $y_j$}
\put(42000,9000){\tiny 0} \put(41500,24500){\tiny 1}

\blacken\path(43750,10750)(44250,10750)(44000,11250)(43750,10750)
\path(44000,10000)(44000,22000) \put(43500,8000){\tiny $y_{N}$}
\put(44000,9000){\tiny 0} \put(43500,22500){\tiny 1}

\end{picture}

\end{minipage}
\end{center}

\caption{Interchanging two vertical lattice lines. The inserted vertex 
is translated vertically through the lattice using the Yang-Baxter equation.}
\label{fig-dwpf-sym}

\end{figure}
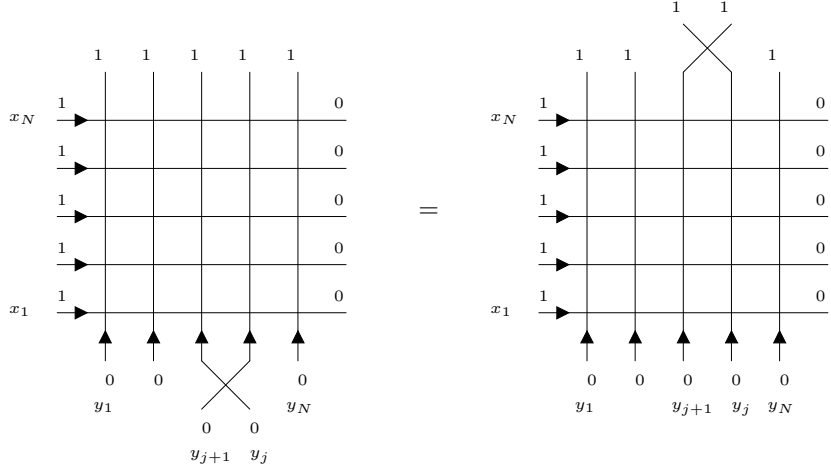

{\bf B.} The symmetry in $\{y_1,\dots,y_N\}$ is proved using the following 
argument for interchanging two vertical lattice lines. Consider multiplying 
the DWPF by the $a$ vertex $a(y_{j+1},y_j) \equiv 1$. This has the graphical 
interpretation on the left of Figure {\bf\ref{fig-dwpf-sym}}. Using 
the Yang-Baxter equation the attached vertex may be threaded vertically 
through the lattice, until it emerges from the top as another $a$ vertex, 
as on the right of Figure {\bf\ref{fig-dwpf-sym}}. The result of this 
procedure is the interchange of the lines carrying the rapidities 
$y_j,y_{j+1}$. Composing such swaps, one finds the lattice is invariant 
under any permutation of $\{y_1,\dots,y_N\}$.

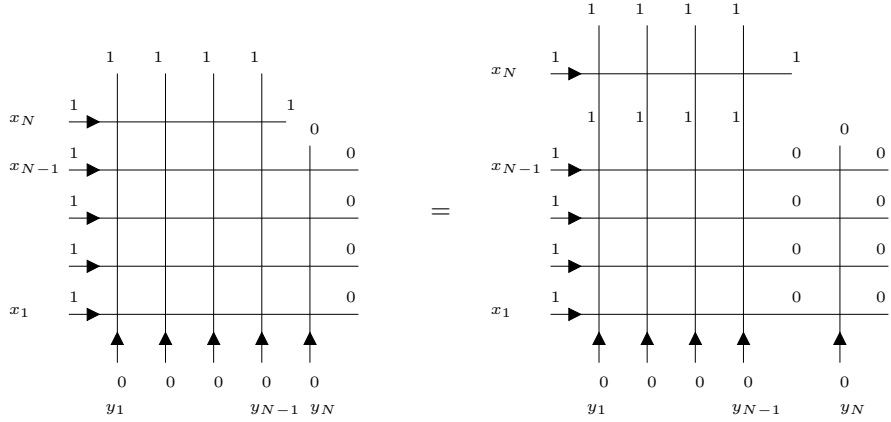
\begin{figure}

\begin{center}
\begin{minipage}{4.3in}

\setlength{\unitlength}{0.00032cm}
\begin{picture}(40000,18000)(14000,8500)



\blacken\path(14750,20250)(14750,19750)(15250,20000)(14750,20250)
\path(14000,20000)(23000,20000) \put(11500,20000){\tiny $x_N$} 
\put(14000,20500){\tiny 1}  
\put(23000,20500){\tiny 1}

\blacken\path(14750,18250)(14750,17750)(15250,18000)(14750,18250)
\path(14000,18000)(26000,18000) \put(11500,18000){\tiny $x_{N-1}$}
\put(14000,18500){\tiny 1}  
\put(25500,18500){\tiny 0}

\blacken\path(14750,16250)(14750,15750)(15250,16000)(14750,16250)
\path(14000,16000)(26000,16000)
\put(14000,16500){\tiny 1}  
\put(25500,16500){\tiny 0}

\blacken\path(14750,14250)(14750,13750)(15250,14000)(14750,14250)
\path(14000,14000)(26000,14000)
\put(14000,14500){\tiny 1}  
\put(25500,14500){\tiny 0}

\blacken\path(14750,12250)(14750,11750)(15250,12000)(14750,12250)
\path(14000,12000)(26000,12000) \put(11500,12000){\tiny $x_1$}
\put(14000,12500){\tiny 1} 
\put(25500,12500){\tiny 0}


\blacken\path(15750,10750)(16250,10750)(16000,11250)(15750,10750)
\path(16000,10000)(16000,22000) \put(15500,8000){\tiny $y_1$}
\put(16000,9000){\tiny 0} \put(15500,22500){\tiny 1}

\blacken\path(17750,10750)(18250,10750)(18000,11250)(17750,10750)
\path(18000,10000)(18000,22000)
\put(18000,9000){\tiny 0} \put(17500,22500){\tiny 1}

\blacken\path(19750,10750)(20250,10750)(20000,11250)(19750,10750)
\path(20000,10000)(20000,22000) 
\put(20000,9000){\tiny 0} \put(19500,22500){\tiny 1}

\blacken\path(21750,10750)(22250,10750)(22000,11250)(21750,10750)
\path(22000,10000)(22000,22000) \put(21500,8000){\tiny $y_{N-1}$}
\put(22000,9000){\tiny 0} \put(21500,22500){\tiny 1}

\blacken\path(23750,10750)(24250,10750)(24000,11250)(23750,10750)
\path(24000,10000)(24000,19000) \put(24000,8000){\tiny $y_{N}$}
\put(24000,9000){\tiny 0} \put(24000,19500){\tiny 0}


\put(29000,16000){$=$}



\blacken\path(34750,22250)(34750,21750)(35250,22000)(34750,22250)
\path(34000,22000)(44000,22000) \put(31500,22000){\tiny $x_N$} 
\put(34000,22500){\tiny 1}  
\put(44000,22500){\tiny 1}

\blacken\path(34750,18250)(34750,17750)(35250,18000)(34750,18250)
\path(34000,18000)(48000,18000) \put(31500,18000){\tiny $x_{N-1}$}
\put(34000,18500){\tiny 1}
\put(44000,18500){\tiny 0}  
\put(47500,18500){\tiny 0}

\blacken\path(34750,16250)(34750,15750)(35250,16000)(34750,16250)
\path(34000,16000)(48000,16000)
\put(34000,16500){\tiny 1}
\put(44000,16500){\tiny 0}  
\put(47500,16500){\tiny 0}

\blacken\path(34750,14250)(34750,13750)(35250,14000)(34750,14250)
\path(34000,14000)(48000,14000)
\put(34000,14500){\tiny 1}
\put(44000,14500){\tiny 0}  
\put(47500,14500){\tiny 0}

\blacken\path(34750,12250)(34750,11750)(35250,12000)(34750,12250)
\path(34000,12000)(48000,12000) \put(31500,12000){\tiny $x_1$}
\put(34000,12500){\tiny 1}
\put(44000,12500){\tiny 0} 
\put(47500,12500){\tiny 0}


\blacken\path(35750,10750)(36250,10750)(36000,11250)(35750,10750)
\path(36000,10000)(36000,24000) \put(35500,8000){\tiny $y_1$}
\put(36000,9000){\tiny 0} \put(35500,20000){\tiny 1} \put(35500,24500){\tiny 1}

\blacken\path(37750,10750)(38250,10750)(38000,11250)(37750,10750)
\path(38000,10000)(38000,24000)
\put(38000,9000){\tiny 0} \put(37500,20000){\tiny 1} \put(37500,24500){\tiny 1}

\blacken\path(39750,10750)(40250,10750)(40000,11250)(39750,10750)
\path(40000,10000)(40000,24000) 
\put(40000,9000){\tiny 0} \put(39500,20000){\tiny 1} \put(39500,24500){\tiny 1}

\blacken\path(41750,10750)(42250,10750)(42000,11250)(41750,10750)
\path(42000,10000)(42000,24000) \put(41500,8000){\tiny $y_{N-1}$}
\put(42000,9000){\tiny 0} \put(41500,20000){\tiny 1} \put(41500,24500){\tiny 1}

\blacken\path(45750,10750)(46250,10750)(46000,11250)(45750,10750)
\path(46000,10000)(46000,19000) \put(46000,8000){\tiny $y_{N}$}
\put(46000,9000){\tiny 0} \put(46000,19500){\tiny 0}

\end{picture}

\end{minipage}
\end{center}

\caption{The splitting of the top-right vertex. Setting $x_N=y_N$ makes the DWPF 
equal to the partition function on the left. The top and right lines give only a 
trivial contribution, and we obtain the DWPF of one size smaller, as on the 
right.}
\label{fig-dwpf-rec}

\end{figure}

{\bf C.} Setting $x_N = y_N$ forces the top-right vertex in Figure 
{\bref{fig-dwpf}} to be a $c_{-}$ vertex, since $b$ vertices vanish 
when their incoming rapidities are equal. In addition, this $c_{-}$ 
vertex has weight 1, since its rapidities are equal. Therefore the 
effect of this evaluation is the splitting of the top-right vertex 
as on the left of Figure {\bref{fig-dwpf-rec}}. Considering 
the top and right-most lines of the new lattice, it is clear that 
they only contribute a common factor of $a$ weights to the partition 
function, which have weight 1. See the right of Figure 
{\bref{fig-dwpf-rec}}. Neglecting these lines altogether, the remainder 
of the lattice is the DWPF $Z(\{x\}_{N-1} | \{y\}_{N-1})$. Hence we have 
proved the recursion relation (\ref{pf-c}).

{\bf D.} It is clear from the definition of the vertices in 
Figure {\bf\ref{fig-6v}} that the DWPF on a $1 \times 1$ lattice 
is a $c_{-}(x_1,y_1)$ vertex.

\subsection{Evaluation of the DWPF} 

Following \cite{izergin}, in the rational parametrization 
of equation (\ref{rat-wt}) the DWPF is given by 

\begin{align}
Z \ll  \{x\}_N \Big| \{y\}_N \rr
=
\frac{
\prod_{i,j=1}^{N} (x_i - y_j)
}
{
\Delta\{x\}_N \Delta\{-y\}_N
}
\det
\left[
\frac{1}{(x_i - y_j)(x_i - y_j + 1)}
\right]_{1\leq i,j \leq N}
\label{IK}
\end{align}

\noindent In the trigonometric parametrization of equation (\ref{trig-wt}) 
it is given by

\begin{align}
Z 
\ll 
\{x\}_N \Big| \{y\}_N 
\rr
=
\frac{
e^{|y|-|x|}
\prod_{i,j=1}^{N} [x_i - y_j]
}
{
\Delta\{x\}_N \Delta\{-y\}_N
}
\det
\left[
\frac{[\gamma]}{[x_i - y_j][x_i - y_j + \gamma]}
\right]_{1\leq i,j \leq N}
\label{IK-trig}
\end{align}

\noindent where we use the notation 
$|x|= \sum_{k=1}^{N} x_k$.

\section{$A_1$ scalar products}
\label{section.scalar.product}

\subsection{The $A_1$ scalar products in vertex model terms}
\label{ssec-int-sp}

We have so far discussed the $A_1$ scalar products in spin-chain 
terms. We can also consider them in vertex model terms. In vertex 
model terms, an $A_1$ scalar product is a partition function 
$S(\{x\}_N, \{b\}_N | \{y\}_L)$ that depends on two sets of 
auxiliary variables, or rapidities that flow in horizontal lattice
lines, $\{x\}_N = \{x_1,\dots,x_N\}$ and $\{b\}_N = \{b_1,\dots,b_N\}$, 
and one set of quantum variables, or inhomogeneities that flow in
vertical lattice lines, $\{y\}_L = \{y_1,\dots,y_L\}$. 
If both sets of variables $\{x\}_N$ and $\{b\}_N$ are free, we obtain
Korepin's off-shell/off-shell scalar product $\cK_1$ \cite{korepin}. 
If the set $\{x\}_N$ is free, while the set $\{b\}_N$ obeys Bethe 
equations, we obtain
Slavnov's off-shell/on-shell scalar product $\cS_1$ \cite{slavnov}.

\subsection{Definition of the scalar product} In the sequel, we say 
\lq the scalar product\rq\ to refer to the $A_1$ vertex model 
partition function that 
evaluates to $\cK_1$ when the set $\{b\}_N$ is free, and 
to $\cS_1$ when it satisfies Bethe equations. 

\begin{figure}

\begin{center}
\begin{minipage}{4.3in}

\setlength{\unitlength}{0.00032cm}
\begin{picture}(40000,22000)(5000,2500)


\blacken\path(14750,20250)(14750,19750)(15250,20000)(14750,20250)
\path(14000,20000)(30000,20000) \put(12000,20000){\tiny $b_N$}
\put(14000,20500){\tiny 0}  
\put(29500,20500){\tiny 1}

\blacken\path(14750,18250)(14750,17750)(15250,18000)(14750,18250)
\path(14000,18000)(30000,18000) 
\put(14000,18500){\tiny 0}  
\put(29500,18500){\tiny 1}

\blacken\path(14750,16250)(14750,15750)(15250,16000)(14750,16250)
\path(14000,16000)(30000,16000) 
\put(14000,16500){\tiny 0}  
\put(29500,16500){\tiny 1}

\blacken\path(14750,14250)(14750,13750)(15250,14000)(14750,14250)
\path(14000,14000)(30000,14000) \put(12000,14000){\tiny $b_1$}
\put(14000,14500){\tiny 0}  
\put(29500,14500){\tiny 1}


\blacken\path(14750,12250)(14750,11750)(15250,12000)(14750,12250)
\path(14000,12000)(30000,12000) \put(12000,12000){\tiny $x_N$}
\put(14000,12500){\tiny 1} 
\put(29500,12500){\tiny 0}

\blacken\path(14750,10250)(14750,9750)(15250,10000)(14750,10250)
\path(14000,10000)(30000,10000)  
\put(14000,10500){\tiny 1}  
\put(29500,10500){\tiny 0}

\blacken\path(14750,8250)(14750,7750)(15250,8000)(14750,8250)
\path(14000,8000)(30000,8000)
\put(14000,8500){\tiny 1}  
\put(29500,8500){\tiny 0}

\blacken\path(14750,6250)(14750,5750)(15250,6000)(14750,6250)
\path(14000,6000)(30000,6000) \put(12000,6000){\tiny $x_1$} 
\put(14000,6500){\tiny 1}  
\put(29500,6500){\tiny 0}


\blacken\path(15750,4750)(16250,4750)(16000,5250)(15750,4750)
\path(16000,4000)(16000,22000) \put(15500,2000){\tiny $y_1$}
\put(16000,3000){\tiny 0} \put(15500,22500){\tiny 0}

\blacken\path(17750,4750)(18250,4750)(18000,5250)(17750,4750)
\path(18000,4000)(18000,22000)
\put(18000,3000){\tiny 0} \put(17500,22500){\tiny 0}

\blacken\path(19750,4750)(20250,4750)(20000,5250)(19750,4750)
\path(20000,4000)(20000,22000)
\put(20000,3000){\tiny 0} \put(19500,22500){\tiny 0}

\blacken\path(21750,4750)(22250,4750)(22000,5250)(21750,4750)
\path(22000,4000)(22000,22000)
\put(22000,3000){\tiny 0} \put(21500,22500){\tiny 0}

\blacken\path(23750,4750)(24250,4750)(24000,5250)(23750,4750)
\path(24000,4000)(24000,22000) 
\put(24000,3000){\tiny 0} \put(23500,22500){\tiny 0}

\blacken\path(25750,4750)(26250,4750)(26000,5250)(25750,4750)
\path(26000,4000)(26000,22000) 
\put(26000,3000){\tiny 0} \put(25500,22500){\tiny 0}

\blacken\path(27750,4750)(28250,4750)(28000,5250)(27750,4750)
\path(28000,4000)(28000,22000) \put(27500,2000){\tiny $y_L$}
\put(28000,3000){\tiny 0} \put(27500,22500){\tiny 0}

\end{picture}

\end{minipage}
\end{center}

\caption{Lattice representation of $S(\{x\}_N, \{b\}_N | \{y\}_L)$. 
In the algebraic Bethe Ansatz scheme, each horizontal line represents 
a monodromy matrix operator, see \cite{wheeler}. The vertical lines 
represent the sites in a spin-chain of length $L$, with local 
inhomogeneities.}

\label{fig-sp}

\end{figure}
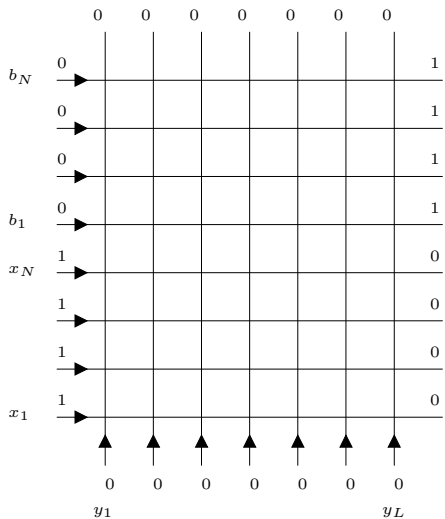

We define the scalar product $S(\{x\}_N, \{b\}_N | \{y\}_L)$ 
as the partition function of the lattice configuration in 
Figure {\bf\ref{fig-sp}}.

\begin{figure}

\begin{center}
\begin{minipage}{4.3in}

\setlength{\unitlength}{0.00032cm}
\begin{picture}(40000,22000)(15000,2500)



\blacken\path(14750,20250)(14750,19750)(15250,20000)(14750,20250)
\path(14000,20000)(30000,20000) \put(12500,20000){\tiny $b_{\m{m}}$}
\put(14000,20500){\tiny 0}  
\put(29500,20500){\tiny 1}

\blacken\path(14750,18250)(14750,17750)(15250,18000)(14750,18250)
\path(14000,18000)(30000,18000) 
\put(14000,18500){\tiny 0}  
\put(29500,18500){\tiny 1}

\blacken\path(14750,16250)(14750,15750)(15250,16000)(14750,16250)
\path(14000,16000)(30000,16000) \put(12500,16000){\tiny $b_1$}
\put(14000,16500){\tiny 0}  
\put(29500,16500){\tiny 1}


\blacken\path(14750,14250)(14750,13750)(15250,14000)(14750,14250)
\path(14000,14000)(30000,14000) \put(12500,14000){\tiny $x_N$}
\put(14000,14500){\tiny 1}  
\put(29500,14500){\tiny 0}

\blacken\path(14750,12250)(14750,11750)(15250,12000)(14750,12250)
\path(14000,12000)(30000,12000) 
\put(14000,12500){\tiny 1} 
\put(29500,12500){\tiny 0}

\blacken\path(14750,10250)(14750,9750)(15250,10000)(14750,10250)
\path(14000,10000)(30000,10000)  
\put(14000,10500){\tiny 1}  
\put(29500,10500){\tiny 0}

\blacken\path(14750,8250)(14750,7750)(15250,8000)(14750,8250)
\path(14000,8000)(30000,8000)
\put(14000,8500){\tiny 1}  
\put(29500,8500){\tiny 0}

\blacken\path(14750,6250)(14750,5750)(15250,6000)(14750,6250)
\path(14000,6000)(30000,6000) \put(12500,6000){\tiny $x_1$} 
\put(14000,6500){\tiny 1}  
\put(29500,6500){\tiny 0}


\blacken\path(15750,4750)(16250,4750)(16000,5250)(15750,4750)
\path(16000,4000)(16000,22000) \put(15500,2000){\tiny $y_{N-\m{m}+1}$}
\put(16000,3000){\tiny 0} \put(15500,22500){\tiny 0}

\blacken\path(17750,4750)(18250,4750)(18000,5250)(17750,4750)
\path(18000,4000)(18000,22000)
\put(18000,3000){\tiny 0} \put(17500,22500){\tiny 0}

\blacken\path(19750,4750)(20250,4750)(20000,5250)(19750,4750)
\path(20000,4000)(20000,22000)
\put(20000,3000){\tiny 0} \put(19500,22500){\tiny 0}

\blacken\path(21750,4750)(22250,4750)(22000,5250)(21750,4750)
\path(22000,4000)(22000,22000)
\put(22000,3000){\tiny 0} \put(21500,22500){\tiny 0}

\blacken\path(23750,4750)(24250,4750)(24000,5250)(23750,4750)
\path(24000,4000)(24000,22000) \put(23500,2000){\tiny $y_L$}
\put(24000,3000){\tiny 0} \put(23500,22500){\tiny 0}

\blacken\path(25750,4750)(26250,4750)(26000,5250)(25750,4750)
\path(26000,4000)(26000,22000) \put(25500,2000){\tiny $y_1$}
\put(26000,3000){\tiny 0} \put(25500,22500){\tiny 1}

\blacken\path(27750,4750)(28250,4750)(28000,5250)(27750,4750)
\path(28000,4000)(28000,22000) \put(27500,2000){\tiny $y_{N-\m{m}}$}
\put(28000,3000){\tiny 0} \put(27500,22500){\tiny 1}


\put(31000,13000){$=$}



\blacken\path(34750,20250)(34750,19750)(35250,20000)(34750,20250)
\path(34000,20000)(45000,20000) \put(32500,20000){\tiny $b_{\m{m}}$}
\put(34000,20500){\tiny 0}  
\put(44500,20500){\tiny 1}

\blacken\path(34750,18250)(34750,17750)(35250,18000)(34750,18250)
\path(34000,18000)(45000,18000) 
\put(34000,18500){\tiny 0}  
\put(44500,18500){\tiny 1}

\blacken\path(34750,16250)(34750,15750)(35250,16000)(34750,16250)
\path(34000,16000)(45000,16000) \put(32500,16000){\tiny $b_1$}
\put(34000,16500){\tiny 0}  
\put(44500,16500){\tiny 1}


\blacken\path(34750,14250)(34750,13750)(35250,14000)(34750,14250)
\path(34000,14000)(50000,14000) \put(32500,14000){\tiny $x_N$}
\put(34000,14500){\tiny 1}  
\put(49500,14500){\tiny 0}

\blacken\path(34750,12250)(34750,11750)(35250,12000)(34750,12250)
\path(34000,12000)(50000,12000) 
\put(34000,12500){\tiny 1} 
\put(49500,12500){\tiny 0}

\blacken\path(34750,10250)(34750,9750)(35250,10000)(34750,10250)
\path(34000,10000)(50000,10000)  
\put(34000,10500){\tiny 1}  
\put(49500,10500){\tiny 0}

\blacken\path(34750,8250)(34750,7750)(35250,8000)(34750,8250)
\path(34000,8000)(50000,8000)
\put(34000,8500){\tiny 1}  
\put(49500,8500){\tiny 0}

\blacken\path(34750,6250)(34750,5750)(35250,6000)(34750,6250)
\path(34000,6000)(50000,6000) \put(32500,6000){\tiny $x_1$} 
\put(34000,6500){\tiny 1}  
\put(49500,6500){\tiny 0}


\blacken\path(35750,4750)(36250,4750)(36000,5250)(35750,4750)
\path(36000,4000)(36000,22000) \put(35500,2000){\tiny $y_{N-\m{m}+1}$}
\put(36000,3000){\tiny 0} \put(35500,22500){\tiny 0}

\blacken\path(37750,4750)(38250,4750)(38000,5250)(37750,4750)
\path(38000,4000)(38000,22000)
\put(38000,3000){\tiny 0} \put(37500,22500){\tiny 0}

\blacken\path(39750,4750)(40250,4750)(40000,5250)(39750,4750)
\path(40000,4000)(40000,22000)
\put(40000,3000){\tiny 0} \put(39500,22500){\tiny 0}

\blacken\path(41750,4750)(42250,4750)(42000,5250)(41750,4750)
\path(42000,4000)(42000,22000)
\put(42000,3000){\tiny 0} \put(41500,22500){\tiny 0}

\blacken\path(43750,4750)(44250,4750)(44000,5250)(43750,4750)
\path(44000,4000)(44000,22000) \put(43500,2000){\tiny $y_L$}
\put(44000,3000){\tiny 0} \put(43500,22500){\tiny 0}

\blacken\path(45750,4750)(46250,4750)(46000,5250)(45750,4750)
\path(46000,4000)(46000,15000) \put(45500,2000){\tiny $y_1$}
\put(46000,3000){\tiny 0} \put(45500,15500){\tiny 1}

\blacken\path(47750,4750)(48250,4750)(48000,5250)(47750,4750)
\path(48000,4000)(48000,15000) \put(47500,2000){\tiny $y_{N-\m{m}}$}
\put(48000,3000){\tiny 0} \put(47500,15500){\tiny 1}

\end{picture}

\end{minipage}
\end{center}

\caption{The lattice representation of $S(\{x\}_N, \{b\}_{\m{m}} | \{y\}_L)$ is 
on the left. 
The top-right corner of the lattice is constrained to be a product of $a$ 
vertices, which is why we omit it from the lattice on the right.}

\label{fig-sp-int}

\end{figure}
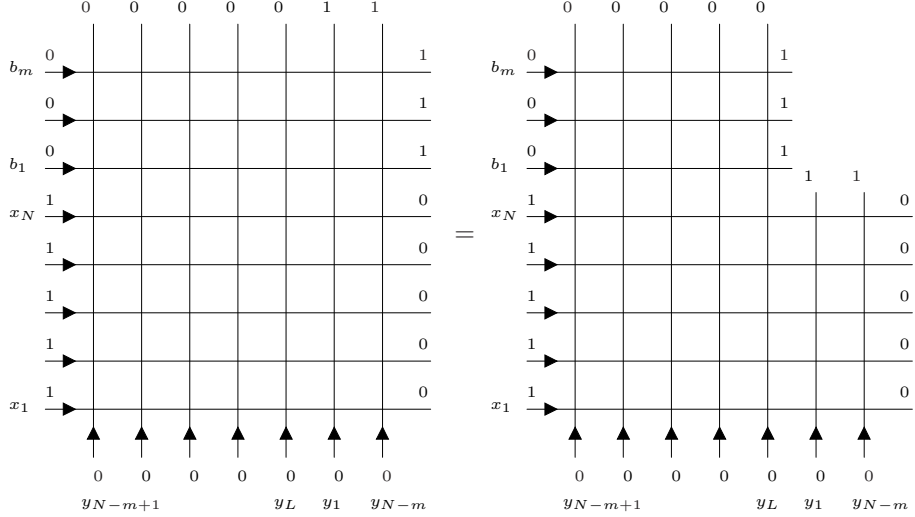

More generally, following \cite{KMT, wheeler}, one can consider 
a family of related objects called {\it restricted} scalar products
\footnote{
In \cite{wheeler}, these were called {\it intermediate} scalar 
products. Note that the convention used in this paper is different, 
in that we restrict the variables $\{b\}_N$ rather than $\{x\}_N$. This 
is completely permissible, since for now the variables $\{b\}_N$ are 
free.}, 
denoted $S(\{x\}_N, \{b\}_{\m{m}} | \{y\}_L)$, where $0 \leq \m{m} \leq N$. 
These quantities interpolate between the domain wall partition 
function $Z(\{x\}_N | \{y\}_N)$, which corresponds to the case 
$\m{m}=0$, and the full scalar product 
$\cS_1 \equiv S(\{x\}_N, \{b\}_N | \{y\}_L)$ (the case $\m{m}=N$). 
The important point regarding the restricted scalar products 
is that they are related to each other by a simple recursion 
relation, see equation (\ref{sp-c}). The lattice version of 
a typical restricted scalar product is shown in Figure 
{\bf\ref{fig-sp-int}}.

\subsection{Properties of the $A_1$ \michaelwrites{restricted} scalar \michaelwrites{products}}

Following \cite{wheeler}, the $A_1$ \michaelwrites{restricted} scalar \michaelwrites{products} 
$S(\{x\}_N, \{b\}_{\m{m}} | \{y\}_L)$ of Section 
{\bf\ref{ssec-int-sp}} \michaelwrites{satisfy} a set of four 
properties which determine \michaelwrites{them} uniquely. 

{\bf A.} $S(\{x\}_N,\{b\}_{\m{m}} | \{y\}_L)$ is a meromorphic 
function of the form

\begin{align}
S\ll\{x\}_N,\{b\}_{\m{m}} \Big| \{y\}_L\rr
=
\frac{P(\{x\}_N,\{b\}_{\m{m}}|\{y\}_L)}
{\prod_{i=1}^{N} \prod_{j=1}^{L} (x_i-y_j+1) 
\prod_{i=1}^{\m{m}} \prod_{j=N-\m{m}+1}^{L} (b_i-y_j+1)}
\label{sp-a}
\end{align}

\noindent where $P(\{x\}_N,\{b\}_{\m{m}} | \{y\}_L)$ is a polynomial 
of degree $L-N+\m{m}-1$ in $b_{\m{m}}$.

{\bf B.} $S(\{x\}_N,\{b\}_{\m{m}} | \{y\}_L)$ is symmetric in the set 
of variables $\{y_{N-\m{m}+1},\dots,y_L\}$.

{\bf C.} By setting $b_{\m{m}} = y_{N-\m{m}+1}$, we obtain the recursion 
relation

\begin{align}
S\ll\{x\}_N,\{b\}_{\m{m}} \Big| \{y\}_L\rr
\Big|_{b_{\m{m}} = y_{N-\m{m}+1}}
=
S\ll\{x\}_{N},\{b\}_{\m{m}-1} \Big| \{y\}_L\rr
\label{sp-c}
\end{align}

{\bf D.} In the case $\m{m}=0$, we have 

\begin{align}
S\ll\{x\}_N,\{b\}_0 \Big| \{y\}_L\rr 
=
\prod_{i=1}^{N}
\prod_{j=N+1}^{L}
\frac{(x_i-y_j)}{(x_i-y_j+1)}
Z\ll\{x\}_N \Big| \{y\}_N\rr
\label{sp-d}
\end{align}

\medskip

\myProof In analogy with Section {\bref{ssec-pf-prop}}, we prove these 
four properties using the lattice representation of the $A_1$ \michaelwrites{restricted} scalar 
product in Figure {\bref{fig-sp-int}}.

{\bf A.} Referring to the right of Figure {\bref{fig-sp-int}}, we see 
that all vertices in the bottom $N$ rows contain the factor 
$\frac{1}{(x_i-y_j+1)}$, while all those in the top $\m{m}$ rows contain 
$\frac{1}{(b_i-y_j+1)}$. This explains the denominator of (\ref{sp-a}), 
where the size of the second product is smaller than the first due to 
the non-rectangular geometry of Figure {\bref{fig-sp-int}}. To obtain 
the numerator, consider the top row of the lattice in 
Figure {\bref{fig-sp-int}}, which contributes all dependence on the 
variable $b_{\m{m}}$. There are $L-N+\m{m}$ vertices in this row, and in every 
configuration exactly one of them is a $c_{+}$ vertex, which has 
degree 0 in $b_{\m{m}}$. Hence it is clear that $P(\{x\}_N,\{b\}_{\m{m}} | \{y\}_L)$ 
has degree $L-N+\m{m}-1$ in $b_{\m{m}}$.

{\bf B.} To prove the symmetry in $\{y_{N-\m{m}+1},\dots,y_L\}$, one follows 
the same procedure used to prove property {\bf B} in Section {\bref{ssec-pf-prop}}. 
Since the argument is identical (albeit applied to a different set of variables), 
we do not repeat it here.

\begin{figure}

\begin{center}
\begin{minipage}{4.3in}

\setlength{\unitlength}{0.00032cm}
\begin{picture}(40000,24000)(16000,0500)



\path(17000,20000)(30000,20000)
\put(17000,19000){\arc{2000}{-3.1416}{-1.5708}} 
\put(29500,20500){\tiny 1}

\blacken\path(14750,18250)(14750,17750)(15250,18000)(14750,18250)
\path(14000,18000)(30000,18000) 
\put(14000,18500){\tiny 0}  
\put(29500,18500){\tiny 1}

\blacken\path(14750,16250)(14750,15750)(15250,16000)(14750,16250)
\path(14000,16000)(30000,16000) \put(12500,16000){\tiny $b_1$}
\put(14000,16500){\tiny 0}  
\put(29500,16500){\tiny 1}


\blacken\path(14750,14250)(14750,13750)(15250,14000)(14750,14250)
\path(14000,14000)(30000,14000) \put(12500,14000){\tiny $x_N$}
\put(14000,14500){\tiny 1}  
\put(29500,14500){\tiny 0}

\blacken\path(14750,12250)(14750,11750)(15250,12000)(14750,12250)
\path(14000,12000)(30000,12000) 
\put(14000,12500){\tiny 1} 
\put(29500,12500){\tiny 0}

\blacken\path(14750,10250)(14750,9750)(15250,10000)(14750,10250)
\path(14000,10000)(30000,10000)  
\put(14000,10500){\tiny 1}  
\put(29500,10500){\tiny 0}

\blacken\path(14750,8250)(14750,7750)(15250,8000)(14750,8250)
\path(14000,8000)(30000,8000)
\put(14000,8500){\tiny 1}  
\put(29500,8500){\tiny 0}

\blacken\path(14750,6250)(14750,5750)(15250,6000)(14750,6250)
\path(14000,6000)(30000,6000) \put(12500,6000){\tiny $x_1$} 
\put(14000,6500){\tiny 1}  
\put(29500,6500){\tiny 0}


\blacken\path(15750,4750)(16250,4750)(16000,5250)(15750,4750)
\path(16000,4000)(16000,19000) \put(15500,2000){\tiny $y_{N-\m{m}+1}$}
\put(16000,3000){\tiny 0} 

\blacken\path(17750,4750)(18250,4750)(18000,5250)(17750,4750)
\path(18000,4000)(18000,22000)
\put(18000,3000){\tiny 0} \put(17500,22500){\tiny 0}

\blacken\path(19750,4750)(20250,4750)(20000,5250)(19750,4750)
\path(20000,4000)(20000,22000)
\put(20000,3000){\tiny 0} \put(19500,22500){\tiny 0}

\blacken\path(21750,4750)(22250,4750)(22000,5250)(21750,4750)
\path(22000,4000)(22000,22000)
\put(22000,3000){\tiny 0} \put(21500,22500){\tiny 0}

\blacken\path(23750,4750)(24250,4750)(24000,5250)(23750,4750)
\path(24000,4000)(24000,22000) \put(23500,2000){\tiny $y_L$}
\put(24000,3000){\tiny 0} \put(23500,22500){\tiny 0}

\blacken\path(25750,4750)(26250,4750)(26000,5250)(25750,4750)
\path(26000,4000)(26000,22000) \put(25500,2000){\tiny $y_1$}
\put(26000,3000){\tiny 0} \put(25500,22500){\tiny 1}

\blacken\path(27750,4750)(28250,4750)(28000,5250)(27750,4750)
\path(28000,4000)(28000,22000) \put(27500,2000){\tiny $y_{N-\m{m}}$}
\put(28000,3000){\tiny 0} \put(27500,22500){\tiny 1}


\put(31000,13000){$=$}



\blacken\path(34750,20250)(34750,19750)(35250,20000)(34750,20250)
\path(34000,20000)(50000,20000) 
\put(34000,20500){\tiny 0}  
\put(49500,20500){\tiny 1}

\blacken\path(34750,18250)(34750,17750)(35250,18000)(34750,18250)
\path(34000,18000)(50000,18000) \put(32500,18000){\tiny $b_1$}
\put(34000,18500){\tiny 0}  
\put(49500,18500){\tiny 1}


\blacken\path(34750,16250)(34750,15750)(35250,16000)(34750,16250)
\path(34000,16000)(50000,16000) \put(32500,16000){\tiny $x_N$}
\put(34000,16500){\tiny 1}  
\put(49500,16500){\tiny 0}

\blacken\path(34750,14250)(34750,13750)(35250,14000)(34750,14250)
\path(34000,14000)(50000,14000) 
\put(34000,14500){\tiny 1}  
\put(49500,14500){\tiny 0}

\blacken\path(34750,12250)(34750,11750)(35250,12000)(34750,12250)
\path(34000,12000)(50000,12000) 
\put(34000,12500){\tiny 1} 
\put(49500,12500){\tiny 0}

\blacken\path(34750,10250)(34750,9750)(35250,10000)(34750,10250)
\path(34000,10000)(50000,10000)  
\put(34000,10500){\tiny 1}  
\put(49500,10500){\tiny 0}

\blacken\path(34750,8250)(34750,7750)(35250,8000)(34750,8250)
\path(34000,8000)(50000,8000) \put(32500,8000){\tiny $x_1$}
\put(34000,8500){\tiny 1}  
\put(49500,8500){\tiny 0}

\blacken\path(34750,4250)(34750,3750)(35250,4000)(34750,4250)
\path(34000,4000)(47000,4000) \put(32500,3400){\tiny $y_{N-\m{m}+1}$} 
\put(34000,4500){\tiny 0}  


\blacken\path(35750,2750)(36250,2750)(36000,3250)(35750,2750)
\path(36000,2000)(36000,22000) 
\put(36000,1000){\tiny 0}
\put(36200,6000){\tiny 0} 
\put(35500,22500){\tiny 0}

\blacken\path(37750,2750)(38250,2750)(38000,3250)(37750,2750)
\path(38000,2000)(38000,22000)
\put(38000,1000){\tiny 0} 
\put(38200,6000){\tiny 0}
\put(37500,22500){\tiny 0}

\blacken\path(39750,2750)(40250,2750)(40000,3250)(39750,2750)
\path(40000,2000)(40000,22000)
\put(40000,1000){\tiny 0} 
\put(40200,6000){\tiny 0}
\put(39500,22500){\tiny 0}

\blacken\path(41750,2750)(42250,2750)(42000,3250)(41750,2750)
\path(42000,2000)(42000,22000) \put(41500,0000){\tiny $y_L$}
\put(42000,1000){\tiny 0} 
\put(42200,6000){\tiny 0}
\put(41500,22500){\tiny 0} 

\blacken\path(43750,2750)(44250,2750)(44000,3250)(43750,2750)
\path(44000,2000)(44000,22000) \put(43500,0000){\tiny $y_1$}
\put(44000,1000){\tiny 0} 
\put(44200,6000){\tiny 0}
\put(43500,22500){\tiny 1}

\blacken\path(45750,2750)(46250,2750)(46000,3250)(45750,2750)
\path(46000,2000)(46000,22000) \put(45500,0000){\tiny $y_{N-\m{m}}$}
\put(46000,1000){\tiny 0} 
\put(46200,6000){\tiny 0}
\put(45500,22500){\tiny 1}

\path(48000,5000)(48000,22000) 
\put(47000,5000){\arc{2000}{0}{1.5708}}
\put(48200,6000){\tiny 0} 
\put(47500,22500){\tiny 1}

\end{picture}

\end{minipage}
\end{center}

\caption{Proving the recursion relation (\ref{sp-c}). The partition 
function which results from setting $b_{\m{m}} = y_{N-\m{m}+1}$ is on the left. 
An equivalent version of this partition function, obtained using the 
Yang-Baxter equation, is on the right. Up to a factor of $a$ weights, 
the lattice on the right is a restricted scalar product of one size 
smaller. }

\label{fig-sp-rec}

\end{figure}
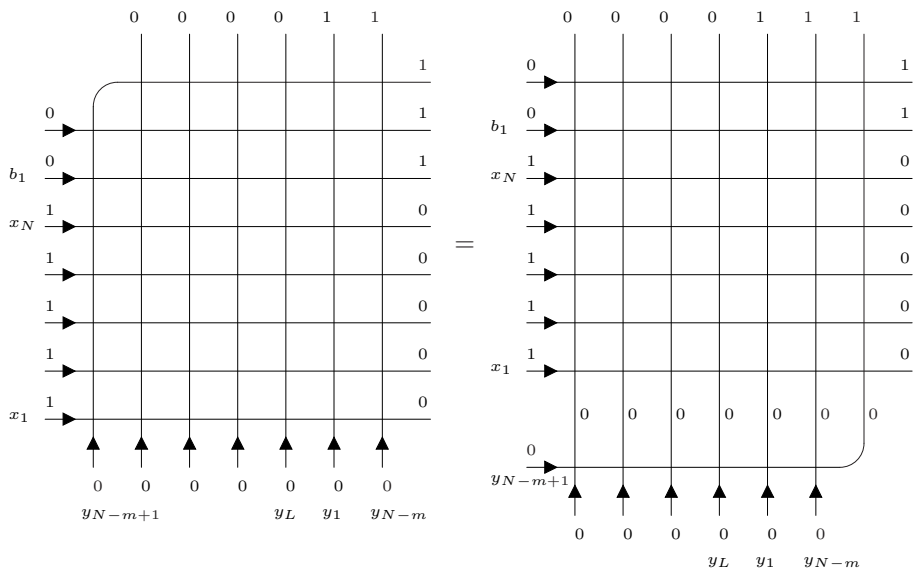

{\bf C.} Setting $b_{\m{m}} = y_{N-\m{m}+1}$ causes the top-left vertex in 
Figure {\bf\ref{fig-sp-int}} to have weight 1, regardless of whether it is 
an $a$ or $c_{+}$ vertex. Since the rapidities on the top and left-most lines 
are now both $y_{N-\m{m}+1}$, the effect of this evaluation is the same as splitting 
the top-left vertex as on the left of Figure {\bf\ref{fig-sp-rec}}. By repeated 
application of the Yang-Baxter equation, it is possible to move the curved through 
the lattice, so that it emerges as on the right of Figure {\bf\ref{fig-sp-rec}}. 
After making this transformation it is clear that (the horizontal part of) the curved 
line contributes only a common factor of $a$ weights to the overall sum. Hence this 
part of the line can be neglected, and the remainder of the lattice is simply 
$S(\{x\}_{N},\{b\}_{\m{m}-1} | \{y\}_{L})$. Hence we have proved the recursion relation 
(\ref{sp-c}).

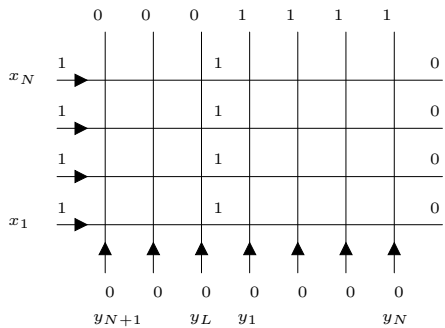
\begin{figure}

\begin{center}
\begin{minipage}{4.3in}

\setlength{\unitlength}{0.00032cm}
\begin{picture}(40000,14000)(5000,2500)


\blacken\path(14750,12250)(14750,11750)(15250,12000)(14750,12250)
\path(14000,12000)(30000,12000) \put(12000,12000){\tiny $x_N$}
\put(14000,12500){\tiny 1}
\put(20500,12500){\tiny 1}  
\put(29500,12500){\tiny 0}

\blacken\path(14750,10250)(14750,9750)(15250,10000)(14750,10250)
\path(14000,10000)(30000,10000)  
\put(14000,10500){\tiny 1}
\put(20500,10500){\tiny 1}    
\put(29500,10500){\tiny 0}

\blacken\path(14750,8250)(14750,7750)(15250,8000)(14750,8250)
\path(14000,8000)(30000,8000)
\put(14000,8500){\tiny 1}
\put(20500,8500){\tiny 1}    
\put(29500,8500){\tiny 0}

\blacken\path(14750,6250)(14750,5750)(15250,6000)(14750,6250)
\path(14000,6000)(30000,6000) \put(12000,6000){\tiny $x_1$} 
\put(14000,6500){\tiny 1}
\put(20500,6500){\tiny 1}    
\put(29500,6500){\tiny 0}


\blacken\path(15750,4750)(16250,4750)(16000,5250)(15750,4750)
\path(16000,4000)(16000,14000) \put(15500,2000){\tiny $y_{N+1}$}
\put(16000,3000){\tiny 0} \put(15500,14500){\tiny 0}

\blacken\path(17750,4750)(18250,4750)(18000,5250)(17750,4750)
\path(18000,4000)(18000,14000)
\put(18000,3000){\tiny 0} \put(17500,14500){\tiny 0}

\blacken\path(19750,4750)(20250,4750)(20000,5250)(19750,4750)
\path(20000,4000)(20000,14000) \put(19500,2000){\tiny $y_L$}
\put(20000,3000){\tiny 0} \put(19500,14500){\tiny 0}

\blacken\path(21750,4750)(22250,4750)(22000,5250)(21750,4750)
\path(22000,4000)(22000,14000) \put(21500,2000){\tiny $y_1$}
\put(22000,3000){\tiny 0} \put(21500,14500){\tiny 1}

\blacken\path(23750,4750)(24250,4750)(24000,5250)(23750,4750)
\path(24000,4000)(24000,14000) 
\put(24000,3000){\tiny 0} \put(23500,14500){\tiny 1}

\blacken\path(25750,4750)(26250,4750)(26000,5250)(25750,4750)
\path(26000,4000)(26000,14000) 
\put(26000,3000){\tiny 0} \put(25500,14500){\tiny 1}

\blacken\path(27750,4750)(28250,4750)(28000,5250)(27750,4750)
\path(28000,4000)(28000,14000) \put(27500,2000){\tiny $y_N$}
\put(28000,3000){\tiny 0} \put(27500,14500){\tiny 1}

\end{picture}

\end{minipage}
\end{center}

\caption{Lattice representation of $S(\{x\}_N, \{b\}_0 | \{y\}_L)$. 
The left-most $N \times (L-N)$ block is forced to be a product of 
$b$ vertices, and the remaining part of the lattice is an $N \times N$ DWPF.}

\label{fig-sp-0}

\end{figure}

{\bf D.} In the case $\m{m}=0$, there are no horizontal lines carrying 
the variables $\{b\}$, and we obtain the lattice shown in Figure 
{\bref{fig-sp-0}}.

It is straightforward to see that the vertices of the $L-N$ 
left-most vertical lines are forced to be $b$ vertices. The remaining 
$N \times N$ block of the lattice is just the DWPF. Hence we immediately 
obtain the condition (\ref{sp-d}), which relates the initial restricted 
scalar product directly to the DWPF. 

\subsection{Evaluation of $\cS_1$}

Following \cite{slavnov}, when the variables $\{b\}_N$ satisfy 
the Bethe equations
\begin{align}
\prod_{j\not=i}^{N}
\frac{b_i-b_j+1}{b_i-b_j-1}
=
\prod_{k=1}^{L}
\frac{b_i-y_k+1}{b_i-y_k}
\label{A_1-bethe}
\end{align}
the $A_1$ scalar product $S(\{x\}_N,\{b\}_N | \{y\}_L)$ can be evaluated 
in determinant form
\begin{multline}
\label{slavnov-det}
S\ll \{x\}_N, \{b\}_N \Big| \{y\}_L \rr
=
\Delta^{-1}\{x\}_N \Delta^{-1}\{-b\}_N
\\
\times
\det\ll
\frac{1}{b_j-x_i}
\ll
\prod_{k\not=j}^{N} (b_k-x_i+1)
-
\prod_{k\not=j}^{N} (b_k-x_i-1)
\prod_{k=1}^{L}
\frac{(x_i-y_k)}{(x_i-y_k+1)}
\rr
\rr_{1\leq i,j \leq N}
\end{multline}
We refer to this special case as {\it Slavnov's scalar product}, 
or $\cS_1$, to distinguish it from the preceding analysis where 
all variables are considered free.

It is important to remark that a determinant formula can only be 
found for the scalar product $\cS_1$ $=$ $S(\{x\}_N,\{b\}_N | \{y\}_L)$, 
depending on the full set of Bethe roots $\{b\}_N$. In particular, 
we cannot deduce from (\ref{slavnov-det}) a determinant formula 
for (what we defined to be) restricted scalar products 
$S(\{x\}_N,\{b\}_{\m{m}} | \{y\}_L)$, since these objects arise by 
treating the set of variables $\{b\}_N$ as free and then 
restricting them to be equal to inhomogeneities
\footnote{In contrast, one can define restricted scalar products 
$S(\{x\}_{\m{m}},\{b\}_N | \{y\}_L)$ by restricting the values of the set 
$\{x\}_N$, which are always free. In this case it is possible to deduce 
a determinant formula for $S(\{x\}_{\m{m}},\{b\}_N | \{y\}_L)$, starting from 
(\ref{slavnov-det}), which is explained in \cite{KMT, wheeler}.}. 
Nevertheless, we shall only require the determinant formula for 
$S(\{x\}_N,\{b\}_N | \{y\}_L)$ in what follows, and have introduced 
restricted scalar products only as a device for uniquely determining 
$S(\{x\}_N,\{b\}_N | \{y\}_L)$.   

\section{Colouring the $A_1$ lattice configurations}
\label{section.colouring.00}

So far we have considered results related to the $A_1$ vertex model, 
and hence all lattice configurations encountered contain only the state
variables labelled $\{0, 1\}$. In this and the following sections we 
remove this restriction, and consider lattice configurations which 
allow all state variables $\{0, \dots, n\}$ to appear. 
We call this process {\it colouring}. The point of these sections 
is to show that, by colouring the $A_1$ DWPF and scalar products 
appropriately, they remain invariant. We describe this invariance by 
calling the relevant configurations {\it colour-independent.}

This short section contains identities which are used throughout the 
rest of the paper, while Sections {\bref{section.colouring.01}} and 
{\bref{section.colouring.02}} study the colouring of the 
of the DWPF and scalar product, respectively.

\subsection{An identity satisfied by vertex weights}

We shortly make use of the identity
\begin{align}
1 = a(x,y) = b_{\pm}(x,y) + c_{\pm}(x,y)
\label{wt-id} 
\end{align}
which is true for both the rational (\ref{rat-wt}) and trigonometric 
(\ref{trig-wt}) parametrizations.

\subsection{A trivial partition function}

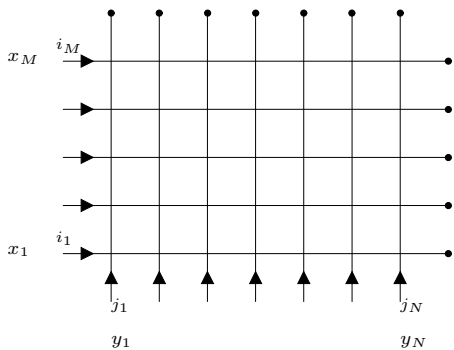
\begin{figure}

\begin{center}
\begin{minipage}{4.3in}

\setlength{\unitlength}{0.00032cm}
\begin{picture}(20000,17500)(-11000,-5000)

\blacken\path(-1250,250)(-1250,-250)(-750,000)(-1250,250)
\path(-2000,0)(14000,0)
\put(14000,0000){\circle*{250}}
\put(-2300,500){\tiny $i_1$}
\put(-4300,0){\scriptsize$x_1$}

\blacken\path(-1250,2250)(-1250,1750)(-750,2000)(-1250,2250)
\path(-2000,2000)(14000,2000)
\put(14000,2000){\circle*{250}}

\blacken\path(-1250,4250)(-1250,3750)(-750,4000)(-1250,4250)
\path(-2000,4000)(14000,4000)
\put(14000,4000){\circle*{250}}

\blacken\path(-1250,6250)(-1250,5750)(-750,6000)(-1250,6250)
\path(-2000,6000)(14000,6000)
\put(14000,6000){\circle*{250}}

\blacken\path(-1250,8250)(-1250,7750)(-750,8000)(-1250,8250)
\path(-2000,8000)(14000,8000)
\put(14000,8000){\circle*{250}}
\put(-2300,8500){\tiny $i_M$}
\put(-4300,8000){\scriptsize$x_M$}


\blacken\path(-250,-1250)(250,-1250)(000,-750)(-250,-1250)
\path(0,-2000)(0,10000)
\put(0,10000){\circle*{250}}
\put(0,-2300){\tiny $j_1$}
\put(0,-3750){\scriptsize$y_1$}

\blacken\path(1750,-1250)(2250,-1250)(2000,-750)(1750,-1250)
\path(2000,-2000)(2000,10000)
\put(2000,10000){\circle*{250}}

\blacken\path(3750,-1250)(4250,-1250)(4000,-750)(3750,-1250)
\path(4000,-2000)(4000,10000)
\put(4000,10000){\circle*{250}}

\blacken\path(5750,-1250)(6250,-1250)(6000,-750)(5750,-1250)
\path(6000,-2000)(6000,10000)
\put(6000,10000){\circle*{250}}

\blacken\path(7750,-1250)(8250,-1250)(8000,-750)(7750,-1250)
\path(8000,-2000)(8000,10000)
\put(8000,10000){\circle*{250}}

\blacken\path(9750,-1250)(10250,-1250)(10000,-750)(9750,-1250)
\path(10000,-2000)(10000,10000)
\put(10000,10000){\circle*{250}}

\blacken\path(11750,-1250)(12250,-1250)(12000,-750)(11750,-1250)
\path(12000,-2000)(12000,10000)
\put(12000,10000){\circle*{250}}
\put(12000,-2300){\tiny $j_N$}
\put(12000,-3750){\scriptsize$y_{N}$}

\end{picture}

\end{minipage}
\end{center}

\caption{Lattice for the partition function 
$f_{[i_1,\dots,i_M],[j_1,\dots,j_N]}(\{x\}_M | \{y\}_N)$. Left and bottom 
boundary segments are fixed to the definite colours shown, while the top 
and right boundary segments are summed over all values.}
\label{fig-f}

\end{figure}

Let $f_{[i_1,\dots,i_M],[j_1,\dots,j_N]}(\{x\}_M | \{y\}_N)$ denote 
the partition function of an $M \times N$ lattice with horizontal 
rapidities $\{x_1,\dots,x_M\}$ and vertical rapidities 
$\{y_1,\dots,y_N\}$, whose left and bottom boundary segments are 
fixed
\footnote{\michaelwrites{
Subsequently, whenever we write $\{k_1,\dots,k_N\} \in \{0,1,\dots,n\}$, we mean that 
$k_1,\dots,k_N$ take fixed values in the set $\{0,1,\dots,n\}$. However, it is \emph{not} true 
that $\{k_1,\dots,k_N\} \subseteq \{0,1,\dots,n\}$ since arbitrarily many of the $k_i$ can be equal, and 
furthermore we say nothing about the relationship between the cardinalities $N$ and $n+1$. 
}}
to colours \michaelwrites{$\{i_1,\dots,i_M\}$ $\in \{0,1,\dots,n\}$} and 
\michaelwrites{$\{j_1,\dots,j_N\} \in \{0,1,\dots,n\}$} 
respectively, while the top and right boundary segments are summed 
over all colours. We represent this partition function by the lattice 
shown in Figure {\bf\ref{fig-f}}.
%

\begin{myLemma}
For every $M \times N$ lattice and all choices of \michaelwrites{$\{i_1,\dots,i_M\} \in \{0,1,\dots,n\}$} and \michaelwrites{$\{j_1,\dots,j_N\}$ $\in \{0,1,\dots,n\}$}, we have
\begin{align}
f_{[i_1,\dots,i_M],[j_1,\dots,j_N]}
\ll \{x\}_M \Big| \{y\}_N \rr
=
1
\label{f=1}
\end{align}
\label{lem-id} 
\end{myLemma}

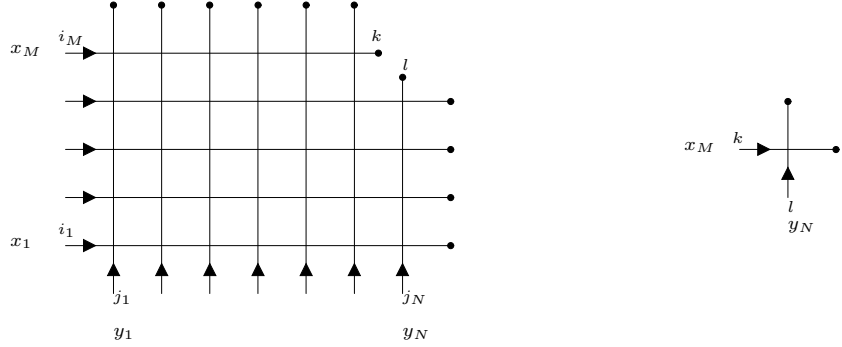
\begin{figure}

\begin{center}
\begin{minipage}{4.3in}

\setlength{\unitlength}{0.00032cm}
\begin{picture}(20000,17500)(-3000,-5000)

\blacken\path(26750,4250)(26750,3750)(27250,4000)(26750,4250)
\path(26000,4000)(30000,4000)
\put(30000,4000){\circle*{250}}
\put(25700,4250){\tiny $k$}
\put(23700,4000){\scriptsize$x_M$}

\blacken\path(27750,2750)(28250,2750)(28000,3250)(27750,2750)
\path(28000,2000)(28000,6000)
\put(28000,6000){\circle*{250}}
\put(28000,1400){\tiny $l$}
\put(28000,700){\scriptsize$y_N$}


\blacken\path(-1250,250)(-1250,-250)(-750,000)(-1250,250)
\path(-2000,0)(14000,0)
\put(14000,0){\circle*{250}}
\put(-2300,500){\tiny $i_1$}
\put(-4300,0){\scriptsize$x_1$}

\blacken\path(-1250,2250)(-1250,1750)(-750,2000)(-1250,2250)
\path(-2000,2000)(14000,2000)
\put(14000,2000){\circle*{250}}

\blacken\path(-1250,4250)(-1250,3750)(-750,4000)(-1250,4250)
\path(-2000,4000)(14000,4000)
\put(14000,4000){\circle*{250}}

\blacken\path(-1250,6250)(-1250,5750)(-750,6000)(-1250,6250)
\path(-2000,6000)(14000,6000)
\put(14000,6000){\circle*{250}}

\blacken\path(-1250,8250)(-1250,7750)(-750,8000)(-1250,8250)
\path(-2000,8000)(11000,8000)
\put(11000,8000){\circle*{250}}
\put(10700,8500){\tiny $k$}
\put(-2300,8500){\tiny $i_M$}
\put(-4300,8000){\scriptsize$x_M$}


\blacken\path(-250,-1250)(250,-1250)(000,-750)(-250,-1250)
\path(0,-2000)(0,10000)
\put(0,10000){\circle*{250}}
\put(0,-2300){\tiny $j_1$}
\put(0,-3750){\scriptsize$y_1$}

\blacken\path(1750,-1250)(2250,-1250)(2000,-750)(1750,-1250)
\path(2000,-2000)(2000,10000)
\put(2000,10000){\circle*{250}}

\blacken\path(3750,-1250)(4250,-1250)(4000,-750)(3750,-1250)
\path(4000,-2000)(4000,10000)
\put(4000,10000){\circle*{250}}

\blacken\path(5750,-1250)(6250,-1250)(6000,-750)(5750,-1250)
\path(6000,-2000)(6000,10000)
\put(6000,10000){\circle*{250}}

\blacken\path(7750,-1250)(8250,-1250)(8000,-750)(7750,-1250)
\path(8000,-2000)(8000,10000)
\put(8000,10000){\circle*{250}}

\blacken\path(9750,-1250)(10250,-1250)(10000,-750)(9750,-1250)
\path(10000,-2000)(10000,10000)
\put(10000,10000){\circle*{250}}

\blacken\path(11750,-1250)(12250,-1250)(12000,-750)(11750,-1250)
\path(12000,-2000)(12000,7000)
\put(12000,7000){\circle*{250}}
\put(12000,7300){\tiny $l$}
\put(12000,-2300){\tiny $j_N$}
\put(12000,-3750){\scriptsize$y_{N}$}

\end{picture}

\end{minipage}
\end{center}

\caption{The top-right vertex of the lattice for a particular choice 
of incoming colours is on the right. When $k=l$, the summation of the 
top and right segments becomes trivial, and we obtain an $a(x_M,y_N)$ 
vertex. When $k\not= l$, we use equation (\ref{wt-id}) to write the 
sum of vertices as a single $a(x_M,y_N)$ vertex. The lattice which 
results from the omission of the top-right vertex is on the right. 
The colours $k$ and $l$ should be considered summed.}
\label{fig-pr-lem1}

\end{figure}

\myProof
Consider the vertex at the top-right corner of Figure {\bf\ref{fig-f}}. 
For a particular choice of its left and lower incoming colours, $k$ 
and $l$ respectively, it has the form on the right of Figure 
{\bf\ref{fig-pr-lem1}}, see the caption.
%
We conclude that the top-right vertex contributes the weight 1 to 
the partition function, regardless of its incoming colours. Hence 
we can ignore its contribution and study the equivalent partition 
function on the left of Figure {\bf\ref{fig-pr-lem1}}. 
Repeating this argument for all vertices in the top row, and then 
for each subsequent row of the lattice, one ultimately removes all 
vertices. The statement (\ref{f=1}) follows.

\proofend 

\section{Introducing colour variables into $A_1$ domain wall 
configurations}
\label{section.colouring.01}

\subsection{Colouring the left and top boundaries}

Recall the domain wall configuration shown in 
Figure {\bf\ref{fig-dwpf}}. The bottom and right boundary segments 
are assigned the colour 0, while the left and top boundary segments 
are assigned the colour 1. One can think of the colours 1 as black 
lines which flow into the lattice from the left and exit from the 
top, while the colours 0 are simply a colourless, or white background.

\begin{figure}

\begin{center}
\begin{minipage}{4.3in}

\setlength{\unitlength}{0.00032cm}
\begin{picture}(40000,15000)(3000,8500)


\blacken\path(14750,20250)(14750,19750)(15250,20000)(14750,20250)
\path(14000,20000)(26000,20000) \put(12000,20000){\tiny $x_N$} 
\put(14000,20500){\tiny $i_N$}  
\put(25500,20500){\tiny 0}

\blacken\path(14750,18250)(14750,17750)(15250,18000)(14750,18250)
\path(14000,18000)(26000,18000) 
\put(25500,18500){\tiny 0}

\blacken\path(14750,16250)(14750,15750)(15250,16000)(14750,16250)
\path(14000,16000)(26000,16000)
\put(25500,16500){\tiny 0}

\blacken\path(14750,14250)(14750,13750)(15250,14000)(14750,14250)
\path(14000,14000)(26000,14000)
\put(12000,14000){\tiny $x_2$}
\put(14000,14500){\tiny $i_{2}$}  
\put(25500,14500){\tiny 0}

\blacken\path(14750,12250)(14750,11750)(15250,12000)(14750,12250)
\path(14000,12000)(26000,12000) \put(12000,12000){\tiny $x_{1}$}
\put(14000,12500){\tiny $i_1$} 
\put(25500,12500){\tiny 0}


\blacken\path(15750,10750)(16250,10750)(16000,11250)(15750,10750)
\path(16000,10000)(16000,22000) \put(15500,8000){\tiny $y_1$}
\put(16000,9000){\tiny 0}
\put(16000,22000){\circle*{250}} 

\blacken\path(17750,10750)(18250,10750)(18000,11250)(17750,10750)
\path(18000,10000)(18000,22000)
\put(18000,9000){\tiny 0}
\put(18000,22000){\circle*{250}} 
 
\blacken\path(19750,10750)(20250,10750)(20000,11250)(19750,10750)
\path(20000,10000)(20000,22000)
\put(20000,9000){\tiny 0}
\put(20000,22000){\circle*{250}} 

\blacken\path(21750,10750)(22250,10750)(22000,11250)(21750,10750) 
\path(22000,10000)(22000,22000)
\put(22000,9000){\tiny 0}
\put(22000,22000){\circle*{250}} 

\blacken\path(23750,10750)(24250,10750)(24000,11250)(23750,10750)  
\path(24000,10000)(24000,22000) \put(23500,8000){\tiny $y_{N}$}
\put(24000,9000){\tiny 0}
\put(24000,22000){\circle*{250}}

\end{picture}

\end{minipage}
\end{center}

\caption{Lattice representation of $Z_{[i_1,\dots,i_N]}(\{x\}_N | \{y\}_N)$. 
The colours on the left boundary segments are fixed to the definite values 
$\{i_1,\dots,i_N\} \in \{1,2,\dots, n\}$, while those on the top boundary 
segments are summed over all values $\{1,2,\dots,n\}$, which is indicated 
by the dots placed on these segments.}
\label{fig-cdwpf}

\end{figure}
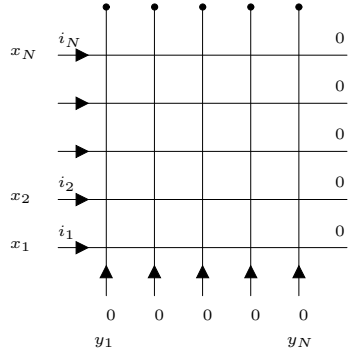

Let us extend our attention to lattice configurations in an $A_n$ 
vertex model, where $n \geq 2$, and assume that some or all of the 
colours which come in from the left take values in $\{2,\dots,n\}$. 
We denote these left-incoming colours by $\{i_1,\dots,i_N\} \in 
\{1,2,\dots,n\}$. Further, assume that the top boundary segments 
are summed over all possible values in $\{1,2,\dots,n\}$. We call 
the resulting lattice a {\it coloured} domain wall configuration, 
denoted by $Z_{[i_1,\dots,i_N]}(\{x\}_N | \{y\}_N)$, and represent 
it as in Figure {\bf\ref{fig-cdwpf}}.

\begin{myLemma}
For all choices of introduced colours \michaelwrites{$\{i_1,\dots,i_N\} \in \{1,2,\dots,n\}$}, we have
\begin{align}
Z_{[i_1,\dots,i_N]}
\ll \{x\}_N \Big| \{y\}_N \rr
=
Z
\ll \{x\}_N \Big| \{y\}_N \rr
\label{zc=z}
\end{align}
In other words, the partition function in Figure {\bref{fig-cdwpf}} is 
colour-independent, and behaves as if all left and top-edge colours 
were 1, which is a black and white configuration.
\label{lem-cdwpf}
\end{myLemma}

\myProof The proof relies on showing that 
$Z_{[i_1,\dots,i_N]}(\{x\}_N | \{y\}_N) $ satisfies the set of properties 
{\bf A}--{\bf D} in Section {\bf\ref{ssec-pf-prop}}, which uniquely 
characterize the DWPF.

{\bf A.} Consider the top row of the lattice in Figure 
{\bf\ref{fig-cdwpf}}, through which $x_N$ flows. Due to the need to conserve 
the colour $i_N \geq 1$ flowing in from the left, it follows that in every 
configuration there must be at least one $c_{-}$ vertex in the top row. As 
we have already argued in Section {\bref{section.dwpf}}, this is sufficient 
to show that

\begin{align}
Z_{[i_1,\dots,i_N]} \ll \{x\}_N \Big| \{y\}_N \rr
=
\frac{P(\{x\}_N | \{y\}_N)}{\prod_{i,j=1}^{N} (x_i-y_j+1)}
\end{align}
where $P(\{x\}_N | \{y\}_N)$ is a polynomial (maximally) of degree $N-1$ in $x_N$. 

{\bf B.} To establish symmetry in $\{y_1,\dots,y_N\}$, one follows the same 
procedure outlined in Section {\bf\ref{ssec-pf-prop}}. Namely, one attaches 
an $a(y_{j+1},y_j)$ vertex at the base of the lattice in 
Figure {\bf\ref{fig-cdwpf}}. Using the Yang-Baxter equation, it is threaded 
vertically through the lattice until it emerges from the top. 

The only difference is that the emerging vertex is not of the type $a(y_{j+1},y_j)$, 
rather it is a sum over a $b_{\pm}(y_{j+1},y_j)$ and $c_{\pm}(y_{j+1},y_j)$ vertex, 
due to the summed boundary condition at the top of the lattice. But from (\ref{wt-id}), 
such a sum produces an $a(y_{j+1},y_j)$ vertex. Hence the $j$-th and $(j+1)$-th lattice 
lines can be freely swapped, and in general the lattice in Figure {\bf\ref{fig-cdwpf}} 
is invariant under permutations of its vertical lines.

\begin{figure}

\begin{center}
\begin{minipage}{4.3in}

\setlength{\unitlength}{0.00032cm}
\begin{picture}(40000,16000)(14000,8500)



\blacken\path(14750,20250)(14750,19750)(15250,20000)(14750,20250)
\path(14000,20000)(23000,20000) \put(11500,20000){\tiny $x_N$} 
\put(14000,20500){\tiny $i_N$}  
\put(23000,20000){\circle*{250}}

\blacken\path(14750,18250)(14750,17750)(15250,18000)(14750,18250)
\path(14000,18000)(26000,18000) \put(11500,18000){\tiny $x_{N-1}$}
\put(14000,18500){\tiny $i_{N-1}$}  
\put(25500,18500){\tiny 0}

\blacken\path(14750,16250)(14750,15750)(15250,16000)(14750,16250)
\path(14000,16000)(26000,16000)  
\put(25500,16500){\tiny 0}

\blacken\path(14750,14250)(14750,13750)(15250,14000)(14750,14250)
\path(14000,14000)(26000,14000)
\put(25500,14500){\tiny 0}

\blacken\path(14750,12250)(14750,11750)(15250,12000)(14750,12250)
\path(14000,12000)(26000,12000) \put(11500,12000){\tiny $x_1$}
\put(14000,12500){\tiny $i_1$} 
\put(25500,12500){\tiny 0}


\blacken\path(15750,10750)(16250,10750)(16000,11250)(15750,10750)
\path(16000,10000)(16000,22000) \put(15500,8000){\tiny $y_1$}
\put(16000,9000){\tiny 0}
\put(16000,22000){\circle*{250}} 

\blacken\path(17750,10750)(18250,10750)(18000,11250)(17750,10750)
\path(18000,10000)(18000,22000)
\put(18000,9000){\tiny 0}
\put(18000,22000){\circle*{250}} 
 
\blacken\path(19750,10750)(20250,10750)(20000,11250)(19750,10750)
\path(20000,10000)(20000,22000) 
\put(20000,9000){\tiny 0}
\put(20000,22000){\circle*{250}} 

\blacken\path(21750,10750)(22250,10750)(22000,11250)(21750,10750) 
\path(22000,10000)(22000,22000) \put(21500,8000){\tiny $y_{N-1}$}
\put(22000,9000){\tiny 0} 
\put(22000,22000){\circle*{250}} 

\blacken\path(23750,10750)(24250,10750)(24000,11250)(23750,10750) 
\path(24000,10000)(24000,19000) \put(24000,8000){\tiny $y_{N}$}
\put(24000,9000){\tiny 0} \put(24000,19500){\tiny 0}


\put(29000,16000){$=$}



\blacken\path(34750,22250)(34750,21750)(35250,22000)(34750,22250)
\path(34000,22000)(44000,22000) \put(31500,22000){\tiny $x_N$} 
\put(34000,22500){\tiny $i_N$}  
\put(44000,22000){\circle*{250}}

\blacken\path(34750,18250)(34750,17750)(35250,18000)(34750,18250)
\path(34000,18000)(48000,18000) \put(31500,18000){\tiny $x_{N-1}$}
\put(34000,18500){\tiny $i_{N-1}$}
\put(44000,18500){\tiny 0}  
\put(47500,18500){\tiny 0}

\blacken\path(34750,16250)(34750,15750)(35250,16000)(34750,16250)
\path(34000,16000)(48000,16000)
\put(44000,16500){\tiny 0}  
\put(47500,16500){\tiny 0}

\blacken\path(34750,14250)(34750,13750)(35250,14000)(34750,14250)
\path(34000,14000)(48000,14000)
\put(44000,14500){\tiny 0}  
\put(47500,14500){\tiny 0}

\blacken\path(34750,12250)(34750,11750)(35250,12000)(34750,12250)
\path(34000,12000)(48000,12000) \put(31500,12000){\tiny $x_1$}
\put(34000,12500){\tiny $i_1$}
\put(44000,12500){\tiny 0} 
\put(47500,12500){\tiny 0}


\blacken\path(35750,10750)(36250,10750)(36000,11250)(35750,10750)
\path(36000,10000)(36000,24000) \put(35500,8000){\tiny $y_1$}
\put(36000,9000){\tiny 0} 
\put(36000,20000){\circle*{250}}
\put(36000,24000){\circle*{250}}

\blacken\path(37750,10750)(38250,10750)(38000,11250)(37750,10750)
\path(38000,10000)(38000,24000)
\put(38000,9000){\tiny 0} 
\put(38000,20000){\circle*{250}}
\put(38000,24000){\circle*{250}} 

\blacken\path(39750,10750)(40250,10750)(40000,11250)(39750,10750)
\path(40000,10000)(40000,24000) 
\put(40000,9000){\tiny 0} 
\put(40000,20000){\circle*{250}}
\put(40000,24000){\circle*{250}}

\blacken\path(41750,10750)(42250,10750)(42000,11250)(41750,10750)
\path(42000,10000)(42000,24000) \put(41500,8000){\tiny $y_{N-1}$}
\put(42000,9000){\tiny 0} 
\put(42000,20000){\circle*{250}}
\put(42000,24000){\circle*{250}}

\blacken\path(45750,10750)(46250,10750)(46000,11250)(45750,10750)
\path(46000,10000)(46000,20000) \put(46000,8000){\tiny $y_{N}$}
\put(46000,9000){\tiny 0} \put(46000,20500){\tiny 0}

\end{picture}

\end{minipage}
\end{center}

\caption{Setting $x_N = y_N$ makes $Z_{[i_1,\dots,i_N]}(\{x\}_N|\{y\}_N)$ 
equal to the partition function on the left. Due to Lemma {\bf\ref{lem-id}}, 
the top line of the lattice gives only a trivial contribution to the partition 
function, and the right line is trivial as before. The remaining part of the 
lattice, on the right, is $Z_{[i_1,\dots,i_{N-1}]}(\{x\}_{N-1}|\{y\}_{N-1})$.}
\label{fig-cdwpf-rec}

\end{figure}
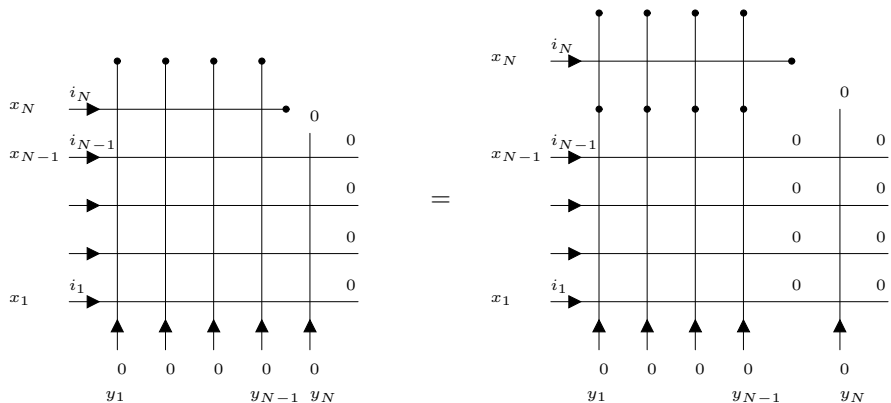

{\bf C.} Consider the effect of setting $x_N = y_N$ in the lattice in 
Figure {\bf\ref{fig-cdwpf}}. As we have seen before, this removes the possibility 
that the top-right vertex can be a $b$ vertex and sets the weight of the resultant 
$c_{-}$ vertex to 1. Hence the vertex at the intersection of the top and right-most 
lines splits, and we obtain the lattice on the left of Figure 
{\bf\ref{fig-cdwpf-rec}}.
%
Observe that the right-most line of the lattice contributes a common factor of $a$ vertices 
to the sum, hence it can be neglected. Similarly, using Lemma {\bref{lem-id}}, the top line 
of vertices contributes a common factor of 1 to the sum, and can be ignored. We obtain the 
lattice on the right of Figure {\bref{fig-cdwpf}}, which represents 
$Z_{[i_1,\dots,i_{N-1}]}(\{x\}_{N-1} | \{y\}_{N-1})$. This proves the required recursion 
relation, namely
\begin{align}
Z_{[i_1,\dots,i_N]} \ll \{x\}_N \Big| \{y\}_N \rr
\Big|_{x_N = y_N}
=
Z_{[i_1,\dots,i_{N-1}]} \ll \{x\}_{N-1} \Big| \{y\}_{N-1} \rr
\end{align} 

{\bf D.} It is easy to check that the correct initial condition is satisfied. Indeed, when 
$N=1$, the colour $i_1$ flowing in from the left must flow out from the top boundary. This 
constrains the sum on the top boundary to one term only, and we obtain a $c_{-}(x_1,y_1)$ vertex.

\proofend

\section{Colouring the $A_1$ scalar product configurations}
\label{section.colouring.02}

\subsection{Colouring the bottom-left and top-right 
boundaries}

Recall the $A_1$ scalar product configuration shown in 
Figure {\bf\ref{fig-sp}}. The lower-left and upper-right boundary 
segments are assigned the colours 1, which effectively flow in from 
the bottom-left and exit from the top-right.

\begin{figure}

\begin{center}
\begin{minipage}{4.3in}

\setlength{\unitlength}{0.00032cm}
\begin{picture}(40000,22000)(5000,2500)


\blacken\path(14750,20250)(14750,19750)(15250,20000)(14750,20250)
\path(14000,20000)(30000,20000) \put(12000,20000){\tiny $b_N$}
\put(14000,20500){\tiny 0}  
\put(30000,20000){\circle*{250}}

\blacken\path(14750,18250)(14750,17750)(15250,18000)(14750,18250)
\path(14000,18000)(30000,18000) 
\put(14000,18500){\tiny 0}  
\put(30000,18000){\circle*{250}}

\blacken\path(14750,16250)(14750,15750)(15250,16000)(14750,16250)
\path(14000,16000)(30000,16000) 
\put(14000,16500){\tiny 0}  
\put(30000,16000){\circle*{250}}

\blacken\path(14750,14250)(14750,13750)(15250,14000)(14750,14250)
\path(14000,14000)(30000,14000) \put(12000,14000){\tiny $b_1$}
\put(14000,14500){\tiny 0}  
\put(30000,14000){\circle*{250}}


\blacken\path(14750,12250)(14750,11750)(15250,12000)(14750,12250)
\path(14000,12000)(30000,12000) \put(12000,12000){\tiny $x_N$}
\put(14000,12500){\tiny $i_N$} 
\put(29500,12500){\tiny 0}

\blacken\path(14750,10250)(14750,9750)(15250,10000)(14750,10250)
\path(14000,10000)(30000,10000)    
\put(29500,10500){\tiny 0}

\blacken\path(14750,8250)(14750,7750)(15250,8000)(14750,8250)
\path(14000,8000)(30000,8000)
\put(29500,8500){\tiny 0}

\blacken\path(14750,6250)(14750,5750)(15250,6000)(14750,6250)
\path(14000,6000)(30000,6000) \put(12000,6000){\tiny $x_1$} 
\put(14000,6500){\tiny $i_1$}  
\put(29500,6500){\tiny 0}


\blacken\path(15750,4750)(16250,4750)(16000,5250)(15750,4750)
\path(16000,4000)(16000,22000) \put(15500,2000){\tiny $y_1$}
\put(16000,3000){\tiny 0} \put(15500,22500){\tiny 0}

\blacken\path(17750,4750)(18250,4750)(18000,5250)(17750,4750)
\path(18000,4000)(18000,22000)
\put(18000,3000){\tiny 0} \put(17500,22500){\tiny 0}

\blacken\path(19750,4750)(20250,4750)(20000,5250)(19750,4750)
\path(20000,4000)(20000,22000)
\put(20000,3000){\tiny 0} \put(19500,22500){\tiny 0}

\blacken\path(21750,4750)(22250,4750)(22000,5250)(21750,4750)
\path(22000,4000)(22000,22000)
\put(22000,3000){\tiny 0} \put(21500,22500){\tiny 0}

\blacken\path(23750,4750)(24250,4750)(24000,5250)(23750,4750)
\path(24000,4000)(24000,22000) 
\put(24000,3000){\tiny 0} \put(23500,22500){\tiny 0}

\blacken\path(25750,4750)(26250,4750)(26000,5250)(25750,4750)
\path(26000,4000)(26000,22000) 
\put(26000,3000){\tiny 0} \put(25500,22500){\tiny 0}

\blacken\path(27750,4750)(28250,4750)(28000,5250)(27750,4750)
\path(28000,4000)(28000,22000) \put(27500,2000){\tiny $y_L$}
\put(28000,3000){\tiny 0} \put(27500,22500){\tiny 0}

\end{picture}

\end{minipage}
\end{center}

\caption{Lattice representation of $S_{[i_1,\dots,i_N]}(\{x\}_N, \{b\}_N | \{y\}_L)$. 
The colours on the lower-left boundary segments are fixed to the definite values 
$\{i_1,\dots,i_N\} \in \{1,2,\dots,n\}$, while those on the upper-right segments are 
summed over all values $\{1,2,\dots,n\}$, as indicated by the dots.}

\label{fig-csp}

\end{figure}
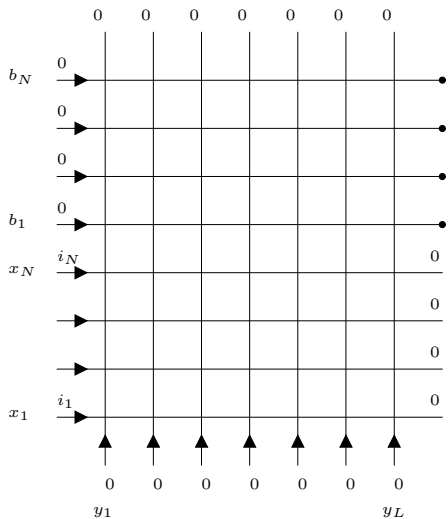

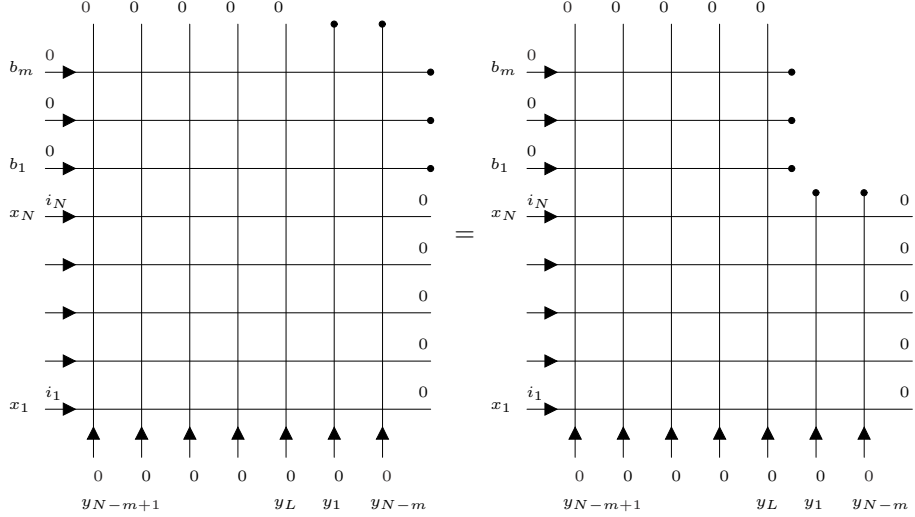
\begin{figure}

\begin{center}
\begin{minipage}{4.3in}

\setlength{\unitlength}{0.00032cm}
\begin{picture}(40000,22000)(15000,2500)



\blacken\path(14750,20250)(14750,19750)(15250,20000)(14750,20250)
\path(14000,20000)(30000,20000) \put(12500,20000){\tiny $b_{\m{m}}$}
\put(14000,20500){\tiny 0}  
\put(30000,20000){\circle*{250}}

\blacken\path(14750,18250)(14750,17750)(15250,18000)(14750,18250)
\path(14000,18000)(30000,18000) 
\put(14000,18500){\tiny 0}  
\put(30000,18000){\circle*{250}}

\blacken\path(14750,16250)(14750,15750)(15250,16000)(14750,16250)
\path(14000,16000)(30000,16000) \put(12500,16000){\tiny $b_1$}
\put(14000,16500){\tiny 0}  
\put(30000,16000){\circle*{250}}


\blacken\path(14750,14250)(14750,13750)(15250,14000)(14750,14250)
\path(14000,14000)(30000,14000) \put(12500,14000){\tiny $x_N$}
\put(14000,14500){\tiny $i_N$}  
\put(29500,14500){\tiny 0}

\blacken\path(14750,12250)(14750,11750)(15250,12000)(14750,12250)
\path(14000,12000)(30000,12000) 
\put(29500,12500){\tiny 0}

\blacken\path(14750,10250)(14750,9750)(15250,10000)(14750,10250)
\path(14000,10000)(30000,10000)  
\put(29500,10500){\tiny 0}

\blacken\path(14750,8250)(14750,7750)(15250,8000)(14750,8250)
\path(14000,8000)(30000,8000)
\put(29500,8500){\tiny 0}

\blacken\path(14750,6250)(14750,5750)(15250,6000)(14750,6250)
\path(14000,6000)(30000,6000) \put(12500,6000){\tiny $x_1$} 
\put(14000,6500){\tiny $i_1$}  
\put(29500,6500){\tiny 0}


\blacken\path(15750,4750)(16250,4750)(16000,5250)(15750,4750)
\path(16000,4000)(16000,22000) \put(15500,2000){\tiny $y_{N-\m{m}+1}$}
\put(16000,3000){\tiny 0} \put(15500,22500){\tiny 0}

\blacken\path(17750,4750)(18250,4750)(18000,5250)(17750,4750)
\path(18000,4000)(18000,22000)
\put(18000,3000){\tiny 0} \put(17500,22500){\tiny 0}

\blacken\path(19750,4750)(20250,4750)(20000,5250)(19750,4750)
\path(20000,4000)(20000,22000)
\put(20000,3000){\tiny 0} \put(19500,22500){\tiny 0}

\blacken\path(21750,4750)(22250,4750)(22000,5250)(21750,4750)
\path(22000,4000)(22000,22000)
\put(22000,3000){\tiny 0} \put(21500,22500){\tiny 0}

\blacken\path(23750,4750)(24250,4750)(24000,5250)(23750,4750)
\path(24000,4000)(24000,22000) \put(23500,2000){\tiny $y_L$}
\put(24000,3000){\tiny 0} \put(23500,22500){\tiny 0}

\blacken\path(25750,4750)(26250,4750)(26000,5250)(25750,4750)
\path(26000,4000)(26000,22000) \put(25500,2000){\tiny $y_1$}
\put(26000,3000){\tiny 0} \put(26000,22000){\circle*{250}}

\blacken\path(27750,4750)(28250,4750)(28000,5250)(27750,4750)
\path(28000,4000)(28000,22000) \put(27500,2000){\tiny $y_{N-\m{m}}$}
\put(28000,3000){\tiny 0} \put(28000,22000){\circle*{250}}


\put(31000,13000){$=$}



\blacken\path(34750,20250)(34750,19750)(35250,20000)(34750,20250)
\path(34000,20000)(45000,20000) \put(32500,20000){\tiny $b_{\m{m}}$}
\put(34000,20500){\tiny 0}  
\put(45000,20000){\circle*{250}}

\blacken\path(34750,18250)(34750,17750)(35250,18000)(34750,18250)
\path(34000,18000)(45000,18000) 
\put(34000,18500){\tiny 0}  
\put(45000,18000){\circle*{250}}

\blacken\path(34750,16250)(34750,15750)(35250,16000)(34750,16250)
\path(34000,16000)(45000,16000) \put(32500,16000){\tiny $b_1$}
\put(34000,16500){\tiny 0}  
\put(45000,16000){\circle*{250}}


\blacken\path(34750,14250)(34750,13750)(35250,14000)(34750,14250)
\path(34000,14000)(50000,14000) \put(32500,14000){\tiny $x_N$}
\put(34000,14500){\tiny $i_N$}  
\put(49500,14500){\tiny 0}

\blacken\path(34750,12250)(34750,11750)(35250,12000)(34750,12250)
\path(34000,12000)(50000,12000)  
\put(49500,12500){\tiny 0}

\blacken\path(34750,10250)(34750,9750)(35250,10000)(34750,10250)
\path(34000,10000)(50000,10000)    
\put(49500,10500){\tiny 0}

\blacken\path(34750,8250)(34750,7750)(35250,8000)(34750,8250)
\path(34000,8000)(50000,8000)  
\put(49500,8500){\tiny 0}

\blacken\path(34750,6250)(34750,5750)(35250,6000)(34750,6250)
\path(34000,6000)(50000,6000) \put(32500,6000){\tiny $x_1$} 
\put(34000,6500){\tiny $i_1$}  
\put(49500,6500){\tiny 0}


\blacken\path(35750,4750)(36250,4750)(36000,5250)(35750,4750)
\path(36000,4000)(36000,22000) \put(35500,2000){\tiny $y_{N-\m{m}+1}$}
\put(36000,3000){\tiny 0} \put(35500,22500){\tiny 0}

\blacken\path(37750,4750)(38250,4750)(38000,5250)(37750,4750)
\path(38000,4000)(38000,22000)
\put(38000,3000){\tiny 0} \put(37500,22500){\tiny 0}

\blacken\path(39750,4750)(40250,4750)(40000,5250)(39750,4750)
\path(40000,4000)(40000,22000)
\put(40000,3000){\tiny 0} \put(39500,22500){\tiny 0}

\blacken\path(41750,4750)(42250,4750)(42000,5250)(41750,4750)
\path(42000,4000)(42000,22000)
\put(42000,3000){\tiny 0} \put(41500,22500){\tiny 0}

\blacken\path(43750,4750)(44250,4750)(44000,5250)(43750,4750)
\path(44000,4000)(44000,22000) \put(43500,2000){\tiny $y_L$}
\put(44000,3000){\tiny 0} \put(43500,22500){\tiny 0}

\blacken\path(45750,4750)(46250,4750)(46000,5250)(45750,4750)
\path(46000,4000)(46000,15000) \put(45500,2000){\tiny $y_1$}
\put(46000,3000){\tiny 0} \put(46000,15000){\circle*{250}}

\blacken\path(47750,4750)(48250,4750)(48000,5250)(47750,4750)
\path(48000,4000)(48000,15000) \put(47500,2000){\tiny $y_{N-\m{m}}$}
\put(48000,3000){\tiny 0} \put(48000,15000){\circle*{250}}

\end{picture}

\end{minipage}
\end{center}

\caption{The lattice representation of $S_{[i_1,\dots,i_N]}(\{x\}_N, \{b\}_{\m{m}} | \{y\}_L)$
is on the left. The colours on the lower-left boundary segments are fixed to the definite 
values $\{i_1,\dots,i_N\} \in \{1,2,\dots,n\}$, while those on the upper-right segments are 
summed over all values $\{1,2,\dots,n\}$, as indicated by the dots. 
Using Lemma {\bf\ref{lem-id}}, the top-right corner of the lattice gives no contribution 
to the partition function, and can be neglected as on the right.}

\label{fig-csp-int}

\end{figure}

In analogy with the previous section, we now generalize to $A_n$ 
configurations with $n \geq 2$. We assume that some or all of the 
colours which come in from the lower-left boundary take values in 
$\{2,\dots,n\}$. As before we denote these incoming colours by 
$\{i_1,\dots,i_N\} \in \{1,2,\dots,n\}$. Further, assume that the 
upper-right boundary segments are summed over all possible values 
in $\{1,2,\dots,n\}$. We call the resulting lattice a coloured 
scalar product configuration, denoted 
$S_{[i_1,\dots,i_N]}(\{x\}_N,\{b\}_N | \{y\}_L)$, and represent 
it as in Figure {\bf\ref{fig-csp}}.
%
More generally, we can colour the restricted scalar products introduced 
in Section {\bf\ref{ssec-int-sp}}. The $N$ lower-left edges are again 
assigned the fixed values $\{i_1,\dots,i_N\} \in \{1,2,\dots,n\}$. 
Summation over all colours $\{1,2,\dots,n\}$ takes place on the edges 
in the upper-right corner of the lattice, see Figure {\bref{fig-csp-int}}.
%

\begin{myLemma}
For all choices of introduced colours \michaelwrites{$\{i_1,\dots,i_N\} \in \{1,2,\dots,n\}$, and 
$0 \leq m \leq N$,}  we have
\begin{align}
S_{[i_1,\dots,i_N]} \ll \{x\}_N, \{b\}_{\m{m}} \Big| \{y\}_L \rr
=
S \ll \{x\}_N, \{b\}_{\m{m}} \Big| \{y\}_L \rr
\label{sc=s}
\end{align}
In other words, the restricted scalar product in Figure {\bref{fig-csp-int}} is 
a colour-independent configuration, and behaves as if its edge colours were the 
same as those in Figure {\bf\ref{fig-sp-int}}. We emphasize that this result 
makes no use of the Bethe equations.
\end{myLemma}

\myProof
We show that $S_{[i_1,\dots,i_N]}(\{x\}_N, \{b\}_{\m{m}} | \{y\}_L)$ obeys 
the properties {\bf A}--{\bf D} in Section {\bref{ssec-int-sp}}, which 
uniquely determine the restricted scalar products.

{\bf A.} Referring to Figure {\bref{fig-csp-int}} and using Lemma 
{\bref{lem-id}}, we conclude that the top-right $\m{m}\times(N-\m{m})$ block 
of the lattice contributes a common factor of 1 to the partition 
function. Therefore we may neglect this part of the lattice, and 
obtain the configuration on the right of Figure {\bref{fig-csp-int}}. 
Now the argument proceeds in the same way as the proof of {\bf A} 
in Section \michaelwrites{{\bref{ssec-int-sp}}}. Considering the top line of the 
lattice (whose right edge is summed over all values $\{1,2,\dots,n\}$), 
it is clear that in every term it contains exactly one $c_{+}$ vertex. 
Hence we conclude that
\begin{multline}
S_{[i_1,\dots,i_N]} \ll \{x\}_N, \{b\}_{\m{m}} \Big| \{y\}_L \rr
=
\\
\frac{P(\{x\}_N,\{b\}_{\m{m}}|\{y\}_L)}
{\prod_{i=1}^{N} \prod_{j=1}^{L} (x_i-y_j+1) 
\prod_{i=1}^{\m{m}} \prod_{j=N-\m{m}+1}^{L} (b_i-y_j+1)}
\label{csp-a}
\end{multline}
where $P(\{x\}_N,\{b\}_{\m{m}} | \{y\}_L)$ is a polynomial of degree $L-N+\m{m}-1$ in $b_{\m{m}}$.

{\bf B.} Since the edge colours on the left-most $L-N+\m{m}$ vertical lines of Figure {\bref{fig-csp-int}} 
have fixed value 0 (rather than being summed), the symmetry in $\{y_{N-\m{m}+1},\dots,y_L\}$ is proved by 
the same argument used to prove {\bf B} in Section {\bref{ssec-pf-prop}}.

{\bf C.} Setting $b_{\m{m}}=y_{N-\m{m}+1}$ in Figure {\bref{fig-csp-int}} causes 
the splitting of the top-left vertex, in the same way described in the 
proof of {\bf C} in Section {\bref{ssec-int-sp}}. In fact, the argument 
illustrated by Figure {\bref{fig-sp-rec}} applies here with just a minor 
change, that each edge 1 colour is replaced by its coloured 
counterpart. Therefore, one obtains the recursion relation
\begin{align}
S_{[i_1,\dots,i_N]} \ll \{x\}_N,\{b\}_{\m{m}} \Big| \{y\}_L \rr
\Big|_{b_{\m{m}} = y_{N-\m{m}+1}}
=
S_{[i_1,\dots,i_N]} \ll \{x\}_{N},\{b\}_{\m{m}-1} \Big| \{y\}_L \rr
\label{csp-c}
\end{align} 

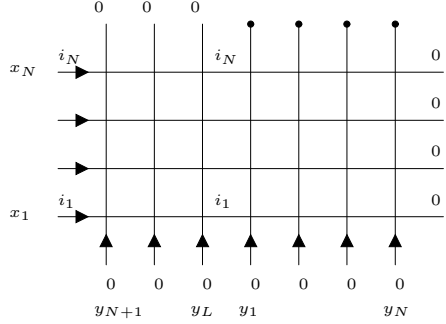
\begin{figure}
\begin{center}
\begin{minipage}{4.3in}

\setlength{\unitlength}{0.00032cm}
\begin{picture}(40000,14000)(5000,2500)


\blacken\path(14750,12250)(14750,11750)(15250,12000)(14750,12250)
\path(14000,12000)(30000,12000) \put(12000,12000){\tiny $x_N$}
\put(14000,12500){\tiny $i_N$}
\put(20500,12500){\tiny $i_N$}  
\put(29500,12500){\tiny 0}

\blacken\path(14750,10250)(14750,9750)(15250,10000)(14750,10250)
\path(14000,10000)(30000,10000)  
\put(29500,10500){\tiny 0}

\blacken\path(14750,8250)(14750,7750)(15250,8000)(14750,8250)
\path(14000,8000)(30000,8000)
\put(29500,8500){\tiny 0}

\blacken\path(14750,6250)(14750,5750)(15250,6000)(14750,6250)
\path(14000,6000)(30000,6000) \put(12000,6000){\tiny $x_1$} 
\put(14000,6500){\tiny $i_1$}
\put(20500,6500){\tiny $i_1$}    
\put(29500,6500){\tiny 0}


\blacken\path(15750,4750)(16250,4750)(16000,5250)(15750,4750)
\path(16000,4000)(16000,14000) \put(15500,2000){\tiny $y_{N+1}$}
\put(16000,3000){\tiny 0} \put(15500,14500){\tiny 0}

\blacken\path(17750,4750)(18250,4750)(18000,5250)(17750,4750)
\path(18000,4000)(18000,14000)
\put(18000,3000){\tiny 0} \put(17500,14500){\tiny 0}

\blacken\path(19750,4750)(20250,4750)(20000,5250)(19750,4750)
\path(20000,4000)(20000,14000) \put(19500,2000){\tiny $y_L$}
\put(20000,3000){\tiny 0} \put(19500,14500){\tiny 0}

\blacken\path(21750,4750)(22250,4750)(22000,5250)(21750,4750)
\path(22000,4000)(22000,14000) \put(21500,2000){\tiny $y_1$}
\put(22000,3000){\tiny 0} \put(22000,14000){\circle*{250}}

\blacken\path(23750,4750)(24250,4750)(24000,5250)(23750,4750)
\path(24000,4000)(24000,14000) 
\put(24000,3000){\tiny 0} \put(24000,14000){\circle*{250}}

\blacken\path(25750,4750)(26250,4750)(26000,5250)(25750,4750)
\path(26000,4000)(26000,14000) 
\put(26000,3000){\tiny 0} \put(26000,14000){\circle*{250}}

\blacken\path(27750,4750)(28250,4750)(28000,5250)(27750,4750)
\path(28000,4000)(28000,14000) \put(27500,2000){\tiny $y_N$}
\put(28000,3000){\tiny 0} \put(28000,14000){\circle*{250}}

\end{picture}

\end{minipage}
\end{center}

\caption{Lattice representation of $S_{[i_1,\dots,i_N]}(\{x\}_N,\{b\}_0|\{y\}_L)$. Since the 
colours $\{i_1,\dots,i_N\}$ must be conserved, the left-most $N \times (L-N)$ block is forced 
to be a product of $b$ vertices. The remaining part of the lattice represents 
$Z_{[i_1,\dots,i_N]}(\{x\}_N|\{y\}_N)$.}

\label{fig-csp-0}

\end{figure}

{\bf D.} In the case $\m{m}=0$, we see that $S_{[i_1,\dots,i_N]}(\{x\}_N,\{b\}_0 |\{y\}_L)$ 
is represented by the lattice in Figure {\bref{fig-csp-0}}.

Since $i_1,\dots,i_N \geq 1$, by colour-conservation it is clear that 
these colours pass horizontally through the left-most $L-N$ vertical 
lines, without interaction. The left $N\times (L-N)$ block of the 
lattice freezes to a product of $b$ vertices, and the remaining part 
is the coloured DWPF 
$Z_{[i_1,\dots,i_N]}(\{x\}_N |\{y\}_N) = Z(\{x\}_N|\{y\}_N)$ (by Lemma 
{\bref{lem-cdwpf}}). Hence
\begin{align}
S_{[i_1,\dots,i_N]} \ll \{x\}_N,\{b\}_0 \Big| \{y\}_L \rr 
=
\prod_{i=1}^{N}
\prod_{j=N+1}^{L}
\frac{(x_i-y_j)}{(x_i-y_j+1)}
Z \ll \{x\}_N \Big| \{y\}_N \rr
\label{csp-d}
\end{align}
 
\proofend

\section{$A_2$ scalar products}
\label{section.applications}

\subsection{Reshetikhin's off-shell/off-shell $A_2$ scalar product}
In \cite{reshetikhin}, Reshetikhin studied scalar products of Bethe 
vectors in a general class of $A_2$ quantum integrable models. 
In this general setup, the Bethe vectors are constructed from 
{\bf 1.} Elements $t_{ij}(x)$, $i < j$, of the $3\times 3$ monodromy 
matrix $T_{\alpha}(x)$, which satisfies the intertwining equation
\begin{align}
R^{(2)}_{\alpha\beta}(x,y) T_{\alpha}(x) T_{\beta}(y)
=
T_{\beta}(y) T_{\alpha}(x) R^{(2)}_{\alpha\beta}(x,y),
\end{align}
{\bf 2.} A pseudo-vacuum state $|0\rangle$, which satisfies
\begin{align}
t_{ii}(x) |0\rangle = a_i(x) |0\rangle,
\quad
t_{ij}(x) |0\rangle = 0\ \text{for all}\ i > j,
\end{align}
{\bf 3.} The nested Bethe Ansatz for building eigenvectors of 
the transfer matrix $\mathcal{T}(x) = \sum_{i=1}^{3} t_{ii}(x)$. 
One of the main results in \cite{reshetikhin} was a sum expression 
for the off-shell/off-shell scalar product $\cK_2$, where the 
dependence on the eigenvalues $a_i$ and the variables of the 
Bethe vectors was made completely explicit.

In this section we restrict our attention to a particular 
type of model with $A_2$-symmetry, namely, a spin chain whose 
monodromy matrix is constructed from a product of fundamental 
and anti-fundamental representations of the universal $R$-matrix,
\begin{align}
\label{f-antif-mon}
T_{\alpha}(x)
=
R^{(2)}_{\alpha 1}(x,y_1)
\dots
R^{(2)}_{\alpha L}(x,y_L)
R^{*(2)}_{\alpha 1^*}(x,z_1)
\dots
R^{*(2)}_{\alpha M^*}(x,z_M)
\end{align}
where
\begin{align}
R^{(2)}_{\alpha i}(x,y_i) 
\in 
{\rm End}(V_{\alpha} \otimes V_i), 
\quad\quad
R^{*(2)}_{\alpha i^*}(x,z_i) 
= 
\ll R^{(2)}_{\alpha i^*}(-x,-z_i) \rr^{{\rm t}_{i^*}} 
\in 
{\rm End}(V_{\alpha} \otimes V_{i^*})
\end{align}
and where we use an asterisk to label the second set of quantum spaces, 
$V_{i^*}$, to distinguish them from the first set $V_i$. This particular 
model was also considered in \cite{reshetikhin}, where it played a central 
role in calculating the off-shell/off-shell scalar product $\cK_2$. Our goal 
is to use the arguments developed earlier in the paper to study the scalar 
product of this model, by representing it as the partition function of 
a certain lattice configuration.

\subsection{Two versions of the off-shell/on-shell scalar product $\cS_2$}

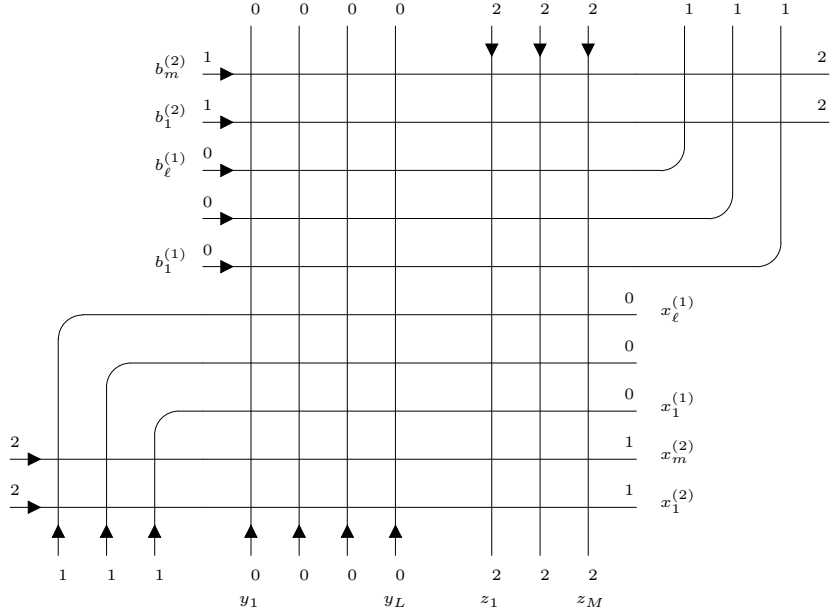
\begin{figure}

\begin{center}
\begin{minipage}{4.3in}

\setlength{\unitlength}{0.00032cm}
\begin{picture}(40000,26000)(6000,-2000)


\blacken\path(14750,20250)(14750,19750)(15250,20000)(14750,20250)
\blacken\path(14750,18250)(14750,17750)(15250,18000)(14750,18250)
\blacken\path(14750,16250)(14750,15750)(15250,16000)(14750,16250)
\blacken\path(14750,14250)(14750,13750)(15250,14000)(14750,14250)
\blacken\path(14750,12250)(14750,11750)(15250,12000)(14750,12250)

\path(14000,20000)(32000,20000)
 \put(12000,20000){\tiny $\bb_m$}
 \put(14000,20500){\tiny 1}
 \put(39500,20500){\tiny 2}

\path(14000,18000)(32000,18000)
 \put(12000,18000){\tiny $\bb_1$}
 \put(14000,18500){\tiny 1}
 \put(39500,18500){\tiny 2}

\path(14000,16000)(32000,16000)
 \put(12000,16000){\tiny $\b_{\ell}$}
 \put(14000,16500){\tiny 0}

\path(14000,14000)(32000,14000)
\put(14000,14500){\tiny 0}

\path(14000,12000)(32000,12000)
 \put(12000,12000){\tiny $\b_{1}$}
 \put(14000,12500){\tiny 0}


\path(32000,20000)(40000,20000)
\path(32000,18000)(40000,18000)

\path(32000,16000)(33000,16000)
\put(33000,17000){\arc{2000}{0}{1.5708}}
\path(34000,17000)(34000,22000)
\put(34000,22500){\tiny 1}

\path(32000,14000)(35000,14000)
\put(35000,15000){\arc{2000}{0}{1.5708}}
\path(36000,15000)(36000,22000)
\put(36000,22500){\tiny 1}

\path(32000,12000)(37000,12000)
\put(37000,13000){\arc{2000}{0}{1.5708}}
\path(38000,13000)(38000,22000)
\put(38000,22500){\tiny 1}


\blacken\path(6750,4250)(6750,3750)(7250,4000)(6750,4250)
\blacken\path(6750,2250)(6750,1750)(7250,2000)(6750,2250)

\path(14000,10000)(32000,10000)
 \put(31500,10500){\tiny 0}
 \put(33000,10000){\tiny $\x_{\ell}$}

\path(14000,8000)(32000,8000)
 \put(31500,8500){\tiny 0}

\path(14000,6000)(32000,6000)
 \put(31500,6500){\tiny 0}
 \put(33000,6000){\tiny $\x_{1}$}

\path(14000,4000)(32000,4000)
 \put(6000,4500){\tiny 2}
 \put(31500,4500){\tiny 1}
 \put(33000,4000){\tiny $\xx_m$}

\path(14000,2000)(32000,2000)
 \put(6000,2500){\tiny 2}
 \put(31500,2500){\tiny 1}
 \put(33000,2000){\tiny $\xx_1$}


\blacken\path(7750,0750)(8250,0750)(8000,1250)(7750,0750)
\blacken\path(9750,0750)(10250,0750)(10000,1250)(9750,0750)
\blacken\path(11750,0750)(12250,0750)(12000,1250)(11750,0750)

\path(9000,10000)(14000,10000)
\put(9000,9000){\arc{2000}{3.142}{4.712}}
\path(8000,9000)(8000,0000)
\put(8000,-1000){\tiny 1}

\path(11000,8000)(14000,8000)
\put(11000,7000){\arc{2000}{3.142}{4.712}}
\path(10000,7000)(10000,0000)
\put(10000,-1000){\tiny 1}

\path(13000,6000)(14000,6000)
\put(13000,5000){\arc{2000}{3.142}{4.712}}
\path(12000,5000)(12000,0000)
\put(12000,-1000){\tiny 1}

\path(6000,4000)(14000,4000)
\path(6000,2000)(14000,2000)


\blacken\path(15750,0750)(16250,0750)(16000,1250)(15750,0750)
\blacken\path(17750,0750)(18250,0750)(18000,1250)(17750,0750)
\blacken\path(19750,0750)(20250,0750)(20000,1250)(19750,0750)
\blacken\path(21750,0750)(22250,0750)(22000,1250)(21750,0750)

\path(16000,0000)(16000,22000)
\put(15500,-2000){\tiny $y_1$}
\put(16000,-1000){\tiny 0}
\put(16000,22500){\tiny 0}

\path(18000,0000)(18000,22000)
\put(18000,-1000){\tiny 0}
\put(18000,22500){\tiny 0}

\path(20000,0000)(20000,22000)
\put(20000,-1000){\tiny 0}
\put(20000,22500){\tiny 0}

\path(22000,0000)(22000,22000)
\put(21500,-2000){\tiny $y_L$}
\put(22000,-1000){\tiny 0}
\put(22000,22500){\tiny 0}


\blacken\path(25750,21250)(26250,21250)(26000,20750)(25750,21250)
\blacken\path(27750,21250)(28250,21250)(28000,20750)(27750,21250)
\blacken\path(29750,21250)(30250,21250)(30000,20750)(29750,21250)

\path(26000,0000)(26000,22000)
\put(25500,-2000){\tiny $z_1$}
\put(26000,-1000){\tiny 2}
\put(26000,22500){\tiny 2}

\path(28000,0000)(28000,22000)
\put(28000,-1000){\tiny 2}
\put(28000,22500){\tiny 2}

\path(30000,0000)(30000,22000)
\put(29500,-2000){\tiny $z_M$}
\put(30000,-1000){\tiny 2}
\put(30000,22500){\tiny 2}

\end{picture}

\end{minipage}
\end{center}

\caption{First lattice representation of
$S(\{\xx\}, \{\x\}, \{\b\}, \{\bb\} |
\{y\}, \{z\})$.}

\label{fig-1a}

\end{figure}

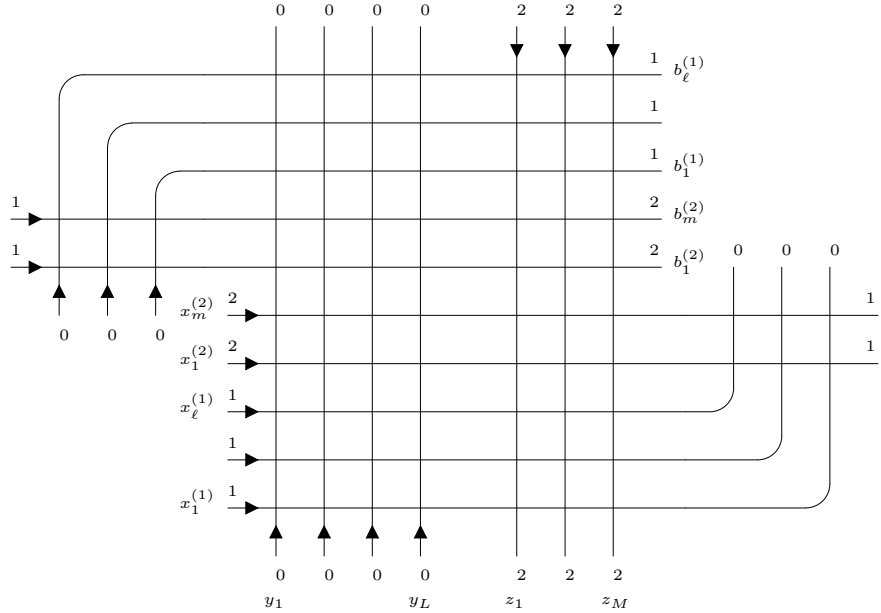
\begin{figure}

\begin{center}
\begin{minipage}{4.3in}

\setlength{\unitlength}{0.00032cm}
\begin{picture}(40000,26000)(6000,-2000)


\blacken\path(5750,14250)(5750,13750)(6250,14000)(5750,14250)
\blacken\path(5750,12250)(5750,11750)(6250,12000)(5750,12250)

\path(13000,20000)(32000,20000)
\put(31500,20500){\tiny 1}
 \put(32500,20000){\tiny $\b_{\ell}$}

\path(13000,18000)(32000,18000)
\put(31500,18500){\tiny 1}

\path(13000,16000)(32000,16000)
\put(31500,16500){\tiny 1}
 \put(32500,16000){\tiny $\b_1$}

\path(13000,14000)(32000,14000)
\put(5000,14500){\tiny 1}
 \put(31500,14500){\tiny 2}
 \put(32500,14000){\tiny $\bb_m$}

\path(13000,12000)(32000,12000)
\put(5000,12500){\tiny 1}
 \put(31500,12500){\tiny 2}
 \put(32500,12000){\tiny $\bb_1$}


\path(8000,20000)(13000,20000)
\put(8000,19000){\arc{2000}{3.142}{4.712}}
\path(7000,19000)(7000,10000)
\put(7000, 9000){\tiny 0}

\path(10000,18000)(13000,18000)
\put(10000,17000){\arc{2000}{3.142}{4.712}}
\path(9000,17000)(9000,10000)
\put(9000, 9000){\tiny 0}

\path(12000,16000)(13000,16000)
\put(12000,15000){\arc{2000}{3.142}{4.712}}
\path(11000,15000)(11000,10000)
\put(11000, 9000){\tiny 0}

\path(5000,14000)(13000,14000)
\path(5000,12000)(13000,12000)


\blacken\path(14750,10250)(14750,9750)(15250,10000)(14750,10250)
\blacken\path(14750,8250)(14750,7750)(15250,8000)(14750,8250)
\blacken\path(14750,6250)(14750,5750)(15250,6000)(14750,6250)
\blacken\path(14750,4250)(14750,3750)(15250,4000)(14750,4250)
\blacken\path(14750,2250)(14750,1750)(15250,2000)(14750,2250)

\path(14000,10000)(33000,10000)
 \put(12000,10000){\tiny $\xx_m$}
 \put(14000,10500){\tiny 2}
 \put(40500,10500){\tiny 1}

\path(14000,8000)(33000,8000)
\put(12000,8000){\tiny $\xx_1$}
 \put(14000,8500){\tiny 2}
 \put(40500,8500){\tiny 1}

\path(14000,6000)(33000,6000)
\put(12000,6000){\tiny $\x_{\ell}$}
 \put(14000,6500){\tiny 1}

\path(14000,4000)(33000,4000)
 \put(14000,4500){\tiny 1}

\path(14000,2000)(33000,2000)
\put(12000,2000){\tiny $\x_{1}$}
 \put(14000,2500){\tiny 1}


\blacken\path(6750,10750)(7250,10750)(7000,11250)(6750,10750)
\blacken\path(8750,10750)(9250,10750)(9000,11250)(8750,10750)
\blacken\path(10750,10750)(11250,10750)(11000,11250)(10750,10750)

\path(33000,10000)(41000,10000)
\path(33000,8000)(41000,8000)

\path(33000,6000)(34000,6000)
\put(34000,7000){\arc{2000}{0}{1.5708}}
\path(35000,7000)(35000,12000)
\put(35000,12500){\tiny 0}

\path(33000,4000)(36000,4000)
\put(36000,5000){\arc{2000}{0}{1.5708}}
\path(37000,5000)(37000,12000)
\put(37000,12500){\tiny 0}

\path(33000,2000)(38000,2000)
\put(38000,3000){\arc{2000}{0}{1.5708}}
\path(39000,3000)(39000,12000)
\put(39000,12500){\tiny 0}


\blacken\path(15750,0750)(16250,0750)(16000,1250)(15750,0750)
\blacken\path(17750,0750)(18250,0750)(18000,1250)(17750,0750)
\blacken\path(19750,0750)(20250,0750)(20000,1250)(19750,0750)
\blacken\path(21750,0750)(22250,0750)(22000,1250)(21750,0750)

\path(16000,0000)(16000,22000)
\put(15500,-2000){\tiny $y_1$}
\put(16000,-1000){\tiny 0}
\put(16000,22500){\tiny 0}

\path(18000,0000)(18000,22000)
\put(18000,-1000){\tiny 0}
\put(18000,22500){\tiny 0}

\path(20000,0000)(20000,22000)
\put(20000,-1000){\tiny 0}
\put(20000,22500){\tiny 0}

\path(22000,0000)(22000,22000)
\put(21500,-2000){\tiny $y_L$}
\put(22000,-1000){\tiny 0}
\put(22000,22500){\tiny 0}


\blacken\path(25750,21250)(26250,21250)(26000,20750)(25750,21250)
\blacken\path(27750,21250)(28250,21250)(28000,20750)(27750,21250)
\blacken\path(29750,21250)(30250,21250)(30000,20750)(29750,21250)

\path(26000,0000)(26000,22000)
\put(25500,-2000){\tiny $z_1$}
\put(26000,-1000){\tiny 2}
\put(26000,22500){\tiny 2}

\path(28000,0000)(28000,22000)
\put(28000,-1000){\tiny 2}
\put(28000,22500){\tiny 2}

\path(30000,0000)(30000,22000)
\put(29500,-2000){\tiny $z_M$}
\put(30000,-1000){\tiny 2}
\put(30000,22500){\tiny 2}

\end{picture}

\end{minipage}
\end{center}

\caption{Second lattice representation of
$S(\{\xx\}, \{\x\}, \{\b\}, \{\bb\} |
\{y\}, \{z\})$.}

\label{fig-1b}

\end{figure}

For a detailed discussion of the construction of $\cS_2$ for the model 
(\ref{f-antif-mon}), we refer the reader to \cite{reshetikhin}. Here we 
shall use the results of \cite{reshetikhin} without further explanation. 
Once again, we use the term \lq scalar product\rq\ to indicate the 
partition function of certain configurations in the $A_2$ vertex model.
In $A_2$ vertex model terms, the scalar product 
\begin{align}
\cS_2 \equiv 
S\ll \{x^{(2)}\},\{x^{(1)}\},\{b^{(1)}\},\{b^{(2)}\} \Big| \{y\}, \{z\} \rr
=
\Big\langle \{x^{(2)}\}, \{x^{(1)}\} \Big| \{b^{(1)}\}, \{b^{(2)}\} \Big\rangle
\end{align}
depends on six sets of variables. 
$\{x^{(1)}\} = \{x^{(1)}_1,\dots,x^{(1)}_{\ell}\}$ and 
$\{x^{(2)}\} =\{x^{(2)}_1,\dots,$ $x^{(2)}_m\}$ are free auxiliary rapidities, 
$\{b^{(1)}\} = \{b^{(1)}_1,\dots,b^{(1)}_{\ell}\}$ and 
$\{b^{(2)}\} =\{b^{(2)}_1,\dots,b^{(2)}_m\}$ 
are auxiliary rapidities which satisfy Bethe equations, and 
$\{y\} = \{y_1,\dots,y_L\}$ and 
$\{z\} = \{z_1,\dots,z_M\}$ are quantum rapidities. In order for the scalar 
product to be non-zero, we require that the cardinalities satisfy 
$\ell + m \leq L + M$. The scalar product is the partition function of the 
lattice shown in Figure {\bf\ref{fig-1a}}. Using the Yang-Baxter equation 
repeatedly, the lattice in Figure {\bf\ref{fig-1a}} can be transformed to 
that shown in Figure {\bf\ref{fig-1b}}.


Despite recent studies 
\cite{wheeler.su3,belliard.1,belliard.2,belliard.3,belliard.4}, 
no compact expression for $\cS_2$, such as a determinant, is 
known to exist, unless the variables $\{\x\}$ and $\{\xx\}$ also 
satisfy Bethe equations \cite{reshetikhin,belliard.2}. 
Alternatively, one can consider the case where the Bethe \michaelwrites{eigenvector} 
in $\cS_2$ becomes an $A_1$-like Bethe \michaelwrites{eigenvector}, which amounts to 
sending one set of Bethe variables to infinity. While less than
fully general, this case is relevant to studies of 3-point 
functions that involve operators from an $A_2$ sub-sector of 
SYM$_4$. We 
consider this case in the rest of this section, with the aim of 
recovering the results of \cite{wheeler.su3} from a vertex-model 
point of view.

We treat the partition functions in Figures {\bf\ref{fig-1a}} 
and {\bf\ref{fig-1b}} as equivalent expressions for the $A_2$ 
scalar product. When we send $\{b^{(2)}\} \rightarrow \{\infty\}$, 
we find it most useful to start from the representation in Figure 
{\bf\ref{fig-1a}}. Conversely, when we send 
$\{b^{(1)}\} \rightarrow \{\infty\}$, we start from the 
representation in Figure {\bf\ref{fig-1b}}.

\subsection{Bethe equations}

As we have already mentioned, in this section we assume that $\{\b\}$ 
and $\{\bb\}$ are Bethe roots at all times. To be precise, we assume 
that they are solutions of the nested Bethe Ansatz equations, which 
for the model under consideration are given by

\begin{align}
\prod_{j\not=i}^{\ell}
\frac{\b_i-\b_j+1}{\b_i-\b_j-1}
&=
\prod_{j=1}^{L}
\frac{\b_i-y_j+1}{\b_i-y_j}
\prod_{k=1}^{m}
\frac{\b_i-\bb_k}{\b_i-\bb_k-1}
\label{A_2-bethe-1}
\\
\prod_{j\not=i}^{m}
\frac{\bb_i-\bb_j+1}{\bb_i-\bb_j-1}
&=
\prod_{j=1}^{M}
\frac{\bb_i-z_j}{\bb_i-z_j-1}
\prod_{k=1}^{\ell}
\frac{\bb_i-\b_k+1}{\bb_i-\b_k}
\label{A_2-bethe-2}
\end{align}
When we take the limit in which a set of Bethe roots $\{\b\}$ or $\{\bb\}$ tends to infinity, this causes a simplification of the Bethe equations (\ref{A_2-bethe-1}) and 
(\ref{A_2-bethe-2}). Namely, in this limit, one set of equations trivializes and the remaining set becomes of $A_1$-type. We obtain
\begin{align}
\prod_{j\not=i}^{\ell}
\frac{\b_i-\b_j+1}{\b_i-\b_j-1}
&=
\prod_{j=1}^{L}
\frac{\b_i-y_j+1}{\b_i-y_j},
\quad\quad
\text{when}\ \{\bb\} \rightarrow \{\infty\}
\label{A_2-bethe-deg-1}
\\
\prod_{j\not=i}^{m}
\frac{\bb_i-\bb_j+1}{\bb_i-\bb_j-1}
&=
\prod_{j=1}^{M}
\frac{\bb_i-z_j}{\bb_i-z_j-1},
\quad\quad
\text{when}\ \{\b\} \rightarrow \{\infty\}
\label{A_2-bethe-deg-2}
\end{align}
Equations (\ref{A_2-bethe-deg-1}) are precisely the Bethe equations (\ref{A_1-bethe}) for an $A_1$ XXX spin chain built entirely from fundamental representations of the universal $R$-matrix, while (\ref{A_2-bethe-deg-2}) are those for a spin chain built entirely from anti-fundamental representations. 

\subsection{Change of normalization}

In this section it is most convenient to use the normalization in which the $b$ weights are equal to 1, rather than the $a$ weights. Specifically, we assume that
\begin{align}
a(x,y) = \frac{x-y+1}{x-y},
\quad\quad
b(x,y) = 1,
\quad\quad
c(x,y) = \frac{1}{x-y}
\end{align}
All formulae which we use from previous sections, which were based on the normalization with $a(x,y) =1$, must be renormalized appropriately. 

\subsection{Trivializing the $\{\bb\}$ lines}

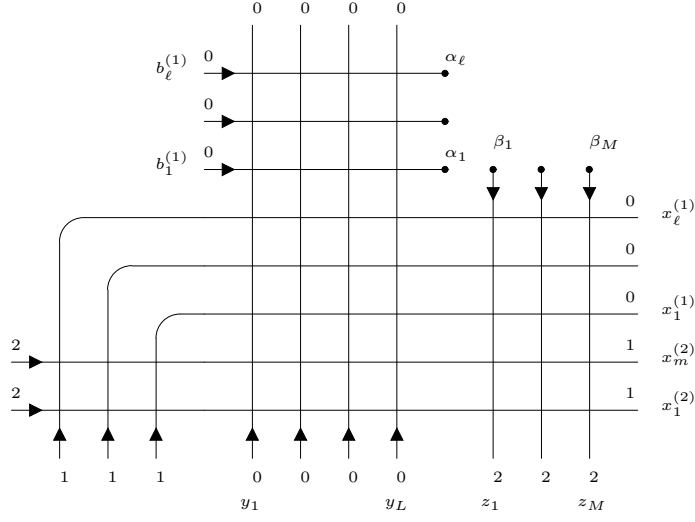
\begin{figure}

\begin{center}
\begin{minipage}{4.3in}

\setlength{\unitlength}{0.00032cm}
\begin{picture}(40000,22000)(4000,-2000)


\blacken\path(14750,16250)(14750,15750)(15250,16000)(14750,16250)
\blacken\path(14750,14250)(14750,13750)(15250,14000)(14750,14250)
\blacken\path(14750,12250)(14750,11750)(15250,12000)(14750,12250)

\path(14000,16000)(24000,16000)
 \put(12000,16000){\tiny $\b_{\ell}$}
 \put(14000,16500){\tiny 0}
\put(24000,16000){\circle*{250}}
\put(24000,16500){\tiny $\alpha_{\ell}$}

\path(14000,14000)(24000,14000)
\put(14000,14500){\tiny 0}
\put(24000,14000){\circle*{250}}

\path(14000,12000)(24000,12000)
 \put(12000,12000){\tiny $\b_{1}$}
 \put(14000,12500){\tiny 0}
\put(24000,12000){\circle*{250}}
\put(24000,12500){\tiny $\alpha_1$}


\blacken\path(6750,4250)(6750,3750)(7250,4000)(6750,4250)
\blacken\path(6750,2250)(6750,1750)(7250,2000)(6750,2250)

\path(14000,10000)(32000,10000)
 \put(31500,10500){\tiny 0}
 \put(33000,10000){\tiny $\x_{\ell}$}

\path(14000,8000)(32000,8000)
 \put(31500,8500){\tiny 0}

\path(14000,6000)(32000,6000)
 \put(31500,6500){\tiny 0}
 \put(33000,6000){\tiny $\x_{1}$}

\path(14000,4000)(32000,4000)
 \put(6000,4500){\tiny 2}
 \put(31500,4500){\tiny 1}
 \put(33000,4000){\tiny $\xx_m$}

\path(14000,2000)(32000,2000)
 \put(6000,2500){\tiny 2}
 \put(31500,2500){\tiny 1}
 \put(33000,2000){\tiny $\xx_1$}


\blacken\path(7750,0750)(8250,0750)(8000,1250)(7750,0750)
\blacken\path(9750,0750)(10250,0750)(10000,1250)(9750,0750)
\blacken\path(11750,0750)(12250,0750)(12000,1250)(11750,0750)

\path(9000,10000)(14000,10000)
\put(9000,9000){\arc{2000}{3.142}{4.712}}
\path(8000,9000)(8000,0000)
\put(8000,-1000){\tiny 1}

\path(11000,8000)(14000,8000)
\put(11000,7000){\arc{2000}{3.142}{4.712}}
\path(10000,7000)(10000,0000)
\put(10000,-1000){\tiny 1}

\path(13000,6000)(14000,6000)
\put(13000,5000){\arc{2000}{3.142}{4.712}}
\path(12000,5000)(12000,0000)
\put(12000,-1000){\tiny 1}

\path(6000,4000)(14000,4000)
\path(6000,2000)(14000,2000)


\blacken\path(15750,0750)(16250,0750)(16000,1250)(15750,0750)
\blacken\path(17750,0750)(18250,0750)(18000,1250)(17750,0750)
\blacken\path(19750,0750)(20250,0750)(20000,1250)(19750,0750)
\blacken\path(21750,0750)(22250,0750)(22000,1250)(21750,0750)

\path(16000,0000)(16000,18000)
\put(15500,-2000){\tiny $y_1$}
\put(16000,-1000){\tiny 0}
\put(16000,18500){\tiny 0}

\path(18000,0000)(18000,18000)
\put(18000,-1000){\tiny 0}
\put(18000,18500){\tiny 0}

\path(20000,0000)(20000,18000)
\put(20000,-1000){\tiny 0}
\put(20000,18500){\tiny 0}

\path(22000,0000)(22000,18000)
\put(21500,-2000){\tiny $y_L$}
\put(22000,-1000){\tiny 0}
\put(22000,18500){\tiny 0}


\blacken\path(25750,11250)(26250,11250)(26000,10750)(25750,11250)
\blacken\path(27750,11250)(28250,11250)(28000,10750)(27750,11250)
\blacken\path(29750,11250)(30250,11250)(30000,10750)(29750,11250)

\path(26000,0000)(26000,12000)
\put(25500,-2000){\tiny $z_1$}
\put(26000,-1000){\tiny 2}
\put(26000,12000){\circle*{250}}
\put(26000,13000){\tiny $\beta_1$}

\path(28000,0000)(28000,12000)
\put(28000,-1000){\tiny 2}
\put(28000,12000){\circle*{250}}

\path(30000,0000)(30000,12000)
\put(29500,-2000){\tiny $z_M$}
\put(30000,-1000){\tiny 2}
\put(30000,12000){\circle*{250}}
\put(30000,13000){\tiny $\beta_M$}

\end{picture}

\end{minipage}
\end{center}

\caption{Lattice representation of $S(\{\xx\},\{\x\},\{\b\},\{\infty\} | \{y\}, \{z\})$. The points marked $\{\alpha_1,\dots,\alpha_{\ell},\beta_1,\dots,\beta_M\}$ are summed over the colours $\{1,2\}$, such that $\#(\alpha_k = 2)$ $+$ $\#(\beta_k = 1)$ $= m$. There is a multiplicative minus sign for every $\beta_k = 1$.}

\label{fig-2a}

\end{figure}

Let us define
\begin{multline}
S \ll \{\xx\},\{\x\},\{\b\},\{\infty\} \Big| \{y\}, \{z\} \rr
=
\\
\frac{1}{m!}
\lim_{\{\bb\} \rightarrow \{\infty\}}
\ll
\bb_m \dots \bb_1
S\ll \{\xx\},\{\x\},\{\b\},\{\bb\} \Big| \{y\}, \{z\} \rr
\rr
\end{multline}
\begin{myLemma}
\label{lemma4}
$
S(\{\xx\},\{\x\},\{\b\},\{\infty\} | \{y\}, \{z\})
$
is equal to the partition function shown in Figure {\bf\ref{fig-2a}}.
\end{myLemma}

\begin{figure}

\begin{center}
\begin{minipage}{4.3in}

\setlength{\unitlength}{0.00032cm}
\begin{picture}(40000,16000)(10000,8000)


\blacken\path(24750,20250)(24750,19750)(25250,20000)(24750,20250)
\blacken\path(24750,18250)(24750,17750)(25250,18000)(24750,18250)
\blacken\path(14750,16250)(14750,15750)(15250,16000)(14750,16250)
\blacken\path(14750,14250)(14750,13750)(15250,14000)(14750,14250)
\blacken\path(14750,12250)(14750,11750)(15250,12000)(14750,12250)

\path(24000,20000)(32000,20000)
 \put(41000,20000){\tiny $\bb_m$}
 \put(24000,20500){\tiny 1}
 \put(39500,20500){\tiny 2}

\path(24000,18000)(32000,18000)
 \put(41000,18000){\tiny $\bb_1$}
 \put(24000,18500){\tiny 1}
 \put(39500,18500){\tiny 2}

\path(14000,16000)(24000,16000)
 \put(12000,16000){\tiny $\b_{\ell}$}
 \put(14000,16500){\tiny 0}

\path(14000,14000)(24000,14000)
\put(14000,14500){\tiny 0}

\path(14000,12000)(24000,12000)
 \put(12000,12000){\tiny $\b_{1}$}
 \put(14000,12500){\tiny 0}


\path(32000,20000)(40000,20000)
\path(32000,18000)(40000,18000)

\path(24000,16000)(25000,16000)
\put(25000,17000){\arc{2000}{0}{1.5708}}
\path(26000,17000)(26000,22000)
\put(26000,22500){\tiny 1}

\path(24000,14000)(27000,14000)
\put(27000,15000){\arc{2000}{0}{1.5708}}
\path(28000,15000)(28000,22000)
\put(28000,22500){\tiny 1}

\path(24000,12000)(29000,12000)
\put(29000,13000){\arc{2000}{0}{1.5708}}
\path(30000,13000)(30000,22000)
\put(30000,22500){\tiny 1}


\blacken\path(15750,10750)(16250,10750)(16000,11250)(15750,10750)
\blacken\path(17750,10750)(18250,10750)(18000,11250)(17750,10750)
\blacken\path(19750,10750)(20250,10750)(20000,11250)(19750,10750)
\blacken\path(21750,10750)(22250,10750)(22000,11250)(21750,10750)

\path(16000,10000)(16000,18000)
\put(15500,8000){\tiny $y_1$}
\put(15850,9000){\vdots}
\put(16000,18500){\tiny 0}

\path(18000,10000)(18000,18000)
\put(18000,18500){\tiny 0}

\path(20000,10000)(20000,18000)
\put(20000,18500){\tiny 0}

\path(22000,10000)(22000,18000)
\put(21500,8000){\tiny $y_L$}
\put(21850,9000){\vdots}
\put(22000,18500){\tiny 0}


\blacken\path(33750,21250)(34250,21250)(34000,20750)(33750,21250)
\blacken\path(35750,21250)(36250,21250)(36000,20750)(35750,21250)
\blacken\path(37750,21250)(38250,21250)(38000,20750)(37750,21250)

\path(34000,10000)(34000,22000)
\put(33500,8000){\tiny $z_1$}
\put(33850,9000){\vdots}
\put(34000,22500){\tiny 2}

\path(36000,10000)(36000,22000)
\put(36000,22500){\tiny 2}

\path(38000,10000)(38000,22000)
\put(37500,8000){\tiny $z_M$}
\put(37850,9000){\vdots}
\put(38000,22500){\tiny 2}

\end{picture}

\end{minipage}
\end{center}

\caption{Top half of the scalar product, modified using the Yang-Baxter 
equation.}

\label{fig-proof-1}

\end{figure}
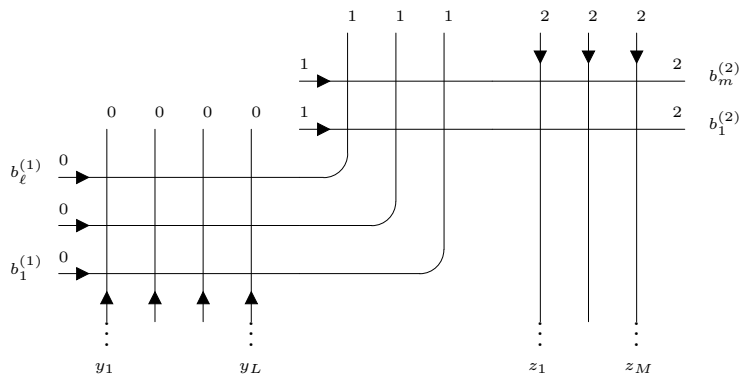

\myProof
We start from the representation of the scalar product in Figure 
{\bf\ref{fig-1a}}. Using the Yang-Baxter equation and deleting all 
frozen blocks of $b$ vertices, we transform the top half of the 
lattice to the form in Figure {\bref{fig-proof-1}}.

Consider sending $\bb_m,\dots,\bb_1$ to infinity, one at a time. $\bb_m$ 
corresponds to the top horizontal line of Figure {\bref{fig-proof-1}}, 
and we are interested in the possible positions of
$c$ vertices along this line. The $c$ vertices are
positioned either at the intersection of the $(\bb_m,\b_i)$ lines or the
$(z_i,\bb_m)$ lines. Let us refer to a $c$ vertex of the type
$1/(\bb_m-\b_i)$ as a \emph{left} $c$ vertex, and a $c$ vertex of the type
$1/(z_i-\bb_m)$ as a \emph{right} $c$ vertex.

Evidently, in the limit being taken the only configurations which contribute 
are those which have one $c$ vertex occurring in the top horizontal line. 
This rules out the possibility of the colour 0 entering this line, to leading 
order. Hence if we multiply by $\bb_m$ and take the limit 
$\bb_m \rightarrow \infty$, we trivialize the top line and produce a sum over 
the colours $\{1,2\}$ along the top of the lattice, such that 
{\bf 1.} The colours $\{\alpha_1,\dots,\alpha_{\ell}\}$ along the top left are 
summed over $\{1,2\}$ with exactly one $\alpha_k=2$, and all colours 
$\{\beta_1,\dots,\beta_M\}$ along the top right are equal to 2 (and the 
configuration is not weighted by a minus sign, because it comes from a left $c$ 
vertex), or 
{\bf 2.} All colours $\{\alpha_1,\dots,\alpha_{\ell}\}$ along the top left are 
equal to 1, and the colours $\{\beta_1,\dots,\beta_M\}$ along the top right are 
summed over $\{1,2\}$ with exactly one $\beta_k=1$ (and the configuration \emph{is} 
weighted by a minus sign, because it comes from a right $c$ vertex).

It is easy to see that by repeating this procedure over all $\{\bb\}$, 
we trivialize this upper block of vertices and arrive at the configuration 
shown in Figure {\bref{fig-2a}}, with precisely the sum indicated in the caption. 
Notice that the factor $1/m!$ eliminates multiple-countings which arise from 
the successive limits. 

\proofend


\subsection{Partitioning of $S(\{\xx\},\{\x\},\{\b\},\{\infty\} | \{y\}, \{z\})$}

Studying Figure {\bf\ref{fig-2a}}, from colour conservation arguments the colours 0 entering at the base of the lattice give a product of $b$ weights, arising from the intersection with the lowest $m$ horizontal lines. Since the $b$ weights have weight 1, we obtain
\begin{multline}
S \ll \{\xx\},\{\x\},\{\b\},\{\infty\} \Big| \{y\}, \{z\} \rr
=
\\
\sum
Z_{[i_{\ell},\dots,i_1],[j_1,\dots,j_M]}
\ll \{\xx\} \Big| \{\x\}, \{z\} \rr
S_{[i_1,\dots,i_{\ell}],[j_1,\dots,j_M]}
\ll \{\x\},\{\b\} \Big| \{y\},\{z\} \rr
\label{towards-fact1}
\end{multline}
where the sum is over all sets of integers
$\{i_1,\dots,i_{\ell}\},\{j_1,\dots,j_M\}$ taking values in $\{1,2\}$, such
that $\#(i_k=2)+\#(j_k=1) = m$. 

\begin{figure}

\begin{center}
\begin{minipage}{4.3in}

\setlength{\unitlength}{0.00032cm}
\begin{picture}(40000,18000)(-4000,2000)



\blacken\path(-5250,12250)(-5250,11750)(-4750,12000)(-5250,12250)
\blacken\path(-5250,10250)(-5250,9750)(-4750,10000)(-5250,10250)

\path(-6000,12000)(10000,12000)
\put(-8000,12000){\tiny $\xx_m$}
\put(-6000,12500){\tiny 2}
\put(9500,12500){\tiny 1}

\path(-6000,10000)(10000,10000)
\put(-8000,10000){\tiny $\xx_1$}
\put(-6000,10500){\tiny 2}
\put(9500,10500){\tiny 1}


\blacken\path(-4250,8750)(-3750,8750)(-4000,9250)(-4250,8750)
\blacken\path(-2250,8750)(-1750,8750)(-2000,9250)(-2250,8750)
\blacken\path(-250,8750)(250,8750)(0000,9250)(-250,8750)

\path(-4000,8000)(-4000,14000)
\put(-4500,6000){\tiny $\x_{\ell}$}
\put(-4000,7000){\tiny 1}
\put(-4000,14500){\tiny $i_{\ell}$}

\path(-2000,8000)(-2000,14000)
\put(-2000,7000){\tiny 1}

\path(0000,8000)(0000,14000)
\put(-500,6000){\tiny $\x_{1}$}
\put(0000,7000){\tiny 1}
\put(0000,14500){\tiny $i_{1}$}


\blacken\path(3750,13250)(4250,13250)(4000,12750)(3750,13250)
\blacken\path(5750,13250)(6250,13250)(6000,12750)(5750,13250)
\blacken\path(7750,13250)(8250,13250)(8000,12750)(7750,13250)

\path(4000,8000)(4000,14000)
\put(3500,6000){\tiny $z_1$}
\put(4000,7000){\tiny 2}
\put(4000,14500){\tiny $j_1$}

\path(6000,8000)(6000,14000)
\put(6000,7000){\tiny 2}

\path(8000,8000)(8000,14000)
\put(7500,6000){\tiny $z_M$}
\put(8000,7000){\tiny 2}
\put(8000,14500){\tiny $j_M$}



\blacken\path(14750,16250)(14750,15750)(15250,16000)(14750,16250)
\blacken\path(14750,14250)(14750,13750)(15250,14000)(14750,14250)
\blacken\path(14750,12250)(14750,11750)(15250,12000)(14750,12250)

\path(14000,16000)(24000,16000)
 \put(12000,16000){\tiny $\b_{\ell}$}
 \put(14000,16500){\tiny 0}
\put(24000,16000){\circle*{250}}
\put(24000,16500){\tiny $\alpha_{\ell}$}

\path(14000,14000)(24000,14000)
\put(14000,14500){\tiny 0}
\put(24000,14000){\circle*{250}}

\path(14000,12000)(24000,12000)
 \put(12000,12000){\tiny $\b_{1}$}
 \put(14000,12500){\tiny 0}
\put(24000,12000){\circle*{250}}
\put(24000,12500){\tiny $\alpha_1$}


\blacken\path(14750,10250)(14750,9750)(15250,10000)(14750,10250)
\blacken\path(14750,8250)(14750,7750)(15250,8000)(14750,8250)
\blacken\path(14750,6250)(14750,5750)(15250,6000)(14750,6250)

\path(14000,10000)(32000,10000)
 \put(12000,10000){\tiny $\x_{\ell}$}
\put(14000,10500){\tiny $i_{\ell}$}
 \put(31500,10500){\tiny 0}

\path(14000,8000)(32000,8000)
 \put(31500,8500){\tiny 0}

\path(14000,6000)(32000,6000)
 \put(12000,6000){\tiny $\x_{1}$}
\put(14000,6500){\tiny $i_{1}$}
 \put(31500,6500){\tiny 0}


\blacken\path(15750,4750)(16250,4750)(16000,5250)(15750,4750)
\blacken\path(17750,4750)(18250,4750)(18000,5250)(17750,4750)
\blacken\path(19750,4750)(20250,4750)(20000,5250)(19750,4750)
\blacken\path(21750,4750)(22250,4750)(22000,5250)(21750,4750)

\path(16000,4000)(16000,18000)
\put(15500,2000){\tiny $y_1$}
\put(16000,3000){\tiny 0}
\put(16000,18500){\tiny 0}

\path(18000,4000)(18000,18000)
\put(18000,3000){\tiny 0}
\put(18000,18500){\tiny 0}

\path(20000,4000)(20000,18000)
\put(20000,3000){\tiny 0}
\put(20000,18500){\tiny 0}

\path(22000,4000)(22000,18000)
\put(21500,2000){\tiny $y_L$}
\put(22000,3000){\tiny 0}
\put(22000,18500){\tiny 0}


\blacken\path(25750,11250)(26250,11250)(26000,10750)(25750,11250)
\blacken\path(27750,11250)(28250,11250)(28000,10750)(27750,11250)
\blacken\path(29750,11250)(30250,11250)(30000,10750)(29750,11250)

\path(26000,4000)(26000,12000)
\put(25500,2000){\tiny $z_1$}
\put(26000,3000){\tiny $j_1$}
\put(26000,12000){\circle*{250}}
\put(26000,13000){\tiny $\beta_1$}

\path(28000,4000)(28000,12000)
\put(28000,12000){\circle*{250}}

\path(30000,4000)(30000,12000)
\put(29500,2000){\tiny $z_M$}
\put(30000,3000){\tiny $j_M$}
\put(30000,12000){\circle*{250}}
\put(30000,13000){\tiny $\beta_M$}

\end{picture}

\end{minipage}
\end{center}

\caption{The partition function 
$Z_{[i_{\ell},\dots,i_1],[j_1,\dots,j_M]}(\{\xx\} | \{\x\}, \{z\})$ 
is on the left. 
The partition function 
$S_{[i_1,\dots,i_{\ell}],[j_1,\dots,j_M]}(\{\x\},\{\b\}|\{y\},\{z\})$
is on the right.}

\label{fig-3a}

\end{figure}
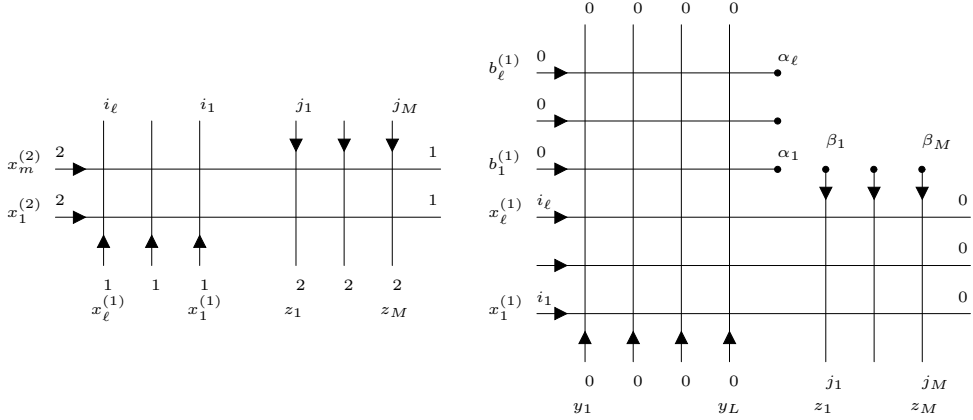

$Z_{[i_{\ell},\dots,i_1],[j_1,\dots,j_M]}$ $(\{\xx\} | \{\x\}, \{z\})$
and
$S_{[i_1,\dots,i_{\ell}],[j_1,\dots,j_M]}$ $(\{\x\},\{\b\} | \{y\}, \{z\})$
are the partition functions on the left and right of Figure {\bf\ref{fig-3a}}, 
respectively.


\subsection{Colour-independence of $S_{[i_1,\dots,i_{\ell}],[j_1,\dots,j_M]}
(\{\x\}, \{\b\} | \{y\}, \{z\})$}

\begin{myLemma}
For all fixed choices of $\{i_1,\dots,i_{\ell}\},\{j_1,\dots,j_M\}$ such that
$\#(i_k=2)+\#(j_k=1)=m$, we have
\begin{align}
\label{colour-inv1}
S_{[i_1,\dots,i_{\ell}],[j_1,\dots,j_M]}
\ll \{\x\},\{\b\} \Big| \{y\},\{z\} \rr
=
(-)^{\#(j_k=1)}
S\ll \{\x\},\{\b\} \Big| \{y\} \rr
\end{align}
where the right hand side of (\ref{colour-inv1}) is an $A_1$ scalar 
product of the type in Figure {\bf\ref{fig-sp}}.
\end{myLemma}

\myProof We consider the colours $\{\beta_1,\dots,\beta_M\}$ which enter 
the right part of the lattice. Since none of these colours are equal to 
0, by colour-conservation we find that 
$\beta_k = j_k$ for all $1\leq k \leq M$. This trivializes the sum over 
$\{\beta_1,\dots,\beta_M\}$ and constrains the right part of the lattice 
to be a product of $b$ weights. We also pick up the multiplicative sign 
$(-)^{\#(\beta_k=1)} = (-)^{\#(j_k=1)}$.

It follows that 
$S_{[i_1,\dots,i_{\ell}],[j_1,\dots,j_M]} (\{\x\},\{\b\}|\{y\},\{z\})$ 
does not genuinely depend on $\{z\}$, and the left part of the lattice 
is a coloured scalar product of the type in Figure {\bref{fig-csp}}. 
As we have already seen, this coloured scalar product is 
colour-invariant, and equal to $S(\{\x\},\{\b\} | \{y\})$.

\proofend

Returning to the sum (\ref{towards-fact1}), 
$S_{[i_1,\dots,i_{\ell}],[j_1,\dots,j_M]}(\{\x\},\{\b\}|\{y\},\{z\})$ 
is virtually a constant with respect to the summation, since 
$S(\{\x\},\{\b\}|\{y\})$ is a common factor to all terms. Therefore 
we turn to computing 
$\sum
(-)^{\#(j_k=1)}
Z_{[i_{\ell},\dots,i_1],[j_1,\dots,j_M]}$ $(\{\xx\} | \{\x\},\{z\})$, 
which is described in the next subsection.

\subsection{Calculation of $\sum
(-)^{\#(j_k=1)}
Z_{[i_{\ell},\dots,i_1],[j_1,\dots,j_M]}
(\{\xx\} | \{\x\},\{z\})
$}

We write 
\begin{multline}
\label{remaining-fact1}
\sum
(-)^{\#(j_k=1)}
Z_{[i_{\ell},\dots,i_1],[j_1,\dots,j_M]}
\ll \{\xx\} \Big| \{\x\}, \{z\} \rr
=
\\
\frac{1}{m!}
\lim_{\{\bb\} \rightarrow \{\infty\}}
\ll
\bb_m \dots \bb_1
S\ll \{\xx\},\{\bb\} \Big| \{\x\},\{z\} \rr
\rr
\end{multline}
where the right hand side of (\ref{remaining-fact1}) is the 
degeneration of an $A_1$ scalar product. Full details 
are provided in Appendix {\bf A}, where the right hand side 
is expressed in determinant form. Observe that we have 
reintroduced the variables $\{\bb\}$ for purely aesthetic 
reasons, and could have called them anything since they are 
dummy variables. 

\subsection{Factorization of
$S(\{\xx\},\{\x\},\{\b\},\{\infty\} | \{y\}, \{z\})$}

Combining the results (\ref{colour-inv1}) and (\ref{remaining-fact1}), 
we see that (\ref{towards-fact1}) can be evaluated as
\begin{multline}
S \ll \{\xx\},\{\x\},\{\b\},\{\infty\} \Big| \{y\}, \{z\} \rr
=
\\
S \ll \{\x\},\{\b\} \Big| \{y\} \rr
\frac{1}{m!}
\lim_{\{\bb\} \rightarrow \{\infty\}}
\ll
\bb_m \dots \bb_1
S \ll \{\xx\},\{\bb\} \Big| \{\x\},\{z\} \rr
\rr
\end{multline}
where both factors are $A_1$ scalar products, or a degeneration 
thereof. Due to the Bethe equations in the 
$\{\bb\} \rightarrow \{\infty\}$ regime (\ref{A_2-bethe-deg-1}), the 
first factor can be evaluated as a Slavnov determinant. The second 
factor is also a determinant, given by equation (\ref{partial-1}) in 
Appendix {\bf A}. Putting these results together, we obtain
\begin{multline}
S \ll \{\xx\},\{\x\},\{\b\},\{\infty\} \Big| \{y\}, \{z\} \rr
=
\Delta^{-1}\{\x\} \Delta^{-1}\{-\b\} \Delta^{-1}\{\xx\}
\\
\times
\det\ll
\frac{1}{\b_j - \x_i}
\ll
\prod_{k\not=j}^{\ell}
(\b_k-\x_i+1)
\prod_{k=1}^{L}
\ll
\frac{\x_i-y_k+1}{\x_i-y_k}
\rr
-
\prod_{k\not=j}^{\ell}
(\b_k-\x_i-1)
\rr
\rr_{1 \leq i,j \leq \ell}
\\
\times
\det\ll
(\xx_i)^{j-1} 
\prod_{k=1}^{\ell} \ll \frac{\xx_i-\x_k+1}{\xx_i-\x_k}
\rr
-
(\xx_i+1)^{j-1}
\prod_{k=1}^{M} \ll \frac{\xx_i-z_k-1}{\xx_i-z_k} \rr 
\rr_{1 \leq i,j \leq m}
\label{fact-1}
\end{multline}
This formula is the $A_2$ vertex-model version of equation (106) in 
\cite{wheeler.su3}. Here we have given a derivation based purely on 
colour-independence in coloured vertex model partition functions. 

\subsection{Trivializing the $\{\b\}$ lines}

The second half of this section is devoted to taking the other limit, $\{\b\} \rightarrow \{\infty\}$. Let us define
\begin{multline}
S \ll \{\xx\},\{\x\},\{\infty\},\{\bb\} \Big| \{y\}, \{z\} \rr
=
\\
\frac{1}{\ell!}
\lim_{\{\b\} \rightarrow \{\infty\}}
\ll
\b_{\ell} \dots \b_1
S \ll \{\xx\},\{\x\},\{\b\},\{\bb\} \Big| \{y\}, \{z\} \rr
\rr
\end{multline}
\begin{myLemma}
$
S(\{\xx\},\{\x\},\{\infty\},\{\bb\} | \{y\}, \{z\})
$
is equal to the partition function shown in Figure {\bf\ref{fig-2b}}.

\end{myLemma}

\myProof
The proof is similar in nature to the proof of Lemma {\bf\ref{lemma4}}. 
This time we start from the representation of the scalar product in 
Figure {\bref{fig-1b}}, with its top half modified as in Figure 
{\bf\ref{fig-proof-1}}.

Consider sending $\b_{\ell},\dots,\b_1$ to infinity, one at a time. 
$\b_{\ell}$ corresponds to the top-most of the $\{\b\}$ lines, and 
we are interested in the possible positions of $c$ vertices along 
this line. The $c$ vertices are positioned either at the intersection 
of the $(\b_{\ell},y_i)$ lines or the $(\bb_i,\b_{\ell})$ lines. 
We refer to a $c$ vertex of the type $1/(\b_{\ell}-y_i)$ as a left 
$c$ vertex, and a $c$ vertex of the type $1/(\bb_i-\b_{\ell})$ as 
a right $c$ vertex.

In the limit being taken, the only configurations which contribute 
are those which have a single $c$ vertex occurring along this line. 
This eliminates the possibility of the colour 2 entering this line, 
to leading order. Hence if we multiply by $\b_{\ell}$ and take the 
limit $\b_{\ell} \rightarrow \infty$, we trivialize this line and 
produce a sum over the colours $\{0,1\}$ along the edge that the 
line formerly occupied, such that {\bf 1.} 
All colours $\{\alpha_1,\dots,\alpha_m\}$ are equal to $1$, and 
the colours $\{\beta_1,\dots,\beta_L\}$ are summed over $\{0,1\}$ 
with exactly one $\beta_k = 1$ (and the configuration is not weighted 
by a minus sign, because it comes from a left $c$ vertex), or
{\bf 2.} The colours $\{\alpha_1,\dots,\alpha_m\}$ are summed over 
$\{0,1\}$ with exactly one $\alpha_k = 0$, and all colours 
$\{\beta_1,\dots,\beta_L\}$ are equal to 0 (and the configuration 
\emph{is} weighted by a minus sign, because it comes from a right 
$c$ vertex).

Repeating this procedure over all $\{\b\}$, we ultimately arrive 
at the configuration shown in Figure {\bref{fig-2b}}, with precisely 
the sum indicated in the caption. The factor $1/\ell!$ compensates 
for multiple-countings which arise from the successive limits.

\proofend


\begin{figure}

\begin{center}
\begin{minipage}{4.3in}

\setlength{\unitlength}{0.00032cm}
\begin{picture}(40000,22000)(10000,-2000)


\blacken\path(24750,14250)(24750,13750)(25250,14000)(24750,14250)
\blacken\path(24750,12250)(24750,11750)(25250,12000)(24750,12250)

\path(24000,14000)(32000,14000)
\put(32500,14000){\tiny $\bb_m$}
\put(23500,14500){\tiny $\alpha_m$}
\put(24000,14000){\circle*{250}}
\put(31500,14500){\tiny 2}

\path(24000,12000)(32000,12000)
\put(32500,12000){\tiny $\bb_1$}
\put(23500,12500){\tiny $\alpha_1$}
\put(24000,12000){\circle*{250}}
\put(31500,12500){\tiny 2}


\blacken\path(14750,10250)(14750,9750)(15250,10000)(14750,10250)
\blacken\path(14750,8250)(14750,7750)(15250,8000)(14750,8250)
\blacken\path(14750,6250)(14750,5750)(15250,6000)(14750,6250)
\blacken\path(14750,4250)(14750,3750)(15250,4000)(14750,4250)
\blacken\path(14750,2250)(14750,1750)(15250,2000)(14750,2250)

\path(14000,10000)(33000,10000)
 \put(12000,10000){\tiny $\xx_m$}
 \put(14000,10500){\tiny 2}
 \put(40500,10500){\tiny 1}

\path(14000,8000)(33000,8000)
\put(12000,8000){\tiny $\xx_1$}
 \put(14000,8500){\tiny 2}
 \put(40500,8500){\tiny 1}

\path(14000,6000)(33000,6000)
\put(12000,6000){\tiny $\x_{\ell}$}
 \put(14000,6500){\tiny 1}

\path(14000,4000)(33000,4000)
 \put(14000,4500){\tiny 1}

\path(14000,2000)(33000,2000)
\put(12000,2000){\tiny $\x_{1}$}
 \put(14000,2500){\tiny 1}


\path(32000,10000)(41000,10000)
\path(32000,8000)(41000,8000)

\path(33000,6000)(34000,6000)
\put(34000,7000){\arc{2000}{0}{1.5708}}
\path(35000,7000)(35000,12000)
\put(35000,12500){\tiny 0}

\path(33000,4000)(36000,4000)
\put(36000,5000){\arc{2000}{0}{1.5708}}
\path(37000,5000)(37000,12000)
\put(37000,12500){\tiny 0}

\path(33000,2000)(38000,2000)
\put(38000,3000){\arc{2000}{0}{1.5708}}
\path(39000,3000)(39000,12000)
\put(39000,12500){\tiny 0}


\blacken\path(15750,0750)(16250,0750)(16000,1250)(15750,0750)
\blacken\path(17750,0750)(18250,0750)(18000,1250)(17750,0750)
\blacken\path(19750,0750)(20250,0750)(20000,1250)(19750,0750)
\blacken\path(21750,0750)(22250,0750)(22000,1250)(21750,0750)

\path(16000,0000)(16000,12000)
\put(15500,-2000){\tiny $y_1$}
\put(16000,-1000){\tiny 0}
\put(16000,12000){\circle*{250}}
\put(16000,12500){\tiny $\beta_1$}

\path(18000,0000)(18000,12000)
\put(18000,-1000){\tiny 0}
\put(18000,12000){\circle*{250}}

\path(20000,0000)(20000,12000)
\put(20000,-1000){\tiny 0}
\put(20000,12000){\circle*{250}}

\path(22000,0000)(22000,12000)
\put(21500,-2000){\tiny $y_L$}
\put(22000,-1000){\tiny 0}
\put(22000,12000){\circle*{250}}
\put(22000,12500){\tiny $\beta_L$}


\blacken\path(25750,15250)(26250,15250)(26000,14750)(25750,15250)
\blacken\path(27750,15250)(28250,15250)(28000,14750)(27750,15250)
\blacken\path(29750,15250)(30250,15250)(30000,14750)(29750,15250)

\path(26000,0000)(26000,16000)
\put(25500,-2000){\tiny $z_1$}
\put(26000,-1000){\tiny 2}
\put(26000,16500){\tiny 2}

\path(28000,0000)(28000,16000)
\put(28000,-1000){\tiny 2}
\put(28000,16500){\tiny 2}

\path(30000,0000)(30000,16000)
\put(29500,-2000){\tiny $z_M$}
\put(30000,-1000){\tiny 2}
\put(30000,16500){\tiny 2}

\end{picture}

\end{minipage}
\end{center}

\caption{Lattice representation of $S(\{\xx\},\{\x\},\{\infty\},\{\bb\} | \{y\}, \{z\})$. The points marked $\{\alpha_1,\dots,\alpha_{m},\beta_1,\dots,\beta_L\}$ are summed over the colours $\{0,1\}$, such that $\#(\alpha_k = 0)$ $+$ $\#(\beta_k = 1)$ $= \ell$. There is a multiplicative minus sign for every $\alpha_k = 0$.}

\label{fig-2b}

\end{figure}

\subsection{Partitioning of
$S(\{\xx\},\{\x\},\{\infty\},\{\bb\} | \{y\}, \{z\})$}

\begin{figure}

\begin{center}
\begin{minipage}{4.3in}

\setlength{\unitlength}{0.00032cm}
\begin{picture}(40000,14000)(-5500,6000)



\blacken\path(2750,16250)(2750,15750)(3250,16000)(2750,16250)
\blacken\path(2750,14250)(2750,13750)(3250,14000)(2750,14250)

\path(2000,16000)(10000,16000)
\put(-10000,16000){\tiny $\bb_m$}
\put(1500,16500){\tiny $\alpha_m$}
\put(2000,16000){\circle*{250}}
\put(9500,16500){\tiny 2}

\path(2000,14000)(10000,14000)
\put(-10000,14000){\tiny $\bb_1$}
\put(1500,14500){\tiny $\alpha_1$}
\put(2000,14000){\circle*{250}}
\put(9500,14500){\tiny 2}


\blacken\path(-7250,12250)(-7250,11750)(-6750,12000)(-7250,12250)
\blacken\path(-7250,10250)(-7250,9750)(-6750,10000)(-7250,10250)

\path(-8000,12000)(10000,12000)
\put(-10000,12000){\tiny $\xx_m$}
\put(-8000,12500){\tiny 2}
\put(9500,12500){\tiny $i_m$}

\path(-8000,10000)(10000,10000)
\put(-10000,10000){\tiny $\xx_1$}
\put(-8000,10500){\tiny 2}
\put(9500,10500){\tiny $i_1$}


\blacken\path(-6250,8750)(-5750,8750)(-6000,9250)(-6250,8750)
\blacken\path(-4250,8750)(-3750,8750)(-4000,9250)(-4250,8750)
\blacken\path(-2250,8750)(-1750,8750)(-2000,9250)(-2250,8750)
\blacken\path(-250,8750)(250,8750)(0000,9250)(-250,8750)

\path(-6000,8000)(-6000,14000)
\put(-6000,6000){\tiny $y_1$}
\put(-6000,7000){\tiny $j_1$}
\put(-6000,14000){\circle*{250}}
\put(-6500,14500){\tiny $\beta_1$}

\path(-4000,8000)(-4000,14000)
\put(-4000,14000){\circle*{250}}

\path(-2000,8000)(-2000,14000)
\put(-2000,14000){\circle*{250}}

\path(0000,8000)(0000,14000)
\put(-500,6000){\tiny $y_L$}
\put(0000,7000){\tiny $j_L$}
\put(0000,14000){\circle*{250}}
\put(-500,14500){\tiny $\beta_L$}


\blacken\path(3750,17250)(4250,17250)(4000,16750)(3750,17250)
\blacken\path(5750,17250)(6250,17250)(6000,16750)(5750,17250)
\blacken\path(7750,17250)(8250,17250)(8000,16750)(7750,17250)

\path(4000,8000)(4000,18000)
\put(3500,6000){\tiny $z_1$}
\put(4000,7000){\tiny 2}
\put(4000,18500){\tiny 2}

\path(6000,8000)(6000,18000)
\put(6000,7000){\tiny 2}
\put(6000,18500){\tiny 2}

\path(8000,8000)(8000,18000)
\put(7500,6000){\tiny $z_M$}
\put(8000,7000){\tiny 2}
\put(8000,18500){\tiny 2}



\blacken\path(14750,14250)(14750,13750)(15250,14000)(14750,14250)
\blacken\path(14750,12250)(14750,11750)(15250,12000)(14750,12250)
\blacken\path(14750,10250)(14750,9750)(15250,10000)(14750,10250)

\path(14000,14000)(30000,14000)
\put(12000,14000){\tiny $\x_{\ell}$}
\put(14000,14500){\tiny 1}
\put(29500,14500){\tiny 0}

\path(14000,12000)(30000,12000)
\put(14000,12500){\tiny 1}
\put(29500,12500){\tiny 0}

\path(14000,10000)(30000,10000)
\put(12000,10000){\tiny $\x_1$}
\put(14000,10500){\tiny 1}
\put(29500,10500){\tiny 0}


\blacken\path(15750,8750)(16250,8750)(16000,9250)(15750,8750)
\blacken\path(17750,8750)(18250,8750)(18000,9250)(17750,8750)
\blacken\path(19750,8750)(20250,8750)(20000,9250)(19750,8750)
\blacken\path(21750,8750)(22250,8750)(22000,9250)(21750,8750)

\path(16000,8000)(16000,16000)
\put(15500,6000){\tiny $y_1$}
\put(16000,7000){\tiny 0}
\put(16000,16500){\tiny $j_1$}

\path(18000,8000)(18000,16000)
\put(18000,7000){\tiny 0}

\path(20000,8000)(20000,16000)
\put(20000,7000){\tiny 0}

\path(22000,8000)(22000,16000)
\put(21500,6000){\tiny $y_L$}
\put(22000,7000){\tiny 0}
\put(22000,16500){\tiny $j_L$}


\blacken\path(25750,15250)(26250,15250)(26000,14750)(25750,15250)
\blacken\path(27750,15250)(28250,15250)(28000,14750)(27750,15250)

\path(26000,8000)(26000,16000)
\put(25500,6000){\tiny $\xx_1$}
\put(26000,7000){\tiny 1}
\put(26000,16500){\tiny $i_1$}

\path(28000,8000)(28000,16000)
\put(27500,6000){\tiny $\xx_m$}
\put(28000,7000){\tiny 1}
\put(28000,16500){\tiny $i_m$}

\end{picture}

\end{minipage}
\end{center}

\caption{The partition function $S_{[j_1,\dots,j_{L}],[i_1,\dots,i_m]}
(\{\xx\},\{\bb\}|\{y\},\{z\})$ is on the left. 
The partition function $Z_{[j_1,\dots,j_L],[i_1,\dots,i_m]}(\{\x\} | \{y\}, \{\xx\})$
is on the right.}

\label{fig-3b}

\end{figure}
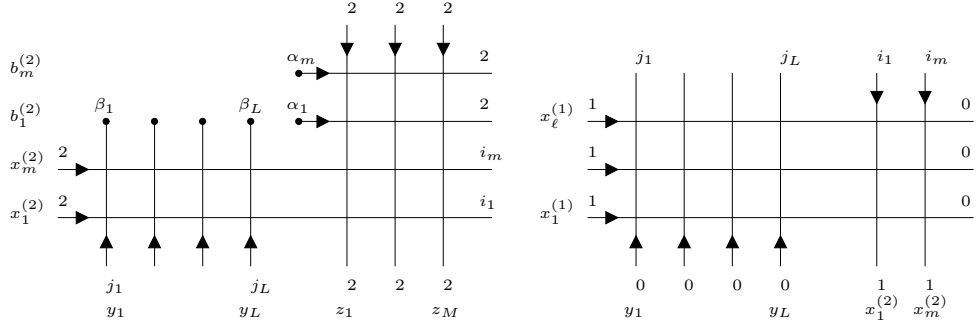

Studying Figure {\bf\ref{fig-2b}}, by colour-conservation arguments the colours 2 leaving at the base of the lattice give rise to a product of $b$ weights, arising from the intersection with the lowest $\ell$ horizontal lines. Since the $b$ weights have weight 1, we obtain
\begin{multline}
S \ll \{\xx\},\{\x\},\{\infty\},\{\bb\} \Big| \{y\}, \{z\} \rr
=
\\
\sum
S_{[j_1,\dots,j_{L}],[i_1,\dots,i_m]}
\ll \{\xx\},\{\bb\} \Big|\{y\},\{z\} \rr
Z_{[j_1,\dots,j_L],[i_1,\dots,i_m]}
\ll \{\x\} \Big| \{y\},\{\xx\} \rr
\label{towards-fact2}
\end{multline}
where the sum is over all sets of integers
$\{j_1,\dots,j_{L}\},\{i_1,\dots,i_m\}$ taking values in $\{0,1\}$, such
that $\#(i_k=0)+\#(j_k=1) = \ell$. 
$S_{[j_1,\dots,j_{L}],[i_1,\dots,i_m]}$ $(\{\xx\},\{\bb\}|\{y\},\{z\})$
and
$Z_{[j_1,\dots,j_L],[i_1,\dots,i_m]} $ $(\{\x\} |\{y\},\{\xx\})$
are the partition functions on the left and right of Figure {\bf\ref{fig-3b}}, 
respectively.


\subsection{Colour-independence of $S_{[j_1,\dots,j_{L}],[i_1,\dots,i_m]}
(\{\xx\},\{\bb\} | \{y\},\{z\})$}

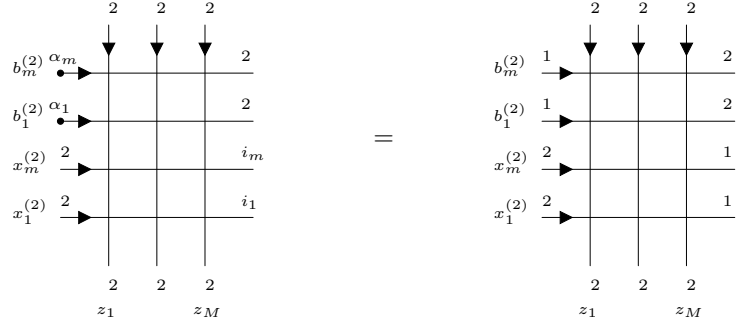
\begin{figure}

\begin{center}
\begin{minipage}{4.3in}

\setlength{\unitlength}{0.00032cm}
\begin{picture}(40000,14000)(-21000,6000)



\blacken\path(-17250,16250)(-17250,15750)(-16750,16000)(-17250,16250)
\blacken\path(-17250,14250)(-17250,13750)(-16750,14000)(-17250,14250)

\path(-18000,16000)(-10000,16000)
\put(-20000,16000){\tiny $\bb_m$}
\put(-18500,16500){\tiny $\alpha_m$}
\put(-18000,16000){\circle*{250}}
\put(-10500,16500){\tiny 2}

\path(-18000,14000)(-10000,14000)
\put(-20000,14000){\tiny $\bb_1$}
\put(-18500,14500){\tiny $\alpha_1$}
\put(-18000,14000){\circle*{250}}
\put(-10500,14500){\tiny 2}


\blacken\path(-17250,12250)(-17250,11750)(-16750,12000)(-17250,12250)
\blacken\path(-17250,10250)(-17250,9750)(-16750,10000)(-17250,10250)

\path(-18000,12000)(-10000,12000)
\put(-20000,12000){\tiny $\xx_m$}
\put(-18000,12500){\tiny 2}
\put(-10500,12500){\tiny $i_m$}

\path(-18000,10000)(-10000,10000)
\put(-20000,10000){\tiny $\xx_1$}
\put(-18000,10500){\tiny 2}
\put(-10500,10500){\tiny $i_1$}


\blacken\path(-16250,17250)(-15750,17250)(-16000,16750)(-16250,17250)
\blacken\path(-14250,17250)(-13750,17250)(-14000,16750)(-14250,17250)
\blacken\path(-12250,17250)(-11750,17250)(-12000,16750)(-12250,17250)

\path(-16000,8000)(-16000,18000)
\put(-16500,6000){\tiny $z_1$}
\put(-16000,7000){\tiny 2}
\put(-16000,18500){\tiny 2}

\path(-14000,8000)(-14000,18000)
\put(-14000,7000){\tiny 2}
\put(-14000,18500){\tiny 2}

\path(-12000,8000)(-12000,18000)
\put(-12500,6000){\tiny $z_M$}
\put(-12000,7000){\tiny 2}
\put(-12000,18500){\tiny 2}


\put(-5000,13000){$=$}




\blacken\path(2750,16250)(2750,15750)(3250,16000)(2750,16250)
\blacken\path(2750,14250)(2750,13750)(3250,14000)(2750,14250)

\path(2000,16000)(10000,16000)
\put(0000,16000){\tiny $\bb_m$}
\put(2000,16500){\tiny 1}
\put(9500,16500){\tiny 2}

\path(2000,14000)(10000,14000)
\put(0000,14000){\tiny $\bb_1$}
\put(2000,14500){\tiny 1}
\put(9500,14500){\tiny 2}


\blacken\path(2750,12250)(2750,11750)(3250,12000)(2750,12250)
\blacken\path(2750,10250)(2750,9750)(3250,10000)(2750,10250)

\path(2000,12000)(10000,12000)
\put(0000,12000){\tiny $\xx_m$}
\put(2000,12500){\tiny 2}
\put(9500,12500){\tiny 1}

\path(2000,10000)(10000,10000)
\put(0000,10000){\tiny $\xx_1$}
\put(2000,10500){\tiny 2}
\put(9500,10500){\tiny 1}


\blacken\path(3750,17250)(4250,17250)(4000,16750)(3750,17250)
\blacken\path(5750,17250)(6250,17250)(6000,16750)(5750,17250)
\blacken\path(7750,17250)(8250,17250)(8000,16750)(7750,17250)

\path(4000,8000)(4000,18000)
\put(3500,6000){\tiny $z_1$}
\put(4000,7000){\tiny 2}
\put(4000,18500){\tiny 2}

\path(6000,8000)(6000,18000)
\put(6000,7000){\tiny 2}
\put(6000,18500){\tiny 2}

\path(8000,8000)(8000,18000)
\put(7500,6000){\tiny $z_M$}
\put(8000,7000){\tiny 2}
\put(8000,18500){\tiny 2}

\end{picture}

\end{minipage}
\end{center}

\caption{Shown on the left hand side, the coloured scalar product 
configuration which 
$S_{[j_1,\dots,j_{L}],[i_1,\dots,i_m]} (\{\xx\},\{\bb\}|\{y\},\{z\})$ 
reduces to. This configuration turns out to be colour-invariant, and 
equal to the $A_1$ scalar product 
$S(\{\xx\},\{\bb\} | \emptyset, \{z\})$ on the right hand side.}

\label{fig-4b}

\end{figure}

\begin{myLemma}
For all fixed choices of 
$\{j_1,\dots,j_L\},\{i_1,\dots,i_m\}$ such that $\#(i_k = 0)+$ $\#(j_k = 1)$ $= \ell$, we have
\begin{equation}
\label{colour-inv2}
S_{[j_1,\dots,j_{L}],[i_1,\dots,i_m]}
\ll \{\xx\},\{\bb\} \Big| \{y\},\{z\} \rr
=
(-)^{\#(i_k=0)}
S\ll \{\xx\},\{\bb\} \Big| \emptyset, \{z\} \rr
\end{equation}
where the right hand side of (\ref{colour-inv2}) is the $A_1$ 
scalar product on the right of Figure {\bf\ref{fig-4b}}.
\end{myLemma}

\myProof We consider the colours $\{\beta_1,\dots,\beta_L\}$ which 
leave from the left part of the lattice. Since none of these colours 
are equal to 2, by colour-conservation we find that $\beta_k = j_k$ 
for all $1\leq k \leq L$. This trivializes the sum over 
$\{\beta_1,\dots,\beta_L\}$ and constrains the left part of the lattice 
to be a product of $b$ weights.

It follows that 
$S_{[j_1,\dots,j_{L}],[i_1,\dots,i_m]} (\{\xx\},\{\bb\}|\{y\},\{z\} )$ 
does not genuinely depend on $\{y\}$, and the right part of the lattice 
is equal to the coloured scalar product on the left of Figure 
{\bref{fig-4b}}. Although we do not prove it here, this coloured 
scalar product is colour-invariant and equal to 
$S( \{\xx\},\{\bb\} | \emptyset, \{z\} )$, on the right of 
Figure {\bref{fig-4b}}. A minus sign of 
$(-)^{\#(\alpha_k=0)} = (-)^{\#(i_k=0)}$ is remnant from the limit 
$\{\b\} \rightarrow \{\infty\}$, taken earlier.

\proofend


Returning to the sum (\ref{towards-fact2}), $S_{[j_1,\dots,j_{L}],[i_1,\dots,i_m]}
(\{\xx\},\{\bb\} |\{y\},\{z\})$ is effectively constant with respect to the summation, 
since we can extract $S(\{\xx\},\{\bb\}| \emptyset, \{z\})$ as a factor common to all 
terms. We calculate 
$\sum (-)^{\#(i_k=0)} Z_{[j_1,\dots,j_L],[i_1,\dots,i_m]}$ $(\{\x\} | \{y\}, \{\xx\})$ 
in the next subsection. 

\subsection{Calculation of
$\sum (-)^{\#(i_k=0)} Z_{[j_1,\dots,j_L],[i_1,\dots,i_m]}(\{\x\} |\{y\},\{\xx\})$}

We write
\begin{multline}
\sum
(-)^{\#(i_k=0)} 
Z_{[j_1,\dots,j_L],[i_1,\dots,i_m]}
\ll \{\x\} \Big| \{y\},\{\xx\} \rr
=
\\
\frac{1}{\ell!}
\lim_{\{\b\} \rightarrow \{\infty\}}
\ll 
\b_{\ell} \dots \b_1
S \ll \{\x\},\{\b\} \Big| \{y\},\{\xx\} \rr
\rr
\label{remaining-fact2}
\end{multline}
where the right hand side of (\ref{remaining-fact2}) is the degeneration 
of an $A_1$-type scalar product. Full details are given in Appendix {\bf A}, 
where the right hand side is expressed in determinant form.

\subsection{Factorization of
$S(\{\xx\},\{\x\},\{\infty\},\{\bb\} | \{y\}, \{z\})$}

Combining the results of (\ref{colour-inv2}) and (\ref{remaining-fact2}), 
we see that (\ref{towards-fact2}) can be evaluated as
\begin{multline}
S \ll \{\xx\},\{\x\},\{\infty\},\{\bb\} \Big| \{y\}, \{z\} \rr
=
\\
S \ll \{\xx\},\{\bb\} \Big| \emptyset, \{z\} \rr
\frac{1}{\ell!}
\lim_{\{\b\} \rightarrow \{\infty\}}
\ll 
\b_{\ell} \dots \b_1
S\ll \{\x\},\{\b\} \Big| \{y\},\{\xx\} \rr
\rr
\end{multline}
where both factors are $A_1$ scalar products, or a degeneration thereof. 
Due to the Bethe equations in the $\{\b\} \rightarrow \{\infty\}$ regime 
(\ref{A_2-bethe-deg-2}), the first factor can be evaluated as a Slavnov 
determinant, that is $\cS_1$. The second factor is also a determinant, 
given by equation (\ref{partial-2}) in Appendix {\bf A}. Using these 
results, we obtain
\begin{multline}
S
\ll \{\xx\},\{\x\},\{\infty\},\{\bb\} \Big| \{y\}, \{z\} \rr
=
\Delta^{-1}\{\xx\} \Delta^{-1}\{-\bb\} \Delta^{-1}\{\x\}
\\
\times
\det\ll
\frac{1}{\bb_j - \xx_i}
\ll
\prod_{k\not=j}^{m}
(\bb_k-\xx_i+1)
-
\prod_{k\not=j}^{m}
(\bb_k-\xx_i-1)
\prod_{k=1}^{M}
\ll
\frac{\xx_i-z_k-1}{\xx_i-z_k}
\rr
\rr
\rr_{1 \leq i,j \leq m}
\\
\times
\det\ll
(\x_i)^{j-1} \prod_{k=1}^{L} \ll \frac{\x_i-y_k+1}{\x_i-y_k} \rr
-
(\x_i+1)^{j-1} \prod_{k=1}^{m} \ll \frac{\x_i-\xx_k-1}{\x_i-\xx_k} \rr
\rr_{1 \leq i,j \leq \ell}
\label{fact-2}
\end{multline}
This formula is the $A_2$ vertex-model version of equation (110) in 
\cite{wheeler.su3}.

\section{Comments}
\label{section.comments}

Because we expect SYM$_4$ to be integrable 
\cite{beisert.review, serban.review}, we expect to be able to compute 
3-point functions, including those that involve SYM$_4$ local composite 
operators with more symmetry than $A_1$, in a tractable form. 
But computing these 3-point functions involves off-shell/on-shell 
scalar products that cannot (to the best of our knowledge) be expressed 
in determinant \michaelwrites{form}, or any other tractable form, when operators with 
more symmetry than $A_1$ are involved.

The results of \cite{caetano.su3, wheeler.su3} show that there are special 
cases of the $A_2$ and higher rank off-shell/on-shell scalar products that 
can be computed in terms of partition functions of $A_1$ vertex model 
configurations that can be expressed as determinants. 

The point of this paper is to understand the results of 
\cite{caetano.su3, wheeler.su3} in combinatorial, vertex model terms, 
and more importantly to \michaelwrites{characterize} the $A_2$ and higher rank vertex 
model configurations that are rank or colour independent and can be 
computed in $A_1$ terms.
Aside from potential relevance to integrable statistical mechanical
models, we expect our results to help in evaluating special cases of 
3-point functions that involve $A_2$ operators such as those classified 
and discussed in \cite{fjks}.

\section*{Acknowledgements} This work was supported by the Australian 
Research Council, Australia, and the Centre National de la Recherche 
Scientifique, France.
Both authors wish to thank the Institut Henri Poincar\'{e} for excellent 
hospitality while part of this work was done.

\vfill
\newpage

\appendix
\section{Partial domain wall partition functions}
\label{appendix}

\subsection{Partial domain wall partition functions as limiting 
cases of scalar products}
\label{appendix.1}

\begin{figure}

\begin{center}
\begin{minipage}{4.3in}

\setlength{\unitlength}{0.00032cm}
\begin{picture}(40000,14000)(5000,2500)


\blacken\path(14750,12250)(14750,11750)(15250,12000)(14750,12250)
\path(14000,12000)(30000,12000) \put(12000,12000){\tiny $x_N$}
\put(14000,12500){\tiny 1} 
\put(29500,12500){\tiny 0}

\blacken\path(14750,10250)(14750,9750)(15250,10000)(14750,10250)
\path(14000,10000)(30000,10000)  
\put(14000,10500){\tiny 1}  
\put(29500,10500){\tiny 0}

\blacken\path(14750,8250)(14750,7750)(15250,8000)(14750,8250)
\path(14000,8000)(30000,8000)
\put(14000,8500){\tiny 1}  
\put(29500,8500){\tiny 0}

\blacken\path(14750,6250)(14750,5750)(15250,6000)(14750,6250)
\path(14000,6000)(30000,6000) \put(12000,6000){\tiny $x_1$} 
\put(14000,6500){\tiny 1}  
\put(29500,6500){\tiny 0}


\blacken\path(15750,4750)(16250,4750)(16000,5250)(15750,4750)
\path(16000,4000)(16000,14000) 
\put(15500,2000){\tiny $y_1$}
\put(16000,3000){\tiny 0} 
\put(16000,14000){\circle*{250}}

\blacken\path(17750,4750)(18250,4750)(18000,5250)(17750,4750)
\path(18000,4000)(18000,14000)
\put(18000,3000){\tiny 0} 
\put(18000,14000){\circle*{250}}

\blacken\path(19750,4750)(20250,4750)(20000,5250)(19750,4750)
\path(20000,4000)(20000,14000)
\put(20000,3000){\tiny 0} 
\put(20000,14000){\circle*{250}}

\blacken\path(21750,4750)(22250,4750)(22000,5250)(21750,4750)
\path(22000,4000)(22000,14000)
\put(22000,3000){\tiny 0} 
\put(22000,14000){\circle*{250}}

\blacken\path(23750,4750)(24250,4750)(24000,5250)(23750,4750)
\path(24000,4000)(24000,14000) 
\put(24000,3000){\tiny 0} 
\put(24000,14000){\circle*{250}}

\blacken\path(25750,4750)(26250,4750)(26000,5250)(25750,4750)
\path(26000,4000)(26000,14000) 
\put(26000,3000){\tiny 0} 
\put(26000,14000){\circle*{250}}

\blacken\path(27750,4750)(28250,4750)(28000,5250)(27750,4750)
\path(28000,4000)(28000,14000) 
\put(27500,2000){\tiny $y_L$}
\put(28000,3000){\tiny 0} 
\put(28000,14000){\circle*{250}}

\end{picture}

\end{minipage}
\end{center}

\caption{Lattice representation of the partial domain wall partition function 
$Z(\{x\}_N | \{y\}_L)$. The state variables at the top boundary are summed over the values $\{0,1\}$.}

\label{fig-pdwpf}

\end{figure}
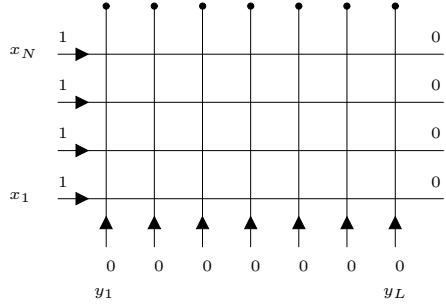

The \lq partial domain wall partition functions\rq\  
$Z(\{x\}_N | \{y\}_L)$ are partition functions of $A_1$-vertex model 
configurations of the form shown in Figure {\bref{fig-pdwpf}}.  More
precisely, they are partition functions on $N \times L$ $A_1$ 
lattice configurations, $N \leq L$, whose top boundary is summed 
over the colours $\{0,1\}$, while the state variables at the 
remaining boundaries are fixed as shown. 
A partial domain wall partition function can be viewed as 
{\bf 1.} A degenerate case of a domain wall partition function 
$Z(\{x\}_L | \{y\}_L)$, by sending 
$x_L, \dots, x_{N+1} \rightarrow \infty$, as in 
\cite{foda.wheeler.partial}, or {\bf 2.} A degenerate case of 
a scalar product $S(\{x\}_N,\{b\}_N | \{y\}_L)$, by sending 
$b_N,\dots,b_1 \rightarrow \infty$, as in 
\cite{kostov.short.paper,kostov.long.paper,foda.wheeler.partial}. 
In this work we are interested in the latter interpretation.

To obtain an explicit formula for $Z(\{x\}_N | \{y\}_L)$ one can 
start from any expression for the $A_1$ scalar product 
$S(\{x\}_N, \{b\}_N | \{y\}_L)$, such as Slavnov's determinant 
(\ref{slavnov-det}), which assumes the Bethe equations for 
$\{b\}_N$, or from the sum expression due to Izergin and Korepin 
\cite{korepin.book.1}, which does not. The final results are 
necessarily equivalent, since both expressions are valid in 
the regime $b_N,\dots, b_1 \rightarrow \infty$. Here we shall 
make use of the sum form, which is given by
\begin{multline}
S \ll \{x\}_N, \{b\}_N \Big| \{y\}_L \rr
=
\\
\sum_{\substack{
\{x\} = \{x_{\rm\tiny I}\} \cup \{x_{\rm\tiny II}\}
\\
\{b\} = \{b_{\rm\tiny I}\} \cup \{b_{\rm\tiny II}\} 
}}
\prod_{b_{\rm\tiny I}} \prod_{k=1}^{L} 
\ll \frac{b_{\rm\tiny I}-y_k+1}{b_{\rm\tiny I}-y_k} \rr
\prod_{x_{\rm\tiny II}} \prod_{k=1}^{L} 
\ll \frac{x_{\rm\tiny II}-y_k+1}{x_{\rm\tiny II}-y_k} \rr
\\
\times
\prod_{x_{\rm\tiny I},x_{\rm\tiny II}}
\ll \frac{x_{\rm\tiny I}-x_{\rm\tiny II}+1}{x_{\rm\tiny I}-x_{\rm\tiny II}} \rr
\prod_{b_{\rm\tiny I},b_{\rm\tiny II}}
\ll \frac{b_{\rm\tiny II}-b_{\rm\tiny I}+1}{b_{\rm\tiny II}-b_{\rm\tiny I}} \rr
Z \ll \{b_{\rm\tiny II}\} \Big| \{x_{\rm\tiny II}\} \rr
Z \ll \{x_{\rm\tiny I}\} \Big| \{b_{\rm\tiny I}\} \rr
\label{IK-sum}
\end{multline}
where we continue to define $S(\{x\}_N,\{b\}_N | \{y\}_L)$ as the partition function of Figure {\bref{fig-sp}}, but in the normalization in which all $b$ weights are equal to 1. 
$Z(\{b_{\rm\tiny II}\} | \{x_{\rm\tiny II}\} )$ and $Z(\{x_{\rm\tiny I}\} | \{b_{\rm\tiny I}\})$ are domain wall partition functions, as given by Figure {\bref{fig-dwpf}}, again in the normalization with all $b$ weights equal to 1. From this, one can readily compute the leading behaviour as $b_N,\dots,b_1 \rightarrow \infty$. The result of the calculation is
\begin{multline}
\frac{1}{N!}
\lim_{\{b\} \rightarrow \{\infty\}}
\ll b_N \dots b_1 S \ll \{x\}_N, \{b\}_N \Big| \{y\}_L \rr \rr
=
\\
\sum_{\{x\} = \{x_{\rm\tiny I}\} \cup \{x_{\rm\tiny II}\}}
(-)^{|\{x_{\rm\tiny I}\}|}
\prod_{x_{\rm\tiny II}} \prod_{k=1}^{L} 
\ll \frac{x_{\rm\tiny II}-y_k+1}{x_{\rm\tiny II}-y_k} \rr
\prod_{x_{\rm\tiny I},x_{\rm\tiny II}}
\ll \frac{x_{\rm\tiny I}-x_{\rm\tiny II}+1}{x_{\rm\tiny I}-x_{\rm\tiny II}} \rr
\end{multline}
The preceding sum can be recognized as the Laplace expansion of the determinant of a sum of two matrices. Therefore we obtain a compact determinant expression for the partial domain wall partition function,
\begin{multline}
Z \ll \{x\}_N \Big| \{y\}_L \rr
=
\frac{1}{N!}
\lim_{\{b\} \rightarrow \{\infty\}}
\ll b_N \dots b_1 S \ll \{x\}_N, \{b\}_N \Big| \{y\}_L \rr \rr
\\
=
\Delta^{-1}\{x\}_N
\det \ll 
x_i^{j-1} \prod_{k=1}^{L} \ll \frac{x_i-y_k+1}{x_i-y_k} \rr
-
(x_i+1)^{j-1} 
\rr_{1 \leq i,j \leq N}
\end{multline}
This expression was obtained by Kostov in 
\cite{kostov.short.paper,kostov.long.paper}, where the starting point 
of the calculation was Slavnov's determinant, rather than the sum 
expression (\ref{IK-sum}). For our purposes, the derivation from 
the sum form is essential, since in the following subsection we study 
the same limit but of scalar products which have no apparent determinant 
form.

\subsection{Scalar products with varying quantum space representations}
\label{appendix.2}

In Section {\bref{section.applications}} we encounter the objects 
(\ref{remaining-fact1}) and (\ref{remaining-fact2}) which resemble 
partial domain wall partition functions, but which arise as the 
limiting cases of more general $A_1$ scalar products. Namely, they 
come from scalar products in $A_1$ spin chains built from fundamental 
\emph{and} anti-fundamental representations of the universal $R$-matrix, 
which are shown in Figure {\bref{fig-appendix}}. 
In sum form
\footnote{In fact we are unable to express them as determinants, 
because they are not candidates for the application of Slavnov's formula. 
This is due to the fact that either $\{\x\}$ or $\{\xx\}$ appear as 
inhomogeneities in these scalar products, whereas these variables make 
no appearance in the Bethe equations.} these scalar products are given by

\begin{figure}

\begin{center}
\begin{minipage}{4.3in}

\setlength{\unitlength}{0.00032cm}
\begin{picture}(40000,17000)(-5000,6000)



\blacken\path(-5250,16250)(-5250,15750)(-4750,16000)(-5250,16250)
\blacken\path(-5250,14250)(-5250,13750)(-4750,14000)(-5250,14250)

\path(-6000,16000)(10000,16000)
\put(-8000,16000){\tiny $\bb_m$}
\put(-6000,16500){\tiny 1}
\put(9500,16500){\tiny 2}

\path(-6000,14000)(10000,14000)
\put(-8000,14000){\tiny $\bb_1$}
\put(-6000,14500){\tiny 1}
\put(9500,14500){\tiny 2}


\blacken\path(-5250,12250)(-5250,11750)(-4750,12000)(-5250,12250)
\blacken\path(-5250,10250)(-5250,9750)(-4750,10000)(-5250,10250)

\path(-6000,12000)(10000,12000)
\put(-8000,12000){\tiny $\xx_m$}
\put(-6000,12500){\tiny 2}
\put(9500,12500){\tiny 1}

\path(-6000,10000)(10000,10000)
\put(-8000,10000){\tiny $\xx_1$}
\put(-6000,10500){\tiny 2}
\put(9500,10500){\tiny 1}


\blacken\path(-4250,8750)(-3750,8750)(-4000,9250)(-4250,8750)
\blacken\path(-2250,8750)(-1750,8750)(-2000,9250)(-2250,8750)
\blacken\path(-250,8750)(250,8750)(0000,9250)(-250,8750)

\path(-4000,8000)(-4000,18000)
\put(-4500,6000){\tiny $\x_{\ell}$}
\put(-4000,7000){\tiny 1}
\put(-4000,18500){\tiny 1}

\path(-2000,8000)(-2000,18000)
\put(-2000,7000){\tiny 1}
\put(-2000,18500){\tiny 1}

\path(0000,8000)(0000,18000)
\put(-500,6000){\tiny $\x_{1}$}
\put(0000,7000){\tiny 1}
\put(0000,18500){\tiny 1}


\blacken\path(3750,17250)(4250,17250)(4000,16750)(3750,17250)
\blacken\path(5750,17250)(6250,17250)(6000,16750)(5750,17250)
\blacken\path(7750,17250)(8250,17250)(8000,16750)(7750,17250)

\path(4000,8000)(4000,18000)
\put(3500,6000){\tiny $z_1$}
\put(4000,7000){\tiny 2}
\put(4000,18500){\tiny 2}

\path(6000,8000)(6000,18000)
\put(6000,7000){\tiny 2}
\put(6000,18500){\tiny 2}

\path(8000,8000)(8000,18000)
\put(7500,6000){\tiny $z_M$}
\put(8000,7000){\tiny 2}
\put(8000,18500){\tiny 2}



\blacken\path(14750,20250)(14750,19750)(15250,20000)(14750,20250)
\blacken\path(14750,18250)(14750,17750)(15250,18000)(14750,18250)
\blacken\path(14750,16250)(14750,15750)(15250,16000)(14750,16250)

\path(14000,20000)(30000,20000)
\put(12000,20000){\tiny $\b_{\ell}$}
\put(14000,20500){\tiny 0}
\put(29500,20500){\tiny 1}

\path(14000,18000)(30000,18000)
\put(14000,18500){\tiny 0}
\put(29500,18500){\tiny 1}

\path(14000,16000)(30000,16000)
\put(12000,16000){\tiny $\b_1$}
\put(14000,16500){\tiny 0}
\put(29500,16500){\tiny 1}


\blacken\path(14750,14250)(14750,13750)(15250,14000)(14750,14250)
\blacken\path(14750,12250)(14750,11750)(15250,12000)(14750,12250)
\blacken\path(14750,10250)(14750,9750)(15250,10000)(14750,10250)

\path(14000,14000)(30000,14000)
\put(12000,14000){\tiny $\x_{\ell}$}
\put(14000,14500){\tiny 1}
\put(29500,14500){\tiny 0}

\path(14000,12000)(30000,12000)
\put(14000,12500){\tiny 1}
\put(29500,12500){\tiny 0}

\path(14000,10000)(30000,10000)
\put(12000,10000){\tiny $\x_1$}
\put(14000,10500){\tiny 1}
\put(29500,10500){\tiny 0}


\blacken\path(15750,8750)(16250,8750)(16000,9250)(15750,8750)
\blacken\path(17750,8750)(18250,8750)(18000,9250)(17750,8750)
\blacken\path(19750,8750)(20250,8750)(20000,9250)(19750,8750)
\blacken\path(21750,8750)(22250,8750)(22000,9250)(21750,8750)

\path(16000,8000)(16000,22000)
\put(15500,6000){\tiny $y_1$}
\put(16000,7000){\tiny 0}
\put(16000,22500){\tiny 0}

\path(18000,8000)(18000,22000)
\put(18000,7000){\tiny 0}
\put(18000,22500){\tiny 0}

\path(20000,8000)(20000,22000)
\put(20000,7000){\tiny 0}
\put(20000,22500){\tiny 0}

\path(22000,8000)(22000,22000)
\put(21500,6000){\tiny $y_L$}
\put(22000,7000){\tiny 0}
\put(22000,22500){\tiny 0}


\blacken\path(25750,21250)(26250,21250)(26000,20750)(25750,21250)
\blacken\path(27750,21250)(28250,21250)(28000,20750)(27750,21250)

\path(26000,8000)(26000,22000)
\put(25500,6000){\tiny $\xx_1$}
\put(26000,7000){\tiny 1}
\put(26000,22500){\tiny 1}

\path(28000,8000)(28000,22000)
\put(27500,6000){\tiny $\xx_m$}
\put(28000,7000){\tiny 1}
\put(28000,22500){\tiny 1}

\end{picture}

\end{minipage}
\end{center}

\caption{The scalar product $S( \{\xx\},\{\bb\} | \{\x\},\{z\} )$ 
is on the left. By taking the limit $\bb_m,\dots,\bb_1 \rightarrow \infty$, 
it reduces to the partition function on the left of Figure {\bref{fig-3a}}, 
with summation implied over the colours 
$\{i_1,\dots,i_{\ell}\}$, $\{j_1,\dots,j_M\}$. On the right, the scalar 
product $S( \{\x\},\{\b\} | \{y\},\{\xx\} )$. By taking the limit 
$\b_{\ell},\dots,\b_1 \rightarrow \infty$, it reduces to the partition 
function on the right of Figure {\bref{fig-3b}}, with summation implied 
over the colours $\{j_1,\dots,j_L\}$, $\{i_1,\dots,i_m\}$.}

\label{fig-appendix}

\end{figure}
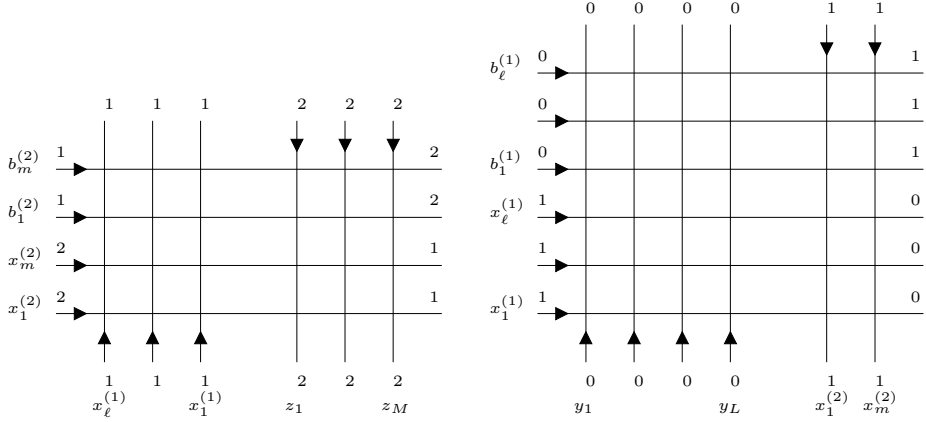
\begin{multline}
S \ll \{\xx\}, \{\bb\} \Big| \{\x\}, \{z\} \rr
=
\sum_{\substack{
\{\xx\} = \{\xx_{\rm\tiny I}\} \cup \{\xx_{\rm\tiny II}\}
\\ 
\{\bb\} = \{\bb_{\rm\tiny I}\} \cup \{\bb_{\rm\tiny II}\} 
}}
\ 
\prod_{\bb_{\rm\tiny I}} 
\prod_{k=1}^{\ell} 
\ll \frac{\bb_{\rm\tiny I}-\x_k+1}{\bb_{\rm\tiny I}-\x_k} \rr
\\
\times
\prod_{\bb_{\rm\tiny II}}
\prod_{k=1}^{M} 
\ll \frac{\bb_{\rm\tiny II}-z_k-1}{\bb_{\rm\tiny II}-z_k} \rr
\prod_{\xx_{\rm\tiny II}} 
\prod_{k=1}^{\ell} 
\ll \frac{\xx_{\rm\tiny II}-\x_k+1}{\xx_{\rm\tiny II}-\x_k} \rr
\prod_{\xx_{\rm\tiny I}}
\prod_{k=1}^{M} 
\ll \frac{\xx_{\rm\tiny I}-z_k-1}{\xx_{\rm\tiny I}-z_k} \rr
\end{multline}

\begin{multline}
\times
\prod_{\xx_{\rm\tiny I},\xx_{\rm\tiny II}}
\ll \frac{\xx_{\rm\tiny I}-\xx_{\rm\tiny II}+1}{\xx_{\rm\tiny I}-\xx_{\rm\tiny II}} \rr
\prod_{\bb_{\rm\tiny I},\bb_{\rm\tiny II}}
\ll \frac{\bb_{\rm\tiny II}-\bb_{\rm\tiny I}+1}{\bb_{\rm\tiny II}-\bb_{\rm\tiny I}} \rr
Z \ll \{\bb_{\rm\tiny II}\} \Big| \{\xx_{\rm\tiny II}\} \rr
Z \ll \{\xx_{\rm\tiny I}\} \Big| \{\bb_{\rm\tiny I}\} \rr
\nonumber
\end{multline}
which is the scalar product in (\ref{remaining-fact1}), and
\begin{multline}
S \ll \{\x\}, \{\b\} \Big| \{y\}, \{\xx\} \rr
=
\sum_{\substack{
\{\x\} = \{\x_{\rm\tiny I}\} \cup \{\x_{\rm\tiny II}\} \\ 
\{\b\} = \{\b_{\rm\tiny I}\} \cup \{\b_{\rm\tiny II}\} 
}
}
\ 
\prod_{\b_{\rm\tiny I}} 
\prod_{k=1}^{L} 
\ll \frac{\b_{\rm\tiny I}-y_k+1}{\b_{\rm\tiny I}-y_k} \rr
\\
\times
\prod_{\b_{\rm\tiny II}}
\prod_{k=1}^{m}
\ll \frac{\b_{\rm\tiny II}-\xx_k-1}{\b_{\rm\tiny II}-\xx_k} \rr
\prod_{\x_{\rm\tiny II}} 
\prod_{k=1}^{L} 
\ll \frac{\x_{\rm\tiny II}-y_k+1}{\x_{\rm\tiny II}-y_k} \rr
\prod_{\x_{\rm\tiny I}}
\prod_{k=1}^{m}
\ll \frac{\x_{\rm\tiny I}-\xx_k-1}{\x_{\rm\tiny I}-\xx_k} \rr
\\
\times
\prod_{\x_{\rm\tiny I},\x_{\rm\tiny II}}
\ll \frac{\x_{\rm\tiny I}-\x_{\rm\tiny II}+1}{\x_{\rm\tiny I}-\x_{\rm\tiny II}} \rr
\prod_{\b_{\rm\tiny I},\b_{\rm\tiny II}}
\ll \frac{\b_{\rm\tiny II}-\b_{\rm\tiny I}+1}{\b_{\rm\tiny II}-\b_{\rm\tiny I}} \rr
Z \ll \{\b_{\rm\tiny II}\} \Big| \{\x_{\rm\tiny II}\} \rr
Z \ll \{\x_{\rm\tiny I}\} \Big| \{\b_{\rm\tiny I}\} \rr
\end{multline}
which is the scalar product in (\ref{remaining-fact2}). Using these formulae, 
and proceeding in analogy with the calculation in Appendix {\bref{appendix.1}}, 
we find that
\begin{multline}
\frac{1}{m!}
\lim_{\{\bb\} \rightarrow \{\infty\}}
\ll
\bb_m \dots \bb_1
S\ll \{\xx\},\{\bb\} \Big| \{\x\},\{z\} \rr
\rr
=
\Delta^{-1}\{\xx\}
\\
\times
\det\ll
(\xx_i)^{j-1} 
\prod_{k=1}^{\ell}
\ll \frac{\xx_i-\x_k+1}{\xx_i-\x_k} \rr
-
(\xx_i+1)^{j-1}
\prod_{k=1}^{M}
\ll \frac{\xx_i-z_k-1}{\xx_i-z_k} \rr
\rr_{1 \leq i,j \leq m}
\label{partial-1}
\end{multline}
which gives rise to the second determinant in (\ref{fact-1}), and
\begin{multline}
\frac{1}{\ell!}
\lim_{\{\b\} \rightarrow \{\infty\}}
\ll
\b_{\ell} \dots \b_1
S\ll \{\x\},\{\b\} \Big| \{y\},\{\xx\} \rr
\rr
=
\Delta^{-1}\{\x\}
\\
\times
\det\ll
(\x_i)^{j-1} 
\prod_{k=1}^{L}
\ll \frac{\x_i-y_k+1}{\x_i-y_k} \rr
-
(\x_i+1)^{j-1}
\prod_{k=1}^{m}
\ll \frac{\x_i-\xx_k-1}{\x_i-\xx_k} \rr
\rr_{1 \leq i,j \leq \ell}
\label{partial-2}
\end{multline}
which gives rise to the second determinant in (\ref{fact-2}).


\vfill
\newpage

\begin{thebibliography}{99}

\bibitem{caetano.su3} 
J Caetano and P Vieira, private communication, March 2012. 

\bibitem{wheeler.su3}
M Wheeler,
{\it Scalar products in generalized models with $SU(3)$-symmetry},
{\tt arXiv:1204.2089} 

\bibitem{korepin.book.1}
V E Korepin, N M Bogoliubov, A G Izergin,
\textit{Quantum inverse scattering method and correlation 
functions},
Cambridge University Press (1993)


\bibitem{korepin.book.2}
F H L Essler, H Frahm, F Gohmann, A Klumper, V E Korepin, 
\textit{One-dimensional Hubbard model},
Cambridge University Press (2005)

\bibitem{baxter.book}
R J Baxter,
\textit{Exactly solved models in statistical mechanics},
Dover (2008)

\bibitem{korepin}
V E Korepin,
\textit{Calculation of norms of Bethe wave functions,}
Commun. Math. Phys. {\bf 86} (1982)
391--418

\bibitem{reshetikhin}
N Yu Reshetikhin,
{\it Calculation of the norm of Bethe vectors in models with 
$SU(3)$-symmetry},
Zap. Nauchn. Sem. {\bf 150} (1986) 196--213

\bibitem{slavnov}
N A Slavnov,
\textit{Calculation of scalar products of wave functions and 
form factors in the framework of the algebraic Bethe Ansatz},
Theor Math Phys {\bf 79} (1989)
502--508

\bibitem{kostov.matsuo}
I Kostov and Y Matsuo, 
\textit{Inner products of Bethe states as partial domain wall 
partition functions},
{\tt arXiv:1207.2562}

\bibitem{foda.wheeler.variations}
O Foda and M Wheeler,
\textit{Variations on Slavnov's scalar product},
JHEP {\bf 10} (2012) 096,
{\tt arXiv:1207.6871}

\bibitem{gaudin.book}
M Gaudin, 
\textit{La fonction d'onde de Bethe}, 
Masson, Paris (1983)

\bibitem{gaudin}
M Gaudin, 
\textit{Bose Gas in One Dimension. II. Orthogonality of the 
Scattering States}, 
J Math Phys {\bf 12} (1971) 1677--1680.

\bibitem{beisert.review}
N Beisert {\it et al.},
\textit{Review of AdS/CFT Integrability: An Overview},
Letters in Mathematical Physics 1 (2011) 163,
{\tt arxiv:1012.3982}, and the reviews that it introduces.

\bibitem{serban.review}
D Serban,
\textit{Integrability and the AdS/CFT correspondence},
J Phys {\bf A44} (2011) 124001
{\tt arXiv:1003.4214}


\bibitem{E1}
J Escobedo, N Gromov, A Sever and P Vieira,
\textit{Tailoring three-point functions and integrability},
J of High Energy Phys {\bf 2011} (2011) Number 9, 28
{\tt arXiv:1012.2475}

\bibitem{E2}
J Escobedo, N Gromov, A Sever and P Vieira,
\textit{Tailoring Three-Point Functions and Integrability II.
Weak/strong coupling match},
J of High Energy Phys {\bf 2011} (2011) Number 9, 29
{\tt arXiv:1104.5501}

\bibitem{GSV}
N Gromov, A Sever and P Vieira,
\textit{Tailoring Three-Point Functions and Integrability III.
Classical Tunneling},
J of High Energy Phys (2012) Number 7, 1--31,
{\tt arXiv:1111.2349}

\bibitem{foda}
O Foda,
\textit{$\N =4$ SYM structure constants as determinants},
J of High Energy Phys {\bf 2012} (2012) Number 3, 96 
{\tt arXiv:1111.4663}

\bibitem{foda.wheeler.jimbo.fest}
O Foda and M Wheeler,
{\it Slavnov determinants, Yang-Mills structure constants, 
and discrete KP}, 
in
{\it Symmetries, Integrable Systems and Representations}, 
K Iohara, S Morier-Genoud and B R\'emy, Editors, 
Springer Proceedings in Mathematics and Statistics, 
Springer, 2012
{\tt arXiv:1203.5621} 


\bibitem{ahn.foda.nepomechie}
C Ahn, O Foda and R I Nepomechie, 
{\it OPE in planar QCD from integrability}, 
J of High Energy Phys (2102) Number 6, 1--25 
{\tt arXiv:1202.6553} 

\bibitem{kostov.short.paper}
I Kostov,
\textit{Classical Limit of the Three-Point Function from Integrability},
{\tt arXiv:1203.6180}

\bibitem{kostov.long.paper}
I Kostov,
\textit{Three-point function of semiclassical states at weak coupling},
{\tt arXiv:1205.4412} 

\bibitem{foda.wheeler.partial}
O Foda, M Wheeler, 
\textit{Partial domain wall partition functions}, 
J of High Energy Phys (2012) Number 7, 186
{\tt arXiv:1205.4400} 

\bibitem{gromov.vieira.theta}
N Gromov and P Vieira,
{\it Quantum integrability for three-point functions,}
{\tt arXiv:1202.4103}

\bibitem{serban}
D Serban, 
\textit{A note on the eigenvectors of long-range spin chains 
and their scalar products}, 
{\tt arXiv:1203.5842}

\bibitem{belliard.1}
S Belliard, S Pakuliak, E Ragoucy, N A Slavnov,
\textit{Highest coefficient of scalar products in SU(3)-invariant 
integrable models}
J Stat Mech (2012) P09003,
{\tt arXiv:1206.4931} 

\bibitem{belliard.2}
S Belliard, S Pakuliak, E Ragoucy, N A Slavnov,
\textit{The algebraic Bethe ansatz for scalar products in 
SU(3)-invariant integrable models},
J Stat Mech (2012) P10017,
{\tt arXiv:1207.0956}

\bibitem{izergin}
A G Izergin,
\textit{Partition function of the six-vertex model 
in a finite volume,}
Sov Phys Dokl {\bf 32} (1987),
878--879

\bibitem{KMT}
N Kitanine, J M Maillet, and V Terras,
\textit{Form factors of the XXZ Heisenberg spin-1/2 finite chain},
Nucl Phys B {\bf 554} [FS] (1999)
647--678,
{\tt arXiv:math-ph/9807020}

\bibitem{wheeler}
M Wheeler,
\textit{An Izergin--Korepin procedure for calculating scalar 
products in six-vertex models},
Nucl Phys {\bf B852} (2011) 468-507,
{\tt arXiv:1104.2113}

\bibitem{belliard.3}
S Belliard, S Pakuliak, E Ragoucy, N A Slavnov,
\textit{Bethe vectors of SU(3)-invariant integrable models},
{\tt arXiv:1210.0768} 

\bibitem{belliard.4}
S Belliard, S Pakuliak, E Ragoucy, N A Slavnov,
\textit{Form factors in SU(3)-invariant integrable models},
{\tt arXiv:1211.3968} 

\bibitem{fjks}
O Foda, Y Jiang, I Kostov and D Serban,
\textit{A tree-level 3-point function in the su(3)-sector of planar 
$\mathcal{N} \! = \! 4$ SYM},
to appear.

\end{thebibliography}
\end{document}